\documentclass[fleqn,usenatbib]{mnras}

\usepackage{newtxtext,newtxmath}

\usepackage[T1]{fontenc}
\usepackage{ae,aecompl}

\usepackage{graphicx}	
\usepackage{subfigure}
\usepackage{bm}
\usepackage{gensymb}
\usepackage{enumitem}
\usepackage{xspace}
\usepackage{multirow}
\usepackage[dvipsnames]{xcolor}
\usepackage[utf8]{inputenc}
\usepackage{lscape}

\newcommand{\msun}{M$_\odot$}
\newcommand{\mc}{\multicolumn}
\newcommand{\mr}{\multirow}
\newcommand{\g}{{\it Gaia }}
\newcommand{\gtwo}{{\it Gaia} DR2 }
\newcommand{\sols}{\textsc{S$_{100}$}\xspace}
\newcommand{\sola}{\textsc{S$_{100}^{12}$}\xspace}
\newcommand{\solb}{\textsc{S$_{100}^{13}$}\xspace}
\newcommand{\solc}{\textsc{S$_{100}^{14}$}\xspace}
\newcommand{\solm}{\textsc{S$_{100}^{15}$}\xspace}
\newcommand{\simn}{{\it Sim N}}
\newcommand{\simm}{{\it Sim M}}

\newcommand{\rfr}[1]{\textcolor{black}{#1}}

\title[Solar neighbourhood SFH from \g]{Star formation history of the solar neighbourhood as told by \g}

\author[J. Alzate, G. Bruzual \& D. D\'iaz-Gonz\'alez]{
Jairo A. Alzate,$^{1}$\thanks{E-mail: j.alzate@irya.unam.mx}
Gustavo Bruzual,$^{1}$
Daniel J. D\'iaz-Gonz\'alez$^{1,2}$
\\
$^{1}$Instituto de Radioastronom{\'i}a y Astrof{\'i}sica, UNAM, Campus Morelia, Michoac\'an, C.P. 58089, M{\'e}xico\\
$^{2}$Shidix Technologies, E-38205, La Laguna, Santa Cruz de Tenerife, Spain
}

\date{Accepted 2020 November 9. Received 2020 October 26; in original form 2020 July 15}

\pubyear{2018}

\begin{document}
\label{firstpage}
\pagerange{\pageref{firstpage}--\pageref{lastpage}}
\maketitle

\begin{abstract}
The \gtwo catalog is the best source of stellar astrometric and photometric data available today.
The history of the Milky Way galaxy is written in stone in this data set.
Parallaxes and photometry tell us where the stars are today, when were they formed, and with what chemical content, i.e. their
star formation history (SFH).
We develop a Bayesian hierarchical model suited to reconstruct the SFH of a resolved stellar population.
We study the stars brighter than $G\,=\,15$ within 100 pc of the Sun in \gtwo and derive a SFH of the solar
neighbourhood in agreement with previous determinations and improving upon them because we
detect chemical enrichment.
Our results show a maximum of star formation activity about 10 Gyr ago, producing large numbers of stars with slightly below solar metallicity ($Z$\,=\,$0.014$), followed by a decrease in star formation up to a minimum level occurring around 8 Gyr ago. After a quiet period, star formation rises to a maximum at about 5 Gyr ago, forming stars of solar metallicity ($Z$\,=\,$0.017$). Finally, star formation has been decreasing until the present, forming stars of $Z$\,=\,$0.03$ at a residual level. We test the effects introduced in the inferred SFH by ignoring the presence of unresolved binary stars in the sample, reducing the apparent limiting magnitude, and modifying the stellar initial mass function.
\end{abstract}




\section{Introduction}

The stellar populations of the Milky Way trace different dynamical components of the Galaxy.
Even though the thin and the thick disks, the inner bar, the bulge and the stellar halo are structurally well characterized,
many open questions remain about their origin, chemical composition, time of formation and posterior evolution, i.e., their star formation history (SFH).\footnote{By SFH we mean the run of the star formation rate (SFR) with time for stars of one or more metallicities.} 
The SFHs of the different Galactic components give us clues about the origin and evolution of the Galaxy.
However, complex internal dynamical processes taking place in galaxies combined with the effects of interactions with other galaxies, e.g. merger events, determine the observed mix of stellar populations and hinder the determination of the Galaxy SFH from this mix.

The SFH of the solar neighbourhood has been derived using different techniques and data sets. 
\citet{verg02} and \citet{cig06} determined the SFH by comparing synthetic CMDs and Hipparcos data, whereas 
\citet{tremb14} and \citet{isern19} analyzed local white dwarf (WD) samples from \citet{giam12} and {\it Gaia}, respectively.
\cite{Mor19} inferred a SFH for the galactic disc using all the stars with $G\leq12$ in the \gtwo catalog and modelling 
the Milky Way with a flexible version of the Besan\c{c}on galaxy model.
From a different perspective, \cite{Sna15} calculate the SFH by fitting a chemical evolution model to the stellar abundances reported by \citet{Adi12}.
Despite the use of different methodologies and data sets, these authors find evidence of two episodes of star formation, one taking place around 10 Gyr ago, and a more recent one between 2 to 5 Gyr ago.

\begin{figure*}
\begin{center}
    \includegraphics[width=0.49\textwidth]{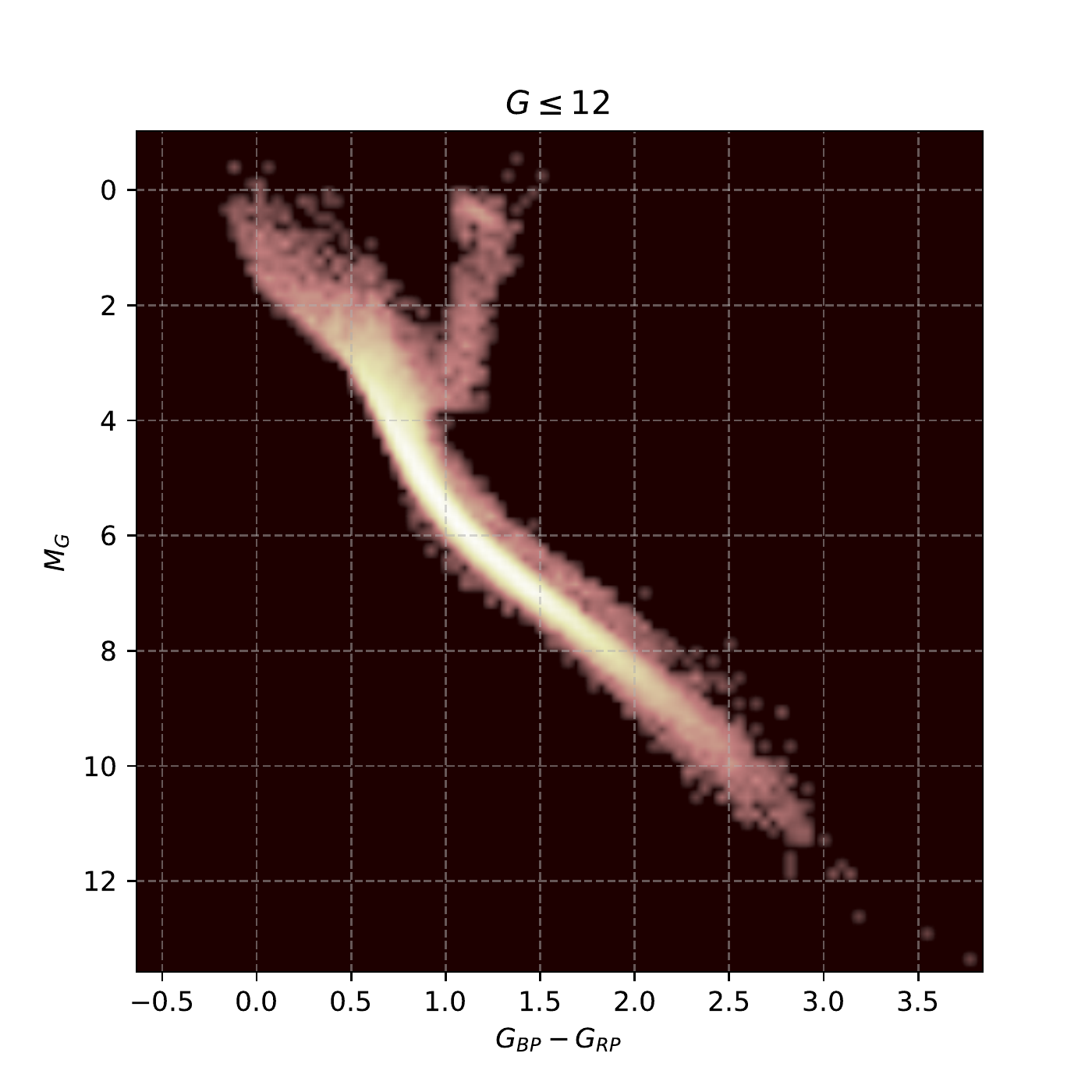}
    \includegraphics[width=0.49\textwidth]{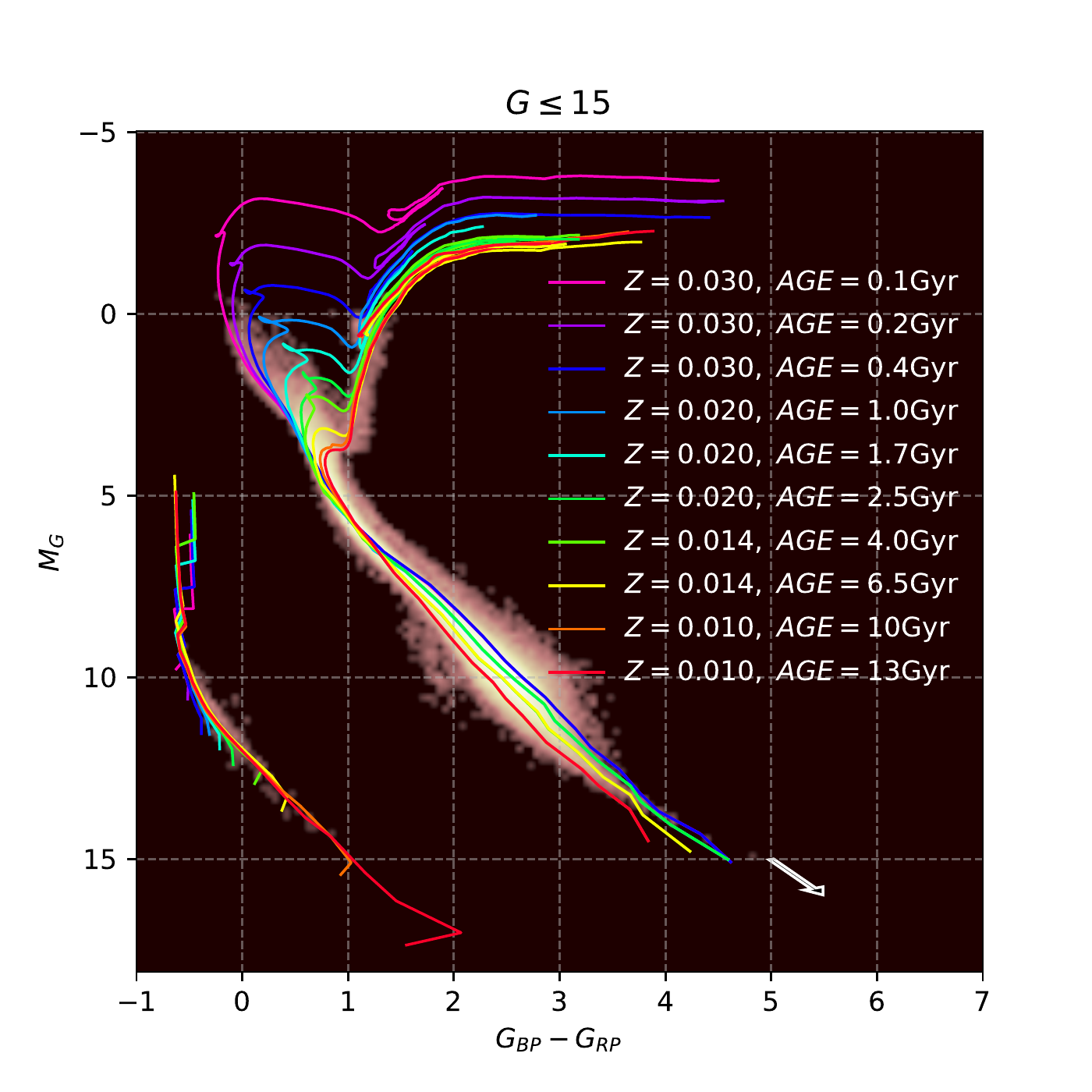}
    \caption{\label{fig:1}\rfr{\gtwo CMD of \sols ({\it gray density area}). The {\it colour lines} correspond to the isochrones used as prior information in our inference analysis (Section\,\ref{samd}).
    The white arrow indicates the reddening vector.}}
\end{center}
\end{figure*}

The \gtwo catalog contains parallaxes and three band photometry for billions of stars in the Galaxy.
The colour-magnitude diagram (CMD) of the solar neighbourhood stars observed by \g unveils a variety of substructure with invaluable information about the SFH of the Galactic thin disc (Fig. \ref{fig:1}).
Stars in a given evolutionary phase are located in a specific region in the CMD.
The relative number of stars in each of these regions provides information about the possible star formation events that 
are represented in the solar neighbourhood.
We will use the \g data in Fig. \ref{fig:1} to re-determine the SFH of the solar neighbourhood and check the consistency of our results with previous determinations.

Processing the large amounts of good quality astronomical observations from modern surveys requires big data analysis techniques that perform unbiased parameter inference, estimate correctly confidence regions, and do not require data binning (to avoid subjectivity) of samples that range from hundreds to millions of data points, often with incomplete or limited distributions.
Statistical modeling and artificial intelligence provide tools to achieve these goals.
In this paper we pose the problem of determining the parameters describing a stellar population as an inference problem, using
a Bayesian statistics framework \citep{Walms13, keith17}.
In stellar population synthesis, the chemical composition (or metallicity $Z$), the evolutionary tracks, the stellar atmospheres, the initial mass function (IMF), and the SFH are input parameters to the model.
The absolute magnitude of each individual star in different photometric bands is the model output, i.e., a synthetic CMD which will be compared to the data. The inference problem can then be solved using a Bayesian hierarchical scheme \citep{keith17}, where input and output represent population and individual properties, respectively.

In this work we infer the SFH of the solar neighbourhood for stars in different metallicity groups using a modern Bayesian approach based on statistical theory \citep{Small13,Bailer15,luri18}. Our method is free of binning, therefore we can deal with small or big numbers of stars. 
A careful error treatment \citep{luri18} provides confidence intervals for each parameter, derived as percentiles of the respective posterior distribution resulting from the Markov chain Monte Carlo (MCMC) process.
We analyze in detail the derivation of the posterior distribution in the case of magnitude-limited samples.
Our results are consistent with previous determinations, but are more general because {\it we include the stellar metallicity} in our analysis.

In Section \ref{sec_dat_desc} we describe the \g data relevant to our work and discuss some caveats. 
The Bayesian hierarchical model is described in Section \ref{sbhm} and Appendix \ref{formalism}.
The resulting age-metallicity distribution (AMD) is presented in Section \ref{samd}.
In Sections \ref{ub}, \ref{slm} and \ref{imf} we discuss the effects on the inferred AMD due to unresolved binaries, sample limiting magnitude, and assumed stellar initial mass function, respectively.
The conclusions are presented in Section \ref{concl}.
In Appendices \ref{mw.mx} and \ref{snmod} we describe and use our Milky Way model.

\section{\gtwo Data}\label{sec_dat_desc}

{\gtwo} became public in April 2018 \citep{brown18, Gaia_ref_2016}. 
Despite some issues, {\gtwo} is the best source of astrometric and photometric data available today.
Of the 1,692,919,135 observed stars: 
78.6\% have parallax and proper motion measurements.
100, 81.6, and 81.7\% of the stars have photometry in the $G,\,G_{BP}$, and $G_{RP}$ bands, respectively. 
Stellar parameters like $G$ band bolometric luminosity, surface gravity and effective temperature are also listed. For 5\% of the sample an estimate of the interstellar reddening is provided.

The following caveats when using {\gtwo} are worth mentioning.
Performing astrometric and photometric measurements in crowded fields is challenging, resulting in spurious parallaxes \citep{Lind18} and overestimated blue and red fluxes \citep{Evas18}.
This is relevant for areas of the sky with high stellar density like the Galactic plane, or when considering low mass stars, which appear in large numbers even in small regions of the sky.
Incorrect matches of the same star over different observations translate into incorrect astrometric solutions which produce inconsistent parallaxes which at first glance look reliable. 
The re-normalised unit weight error RUWE defined by \cite{Lind18} is related directly to the astrometric 
chi-square.
RUWE should be near or equal 1 for high goodness of fit astrometrical solutions.
The distribution of RUWE for the stars within 100 pc of the sun (solar sample or \sols hereafter, Fig. \ref{fig:1}) 
shows a breakpoint at 1.4, such that stellar parallaxes with RUWE $> 1.4$ should be considered spurious \citep{Lind18}\footnote{Visit \url{https://gea.esac.esa.int/archive/documentation/GDR2/Gaia_archive/chap_datamodel/sec_dm_main_tables/ssec_dm_ruwe.html} for more information.}. 
The excess flux problem \citep{Evas18} is noticeable in crowded fields and is due to a conflict between the low resolution $G_{BP}$ and $G_{RP}$ band CCDs and the high resolution $G$ band CCD, and affects mainly the faint-end of the CMD. 
When Gaia observes a small and crowded region of the sky, the blue and red detectors integrate the light of many unresolved stars falling within the $3.5\times 2.1$ arcsec$^{2}$ CCD pixel, whereas the $G$ band astrometric field detector is able to resolve the position of each star in the same region, measuring their respective $G$ magnitude. 
In consequence, overestimated blue and red flux measurements are assigned. 
Following \cite{Evas18}, it is possible to filter these stars using 
\begin{equation}\label{flx_excs}
E<A+B(G_{BP}-G_{RP})^2+C(G_{BP}-G_{RP})^3,
\end{equation}
where $E=(I_{BP}+I_{RP})/I_{G}$ is the flux excess, $I_{G},\,I_{BP}$ and $I_{RP}$ are the fluxes, and $G,\,G_{BP}$ and $G_{RP}$ 
the magnitudes in the $G,\,G_{BP}$ and $G_{RP}$ bands, respectively. For good photometric measurements, $E$ should be near 1. Eq. (\ref{flx_excs}) is used to reject stars with $E$ far from 1.

\subsection{\gtwo solar sample: \sols}\label{samp_sel_sol}

\begin{figure}
\begin{center}
    \includegraphics[width=\columnwidth]{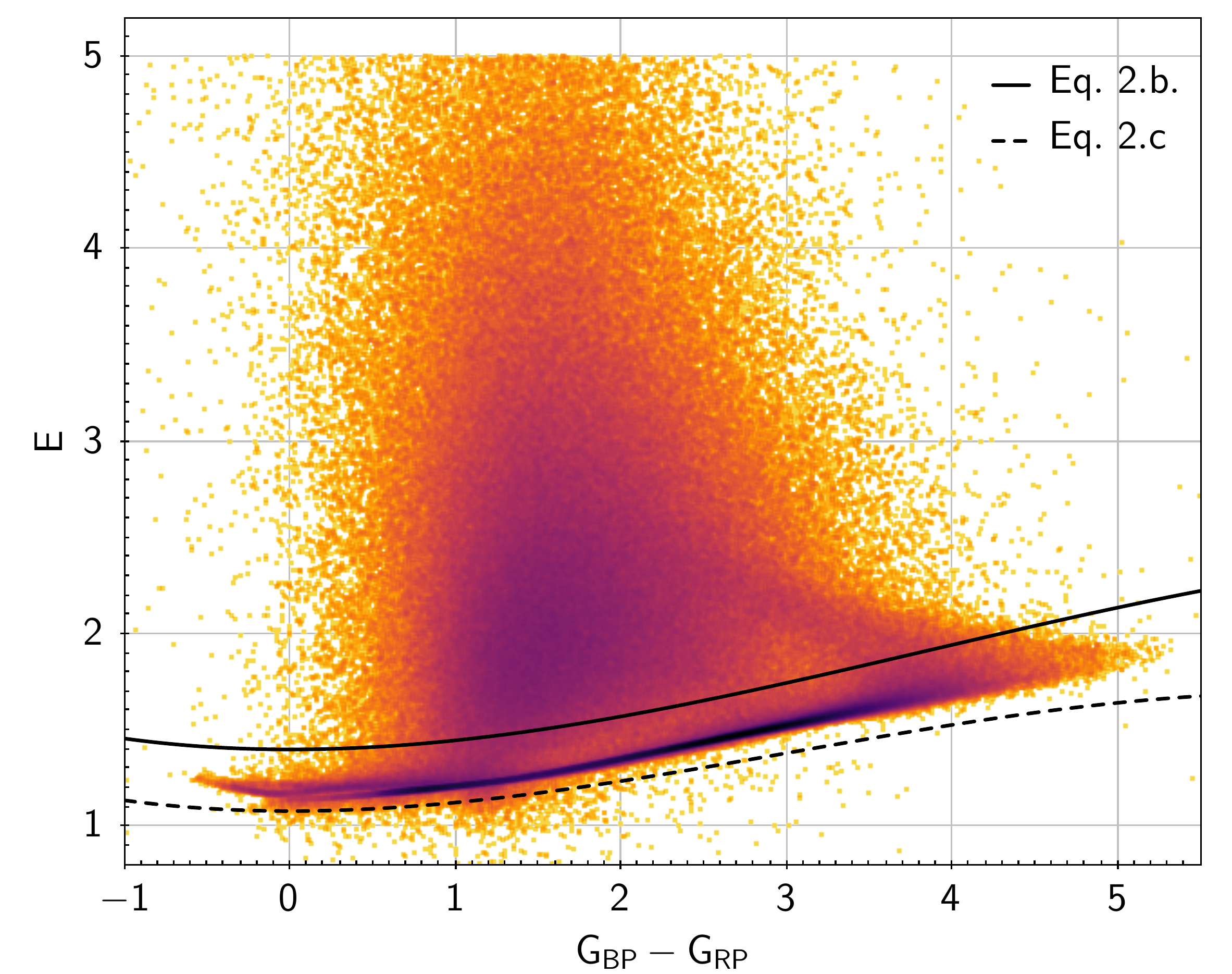}
    \caption{\label{fig:2}\gtwo flux excess plot for the stars in \sols ({\it orange density area)}.
     The {\it solid} and {\it dashed-lines} follow Eqs.\,(2b) and (2c), respectively.}
\end{center}
\end{figure}

As indicated by \cite{Lind18}, the \sols\ sample in Fig.\,\ref{fig:1} shows beautifully a main sequence (MS) belly, 
a parallel MS of unresolved binaries, and different WD cooling sequences.
These features provide useful information about the SFH of the near {\it thin} and {\it thick} discs, which makes \sols a very interesting sample.
In practice, \sols contains the \gtwo stars fulfilling:
\begin{subequations}\label{rules}
\begin{align}
\varpi \geq 10\ \rm{mas},\\
E <1.4+0.052(G_{BP}-G_{RP})^2-0.0045(G_{BP}-G_{RP})^3,\\
E >1.08+0.05(G_{BP}-G_{RP})^2-0.0055(G_{BP}-G_{RP})^3,\\
RUWE \leq 1.4,
\end{align}
\end{subequations}
\noindent
where $\varpi$ is the parallax. 
Flux excess constraints are illustrated in Fig. \ref{fig:2}. \g parallaxes and 3 band photometry are available for all the stars in \sols.
96\% of the stars have relative error in $\varpi \le$ 20\%, and 99\% error $\le$ 40\%. 
$(99,90,99)$\% of the stars have photometric error $\le$\,$(0.005,0.04,0.035)$ mag in the $(G,G_{BP},G_{RP})$ band, respectively.

\begin{figure}
\begin{center}
    \includegraphics[width=1.15\columnwidth]{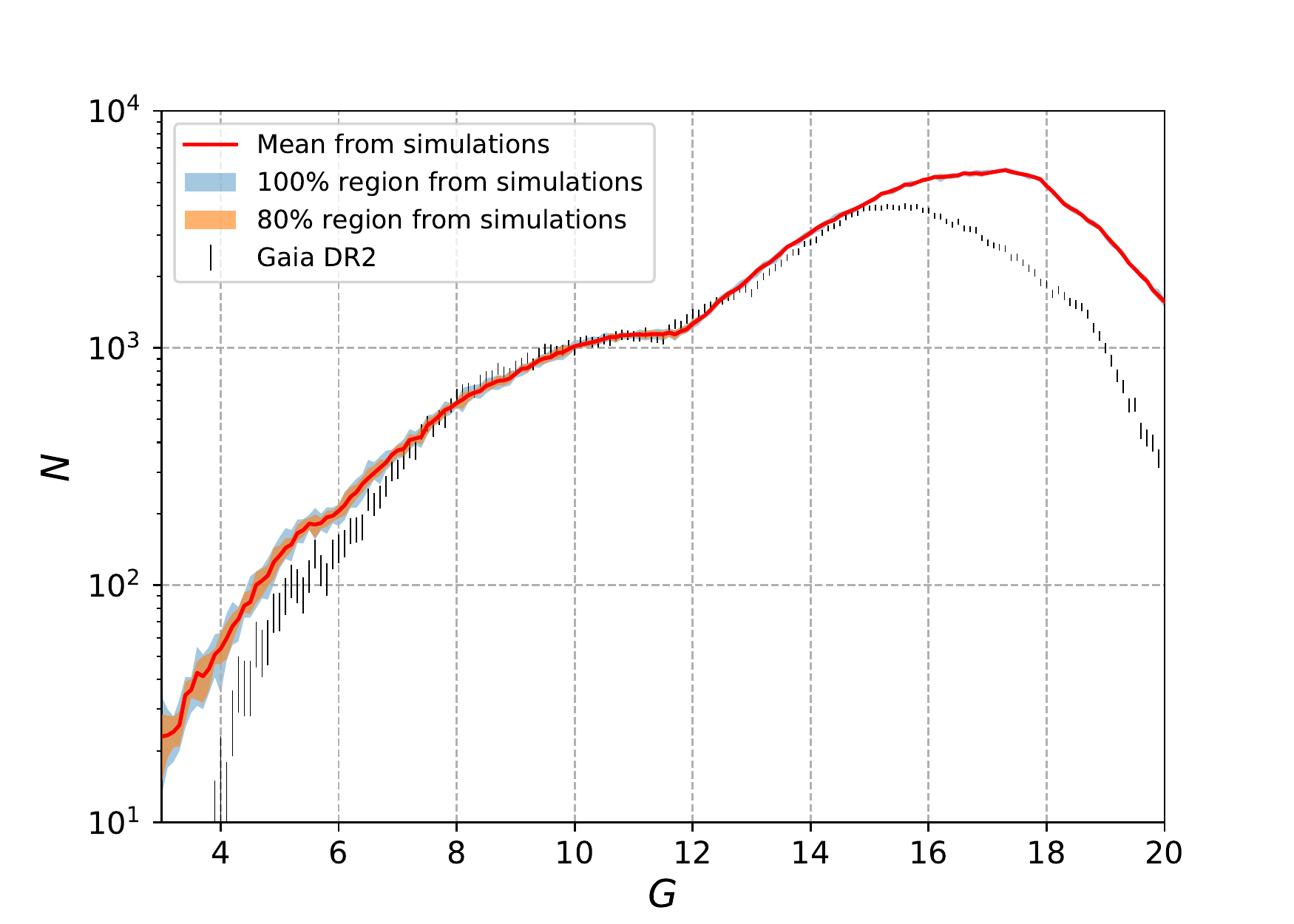}
    \caption{\label{fig:3} $G$ band counts for the \gtwo \sols sample. The error bars were obtained by a bootstrap process
    over the observed counts \citep{wall12}.
    The {\it red} line represents the mean of 10 simulations of \sols with our Milky Way model. The {\it yellow} and {\it blue} filled
    areas indicate the 80\% and 100\% confidence regions, respectively.}
\end{center}
\end{figure}

We investigate the completeness of \sols using the MW.mx model described in Appendix\,\ref{mw.mx} to compute the expected star counts in the \g bands.
We perform 10 simulations of \sols to evaluate the importance of fluctuations in the counts at the bright end, where the number of stars is low, and to estimate the completeness of \sols at the faint end. 
The expected and observed $G$-band counts are shown in Fig. \ref{fig:3}.
From the figure we see that whereas the simulated and observed counts agree well in the range $G=[7.5,12.5]$ and differ by a small and constant amount in $G=[12.5,15]$, the observed counts underestimate the expected counts at the bright and faint ends, the differences being much larger than the statistical fluctuations in our model counts.
 
There are several reasons for the \gtwo\ bright end counts to be low.
CCD saturation and calibration problems for bright stars are severe issues.
There are few bright stars which do not provide enough information to build a reliable photometric calibration at the bright end.
Bright stars quickly saturate the CCDs, forcing to stop the integration and increasing the complexity of the calibration process. 
Many photometric measurements must then be rejected, resulting in the lack of bright stars seen in Fig. \ref{fig:3}. 
Bright stars which are accepted must have their magnitudes corrected by saturation using equations in \cite{Evas18}.
Fainter than $G$\,=\,$15$ the observed counts fall considerably below the synthetic counts.
Even though many stars are lost due to the \g satellite limitations, many faint stars are rejected by the excess flux selection rule Eq. (\ref{rules}b). 

\rfr{\cite{boubert20} have studied the completeness of the \gtwo catalogue, deriving a detection probability for a star as a function of its angular position and apparent magnitude. They estimate the catalogue is nearly complete in the range $G$\,=\,$[7,20]$ in
most directions, and $G$\,=\,$[7,18]$ in some directions. From Fig.\,\ref{fig:3} we decided to restrict our analysis to the range
$G$\,=\,$[7.5,15]$, enclosed in the \cite{boubert20} completeness interval. The $G\leq15$ complete subsample of \sols (\solm hereafter) contains $N_D$\,=\,$120,452$ stars.}
 
\rfr{We use \solm to {\it derive the SFH of the solar neighbourhood} as a first application of the statistical inference algorithm proposed in this work. In a first approximation and with the aim of {\it testing the capabilities of our statistical model}, in this paper we assume {\it spherical symmetry and ignore the dependence of various quantities on $(l,b)$}. We are aware that this is not a good approximation for the youngest thin disk population since the vertical scale height, $h_z$, for these stars is of the order 
of $100$\,pc, the radius of the sphere defining \sols. Likewise, we do not attempt to characterize the transition between the thin and thick disk even though the fraction of thick disk stars at the poles in \sols\ increases with limiting magnitude,
e.g., from $G$\,=\,$12$ to $G$\,=\,$15$ \citep[][their figure 9]{haywood97}. We warn the reader about these {\it limitations of
our model} and remark that our results {\it should not be over-interpreted}. In a future paper we will include an adequate treatment of galactic structure in the solar volume, explore regions beyond $100$\,pc in the direction of the poles, and model unresolved binary stars.}

\section{Bayesian inference of the SFH}\label{sbhm}

\cite{Bailer15} and \cite{luri18} have shown that stellar distances and absolute magnitudes inferred using Bayesian statistics are more reliable than those determined with traditional methods.
This is a motivation to use Bayesian statistics to infer the properties (mass, age, metallicity) of each component of a complex stellar population, and hence the SFH of the system. 

As discussed in Appendix\,\ref{mw.mx}, the IMF, the SFH, and the mass density law $\rho(R,z)$ act as probability distribution functions which determine the mass, age, metallicity, and location of each star in the Galaxy, as well as the loci described by the stars in the HR and CMDs.
Simple stellar populations formed in instantaneous bursts trace an isochrone in the CMD.
Complex populations formed during long periods of continuous star formation or in several bursts trace multiple isochrones in the CMD, one for each star formation event.
The CMD of a complex population is then a {\it linear combination} of $N_{iso}$ isochrones \citep{Dolphin97, Small13}.
The SFH, $\rho(R,z)$ and the IMF determine the relative weight $a_i$ of the $i^{th}$ isochrone, the $i^{th}$ component of the {\it stellar mass fraction unit vector}
\begin{equation}\label{avec}
\bm{a}=\{a_{1},a_{2}, ..., a_{N_{iso}}\},\\
{\rm with}\ a_{i}\geq 0\ \ \ {\rm and}\ \ \ \sum_{i=1}^{N_{iso}}a_{i}=1.
\end{equation}
${\bm a}$ is normalized to 1 since in general the total mass in stars is unknown.
The larger $a_{i}$, the larger the number of stars described by the $i^{th}$ isochrone, and conversely.

\subsection{Bayesian hierarchical model}\label{bhm}

Under these premises we develop a Bayesian hierarchical model \citep{keith17, Wid19} designed to infer the SFH, i.e., the vector ${\bm a}$, of resolved stellar populations from their CMD. The model must be hierarchical to include the SFH as an {\it hyper-parameter} which rules how stars populate the isochrones. In contrast, the IMF and $\rho(R,z)$ enter the model as {\it priors}.
To infer the vector $\bm{a}$ we proceed as follows.
We start from the two level Bayes theorem \citep{gelman13}, written as
\begin{equation}\label{Bayes_1}
  P(\bm{\alpha}, \bm{\beta} \vert \bm{d}) \propto P(\bm{d} \vert \bm{\beta}) P(\bm{\beta}\vert \bm{\alpha})P(\bm{\alpha})
\end{equation}
\noindent
where $\bm{\alpha}$ are the hyper-parameters, $\bm{\beta}$ the parameters and $\bm{d}$ the observables.
$P(\bm{d} \vert \bm{\beta})$ is the likelihood function or error model which depends on $\bm{\alpha}$ through $\bm{\beta}$.
$P(\bm{\beta}\vert \bm{\alpha})$ is the prior distribution and $P(\bm{\alpha})$ the hyper-prior distribution.
$P(\bm{\alpha}, \bm{\beta} \vert \bm{d})$ is the posterior distribution.
If there are $N_D$ observed stars, $\bm{d}=\{d_{j=1,2, ... ,N_{D}}\}$ with respective model counterparts 
$\bm{\beta}=\{\beta_{j}\}$, Eq. (\ref{Bayes_1}) can be written as
\begin{equation}\label{Bayes_2}
    P(\bm{\alpha}, \bm{\beta} \vert \bm{d}) \propto P(\bm{\alpha}) \prod_{j=1}^{N_{D}} \ \frac{S(d_{j})P(d_{j} \vert \beta_{j}) \ P(\beta_{j}\vert \bm{\alpha})}{\ell(\bm{\alpha},S)}.
\end{equation}
\noindent
$\beta_{j}$ are parameters describing individual stars, whereas $\bm{\alpha}$ are parameters of the whole population.
In Eq. (\ref{Bayes_2}) $S(d_{j})$ is the selection function and $\ell(\bm{\alpha},S)$ the normalization constant.
In the ideal case of a complete sample, i.e., without data loss, magnitude limit, bias or selection issues, 
$S(d_{j})$ is a constant, but when incompleteness effects are non negligible, $S(d_{j})$ is a complicated function
and $\ell(\bm{\alpha},S)$ plays an important role. In the magnitude-limited case the selection function does not satisfy all the conditions to be considered a probability distribution function (PDF), and it is necessary to renormalize 
Eq. (\ref{Bayes_2}) dividing by $\ell(\bm{\alpha},S)$, taking into account the restrictions imposed by $S(d_{j})$ 
to get a formal posterior distribution.
For a given sample, the selection function modifies data through a completeness function $\mathcal{C}(l,b,G^{k}_{j})$ and/or a magnitude truncation using a Heaviside function $\mathcal{H}(G^{k}_{j})$.

\begin{figure}
\begin{center}
    \includegraphics[width=0.7\columnwidth]{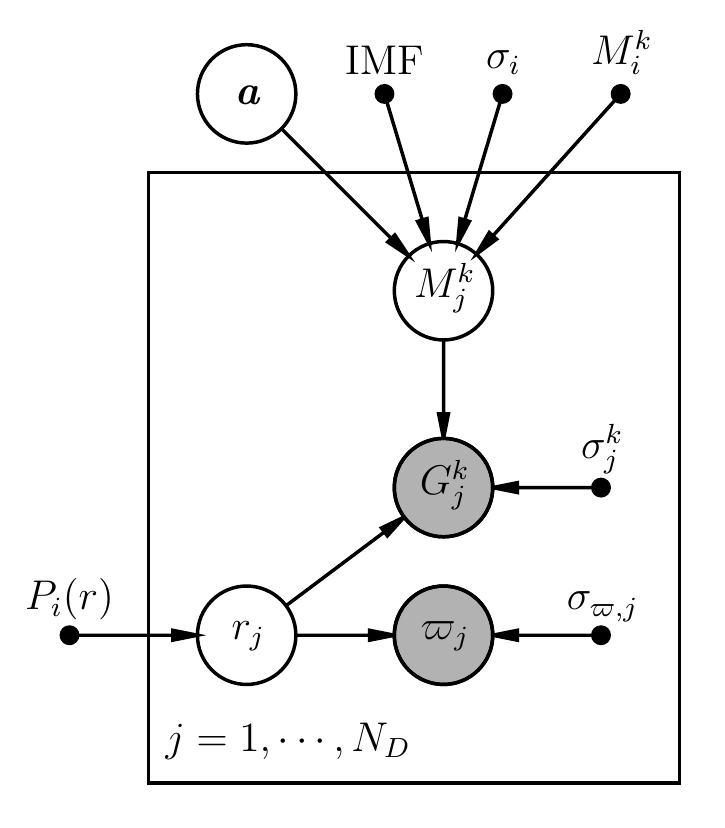}
    \caption{\label{fig:4}Schematic representation of our hierarchical model showing the relation between data ({\it shaded circles}),
    model parameters ({\it open circles}) and fixed quantities ({\it black dots}).}
\end{center}
\end{figure}

\begin{table}
\begin{center}
 \caption{\label{tab:1}Variables entering our hierarchical model.}
 \begin{tabular}{cccc}
 \hline
 \multicolumn{1}{c}{Hyper}         & \multirow{2}{*}{Parameters}      & \multirow{2}{*}{Data}        & \multicolumn{1}{c}{Fixed}      \\
 \multicolumn{1}{c}{Parameters}    &                                  &                              & \multicolumn{1}{c}{Quantities} \\
 \multicolumn{1}{c}{$\bm{\alpha}$} & \multicolumn{1}{c}{$\bm{\beta_j}$} & \multicolumn{1}{c}{$\bm{d_j}$} & \multicolumn{1}{c}{$\bm{Q}$}   \\
 \hline
     $\bm{a}$                      & $r$                              & $\varpi$,\ $\sigma_{\varpi}$     & $P_{i}(r)$                 \\
                                   & $M_{j}^{k}$                      & $G_{j}^{k}$,\ $\sigma_{j}^{k}$   & $M_{i}^{k}$,\ $\sigma_{i}^{k}$  \\
                                   &                                  &                                  & $\phi(m)$ (IMF)                           \\
                                   &                                  &                                  & \sols                      \\
 \hline
 \end{tabular}
\end{center}
\end{table}

A stellar population is a group of stars characterized by global properties like spatial distribution, kinematics, IMF and SFH. These global distributions determine individual attributes of stars, like position and brightness.
Observables, like parallax and apparent magnitude, are modeled as random samplings of the individual parameters (distance and absolute magnitude, cf. Fig. \ref{fig:4}). 
For the $j^{th}$ star in \sols the \g observables are $d_j=\{\varpi_{j}, G_{j}^{k=1,2,3}\}$, where $\varpi_{j}$ is the parallax and $G^{1}, G^{2}, G^{3}\equiv G, G_{BP}, G_{RP}$ are the star apparent magnitudes in the \g photometric system. 
The galactic coordinates ($l,b$) of each star do not enter the statistical inference if we assume that the stars in \sols are isotropically distributed. Including ($l,b$) to study other samples can be done with no problem.
To each star in the sample we associate as $\beta_{j}$ parameters the absolute magnitudes derived from the isochrones, symbolized by $\{r_{j},M_{j}^{k=1,2,3}\}$, where $r_j$ is the heliocentric distance and $k=1,\,2,\,3$ refer to the broad (330-1050 nm), blue (330-680 nm) and red (630-1050 nm) \g photometric bands, respectively,
From these quantities we will infer the true parallax, 
$\varpi_{\rm{true},j}=\frac{1}{r_j}$, 
and the true apparent magnitude, 
$G_{\rm{true},j}^{k}=M_{j}^{k}+5\log(r_{j})-5$. 

After some algebraic manipulation, we can write Eq. (\ref{Bayes_2}) as
\begin{equation}\label{Bayes_3}
    P(\bm{a}, \bm{\beta} \vert {\bm d},\phi) \propto P(\bm{a}) \prod_{j=1}^{N_{D}} \ \frac{S(d_{j})P(d_{j} \vert \beta_{j}) \ P(\bm{\beta_{j}}\vert \bm{a},\phi)}{\ell(\bm{a},S)},
\end{equation}
\noindent
where the IMF $\phi$ appears as a known parameter in the posterior distribution, and ${\bf a}$ remains as the only parameter to be sampled. We now proceed to establish our mathematical formalism. 
Figure \ref{fig:4} illustrates schematically all the parameters and observable quantities in our problem and Table \ref{tab:1} summarizes our notation.

When required, Eq. (\ref{Bayes_3}) must be properly normalized
(for instance, an incorrect normalization would produce spurious results when using MCMC).
From Eq. (\ref{Bayes_3}) the normalization constant is given by 
\begin{equation}\label{norm_main}
    \ell(\bm{a},S) = \prod_{j=1}^{N_{D}} \int S(d'_{j})P(d'_{j} \vert \beta'_{j})\ P(\beta'_{j} \vert \bm{a},\phi)\ d d'_{j}\ d\beta'_{j},
\end{equation}
\noindent
where we have used the prime to emphasize that the integration variables are not data properly but just auxiliary variables. 
The integration limits in Eq. (\ref{norm_main}) will depend on the characteristics of the specific data set. 

\subsubsection{Hyper-prior}

Following \cite{Walms13}), prior information on $a_{i}$ is introduced through a symmetric Dirichlet distribution,
\begin{equation}\label{dirichlet}
    P(\bm{a})=\frac{\Gamma(\xi N_{iso})}{\Gamma(\xi)^{N_{iso}}}\prod_{i=1}^{N_{iso}}a_{i}^{\xi-1},
\end{equation}
\noindent
where $\bm{a}$ must satisfy $a_{i}\geq 0$ and $\sum a_{i}=1$. $\Gamma$ is the gamma function and $\xi$ the concentration parameter.
\rfr{When $\xi$\,=\,$1$ the symmetric Dirichlet distribution is equivalent to a uniform distribution over $\bm{a}$ \citep{gelman13}
which assigns equal density to any vector $\bm{a}$ satisfying $\sum$\,$a_{i}$\,=\,$1$.
In contrast, $\xi$\,>\,$1$ corresponds to evenly distributed $a_i$ (all the values within a single sample are similar to each other).  
$\xi$\,<\,$1$ corresponds to sparsely distributed $a_i$ (most of the values within a single sample will be close to 0, and the vast majority of the $a_i$'s will be concentrated in a few values).
{\it In this paper we use $\xi$\,=\,$1$} to guarantee a uniform prior distribution.}

\subsubsection{Likelihood}

The parallax and photometric magnitudes for a given star over many transits reported in \gtwo can be considered noisy measurements of their true value and follow quite well normal distributions, respectively,
$\varpi_{j}\sim \mathcal{N}(\varpi_{\rm{true},j},\,\sigma_{\varpi,j})$ and 
$G^{k}\sim \mathcal{N}(G^{k}_{\rm{true},j},\,\sigma^{k}_{G,j})$, i.e., the noise is Gaussian \citep{luri18}.
The likelihood function in this case is
\begin{equation}\label{p_like}
 P(d_{j} \vert \beta) \propto \mathcal{N}(\varpi_{j}\vert\varpi_{{\rm true},j},\,\sigma_{\varpi,j}) \prod_{k=1}^{3} \mathcal{N}(G^{k}_{j}\vert G^{k}_{{\rm true},j},\,\sigma^{k}_{G,j}).
\end{equation}
\noindent
$\mathcal{N}(l_j\vert l_{{\rm true},j})$ and $\mathcal{N}(b_j\vert b_{{\rm true},j})$ are simplified to unity through integration 
and are omitted in Eq. (\ref{p_like}).
This is possible because, due to the small errors in the $(l_j,b_j)$ measurements, their PDF can be approximated by 
Dirac's delta functions, $\delta(l_j-l_{{\rm true},j})$ and $\delta(b_j-b_{{\rm true},j})$.

\subsubsection{Prior}\label{subsec_prior_r_M}

Each point in the $i^{th}$ isochrone has associated a stellar mass $m$ and three absolute magnitudes $M_{i}^{k=1,2,3}(m)$.
Convolving $M_i^k(m)$ with a Gaussian error function, each magnitude will follow a normal distribution
$\mathcal{N}(M_i^k(m),\,\sigma_i^k(m))$, where as before $M^{k=1,2,3} =M_{G},M_{G_{BP}},M_{G_{RP}}$.
The function $\sigma_i^k(m)$ is ill defined but can be determined from the mean separation between isochrones 
in the CMD.
Its calculation becomes cumbersome when the number of isochrones in the model increases.
\citet{Small13} have shown that for MS stars on isochrones of consecutive metallicity,
$\sigma_i^k(m)$ can be approximated by half the mean separation $\vert M_{i_1}^{k}(m)-M_{i_2}^{k}(m)\vert$ 
over the length of isochrones $i_1$ and $i_2$. Therefore we adopt their definition of $\sigma_i^k(m)$
\rfr{(see Section\,\ref{methods})}.

The differential probability that a star of magnitude $M_{j}^{k}$ belongs to the $i^{th}$ isochrone is
\begin{equation}\label{diff_dP}
dP=a_{i}\,\mathcal{N}(M_{j}^{k}\vert M_{i}^{k},\sigma_i^k)\,\phi(m)\,dm,
\end{equation}
\noindent
where we omit the dependence of $M$ and $\sigma$ on $m$ for simplicity.
If we suspect that the stellar population characterized by the $i^{th}$ isochrone is spatially associated, 
we must multiply $dP$ by the distance prior $P_i(r_j)$. 
Integrating $dP$ over $m$ and adding the contribution of all isochrones, we get the probability that 
the $j^{th}$ star belongs to the $i^{th}$ isochrone,
\begin{equation}\label{prior_MG}
P(\beta_{j}\vert \textbf{a},\phi)\,\propto\,\sum_{i=1}^{N_{iso}}\,P_{i}(r_{j})\,a_{i}\,\int_{m_{l,i}}^{m_{u,i}}\phi(m)\,\prod_{k=1}^{3} \mathcal{N}(M^{k}_{j}\vert M^{k}_{i},\sigma_i^k)dm.
\end{equation}
\noindent
The integration limits in Eq. (\ref{prior_MG}) depend on the isochrone $i$. This is because each isochrone corresponds to a different age and possibly metallicity, therefore the lower and upper stellar mass limits may differ. If the spatial distribution of the stars is not related to age and metallicity, $P_{i}(r_{j})=P(r_{j})$ can be taken out of the sum in Eq. (\ref{prior_MG}).
The distance prior depends on the sample under study.

\subsubsection{Posterior}

Inserting Eqs. (\ref{p_like}) and (\ref{prior_MG}) in Eq. (\ref{Bayes_3}) for the Bayes theorem, we obtain the general expression for the posterior distribution
\begin{align}\label{post_main}
      P(\bm{a}, \bm{\beta} \vert \bm{d},\phi) &\propto P(\bm{a}) \prod_{j=1}^{N_{D}}\Bigg{[}S(d_{j})\ \mathcal{N}(\varpi_{j} \vert \varpi_{\rm{true},j} \sigma_{\varpi,j})\ \times \nonumber \\
   & \times \prod_{k=1}^{3} \mathcal{N} (G^k_j\vert G^k_{\rm{true},j}\sigma^k_{G,j})\ \sum_{i=1}^{N_{iso}}\ P_i(r_j)\ a_i\ \times \ \nonumber \\
   & \times \int_{m_{l,i}}^{m_{u,i}}\phi(m) \prod_{k=1}^{3}\ \mathcal{N}(M^k_j \vert M^k_i \sigma_i^k)\ dm \Bigg{]}.
\end{align}
For computational reasons it is convenient to use the marginalized posterior distribution of $\bm{a}$, 
\begin{equation}\label{marginal_1}
    P(\bm{a}\vert d, \phi) = \int P(\bm{a}, \bm{\beta} \vert \bm{d},\phi) d\bm{\beta}.
\end{equation}
Fig. \ref{fig:4} shows the flow of variables leading to Eqs. (\ref{post_main}) and (\ref{marginal_1}).

\subsubsection{Sample completeness}

In Appendix\,\ref{formalism} we compute the posterior for the cases of a complete and a magnitude-limited sample. The treatment of an incomplete sample will be the subject of future work.

\subsection{Methodology}\label{methods}
\begin{figure}
\begin{center}
    \includegraphics[width=\columnwidth]{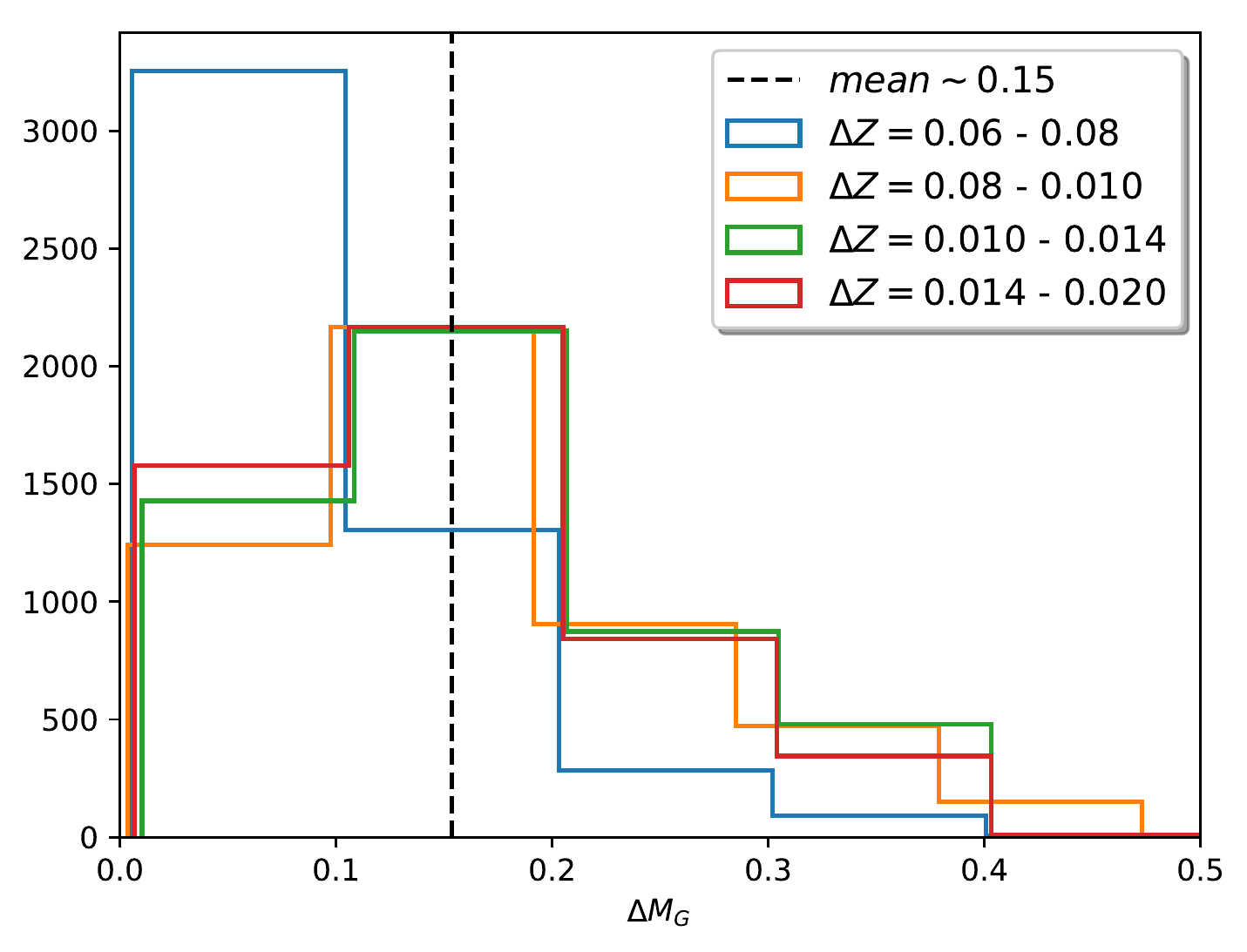}
    \caption{\label{fig:5}Distribution of $\vert M_{i_1}^{k}(m)$\,-\,$M_{i_2}^{k}(m)\vert$, the separation between isochrones of the
    same age (here 0.5 Gyr) for $Z$\,=\,$0.008,0.01,0.014,0.02,0.03$. The {\it dashed line} indicates the average of the individual 
    mean for each histogram. Therefore, following \citet{Small13} we adopt $\sigma_{i}^{k}=0.075$.}
\end{center}
\end{figure}

\begin{figure*}
\begin{center}
    \includegraphics[width=\textwidth]{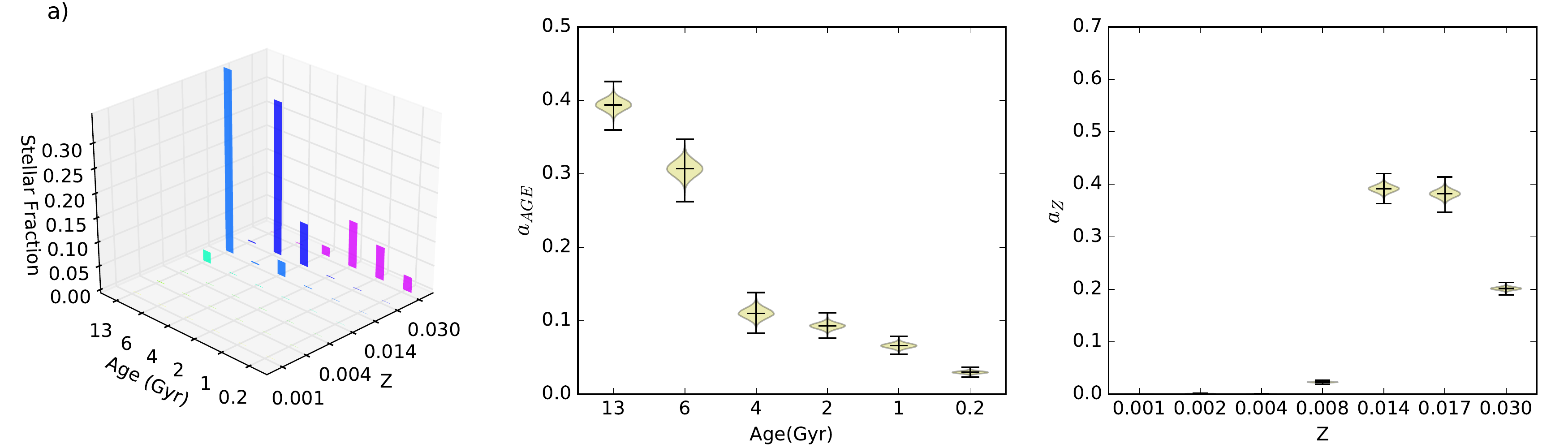}
    \includegraphics[width=\textwidth]{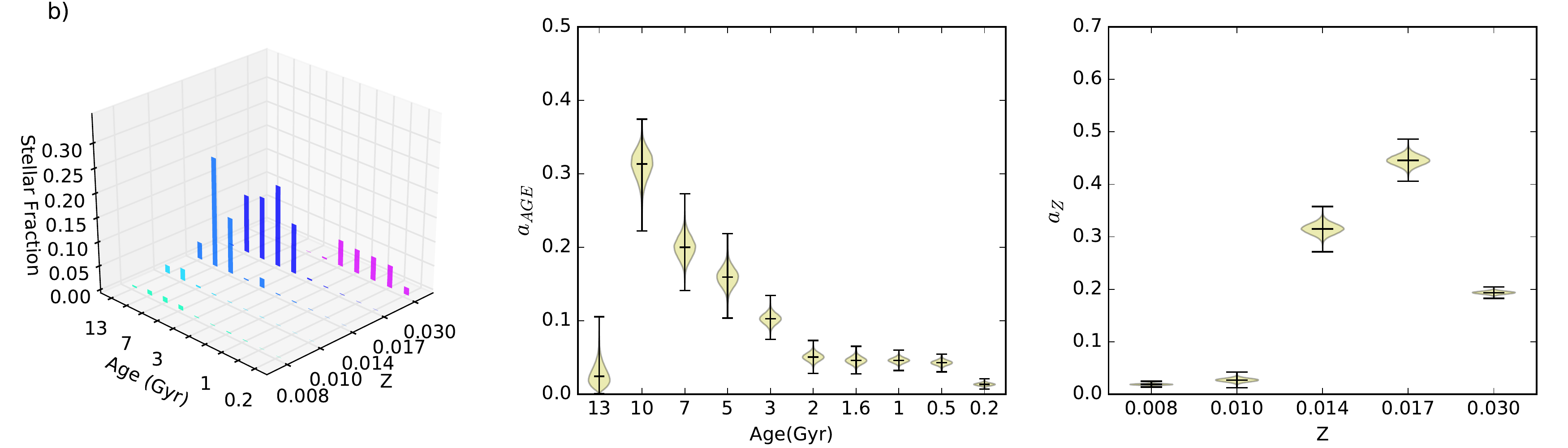}
    \includegraphics[width=\textwidth]{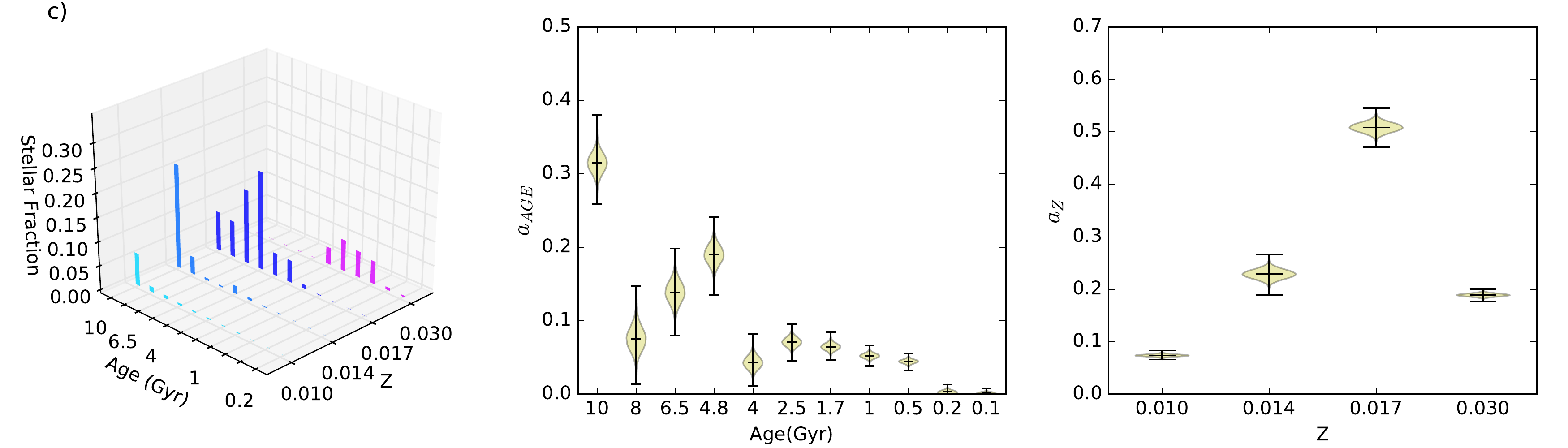}
    \includegraphics[width=\textwidth]{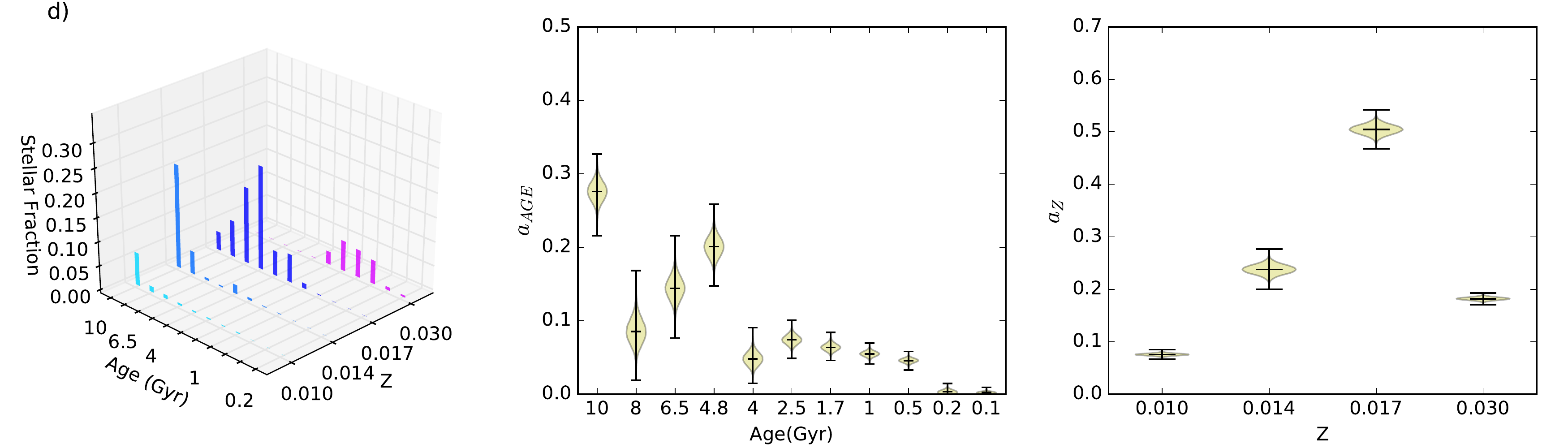}
    \caption{\label{fig:6} AMD for the \solm magnitude limited sample for the three sets of isochrones listed in Tables \ref{tab:2}, \ref{tab:3}, and \ref{tab:4}.
    {\it (a)} Grid A, "low" time resolution. 
    {\it (b)} Grid B, "mid" time resolution. 
    {\it (c)} Grid C, "high" time resolution.
    {\it (d)} Same as {\it (c)} but the Gaia photometry for each star was corrected for extinction using the 3D reddening map of Fig.\,\ref{fig:7} before inferring $\bm{a}$.
    The height of the bars in the 3D plots on the left hand side is the median of the distribution of $a_i$ for the corresponding isochrone.
    The violin plots summarize the marginalized {\it posterior} PDF for age and $Z$.
    The horizontal lines in each violin represent from bottom to top the 0, 50, and 100 percentiles of the distribution.
    The maximum likelihood solution for $a_i$, also provided by the {\it Stan} MCMC platform, is indicated by the solid {\it blue} squares.}
\end{center}
\end{figure*}

Determining exactly the {\it posterior} PDF requires evaluating the matrix elements $P_{ij}$ using Eq. (\ref{post_complete_2}) in the case of a complete sample, or $P_{ij}$ and $C_{ij}$ from Eqs. (\ref{post_trunc_2}) and (\ref{norm_trunc_2}) in the case of a magnitude-limited sample, amounting to calculating $N_{D}\times N_{iso}$ double integrals. 
We use a $10^5$ step MCMC process to sample the $a_i$ parameter space and build an accurate
representation of the marginalized {\it posterior} PDF, from which we infer the distribution, statistics and confidence intervals of $a_i$.
We use the {\it Stan} MCMC platform (\url{https://mc-stan.org}) due to its recognized reliability and acceptance of hard constraints, e.g., $a_i\geq0$ and $\sum a_i = 1$.

Assuming that the sample \solm is complete to $G=15$, the selection function in Eq. (\ref{Bayes_2}) becomes
$S(\bm{d})\approx 1$. For more distant samples, the inclusion of a completeness function is mandatory.
Since $N_{D} = 120,452$ and $N_{iso} \approx 50$, the grid of $N_{iso}$ isochrones must be chosen in a clever
manner to minimize CPU time.\footnote{\rfr{The number of isochrones in these grids is arbitrary but 50 is close to the limit that we
can handle efficiently with local computer resources in a reasonable amount of time.}}
We adopt the following procedure to build our grids of isochrones.
First, we use a small set of isochrones covering coarsely a wide range of age for each stellar metallicity and obtain a first {\it low resolution posterior} PDF. 
Then, we select a larger and finer grid of isochrones centered on the values inferred from the previous solution. This process can be repeated until we are left with only the most statistically significant isochrones.

Each $a_i$ is proportional to the number of stars formed between $(t_i,$\,$t_i$\,+\,$dt)$ with metallicity between
($Z_i$,\,$Z_i$\,+\,$dZ$).  
The SFR is derived by dividing $a_i$ by the intervals $dt$ and $dZ$.
We must keep in mind that the statistical weights $a_i$ cannot provide the complete and true star formation history of the system. The observations provide only a snapshot of the present day stellar population.
We cannot account for stars that have been lost due to internal dynamical processes or interactions with
external systems, or stars that have been gained through mergers. Doing so requires dynamical models
beyond the scope of our work.

We build our isochrones from the PARSEC evolutionary tracks \citep{chen2015,marigo2013} using the isochrone synthesis technique developed by \cite{cb91}.
The stellar photometry in the \g bands in the Vega magnitude system is derived from the \cite{westera02} BaSeL 3.1 spectral library.
Each isochrone defines the locus occupied in the $G$\,vs.\,$G_{BP}-G_{RP}$ CMD by stars of a given age and metallicity
(Fig.\,\ref{fig:1}).
The isochrones are then parametrized by the stellar mass and re-sampled such that the difference in magnitude between consecutive points along the isochrone is $\approx\,0.04$ mag.
Our model requieres that we specify the mean separation between adjacent isochrones, $\sigma_i^k(m)$ in Eq.\,(\ref{post_main}). 
\rfr{In this paper we follow the prescription by \cite{Small13} to derive the value $\sigma_{i}^{k=1,2,3}=0.075$, which applies to MS stars in our isochrones (cf. Fig. \ref{fig:5}).
This is justified because from their position in the CMD we estimate that of the $120,452$ stars in \solm, $906$\,$(0.75)$\% are giants, $437$\,$(0.36)$\% are WDs, and $119,109$\,$(98.89)$\% are MS stars.}

\rfr{Unless indicated otherwise, we populate the isochrones according to the \cite{kr01} IMF, assuming lower and upper mass limits
$(m_l,m_u)$\,=\,$(0.1,100)$\,M$_\odot$. This choice is arbitrary. The stars in the youngest isochrones in our grids have $m$\,$\leq$\,$8$\,M$_\odot$. This does not pose a problem with the adopted normalization as long as the same values of $(m_l,m_u)$ are used for all isochrones.}

\rfr{From the compilation of the solar abundances of chemical elements by \citet[][their table 1]{bressan12}, the present-day solar (photospheric) metallicity is $Z_\odot$\,=\,$0.01524$, which corresponds to a protosolar (before the effects of diffusion) metallicity $Z_\odot^0$\,=\,$0.01774$. From now on, we will refer to $Z$\,=\,$0.017$ as solar metallicity.}

\section{Results}\label{samd}

\begin{table*}
 \caption{\label{tab:2}Inference results for grid A, 42 isochrones (abridged; the full table is available as Table\,\ref{tab:d1} in the supplementary online material).}
 \renewcommand{\arraystretch}{1.25} 
 \begin{tabular}{cccccccc}
 \hline
 Age    &\mc{7}{c}{$a_i \times 100$} \\
 (Gyr)  &\mc{1}{c}{Z=0.001} & \mc{1}{c}{Z=0.002} & \mc{1}{c}{Z=0.004} & \mc{1}{c}{Z=0.008}    & \mc{1}{c}{Z=0.014}      & \mc{1}{c}{Z=0.017}       & \mc{1}{c}{Z=0.030}     \\
 \hline
  0.2   &        $0$        &         $0$        &         $0$        &          $0$          &           $0$           &            $0$           &  $2.9_{-0.2}^{+0.2}$   \\
  1.0   &        $0$        &         $0$        &         $0$        &          $0$          &           $0$           &            $0$           &  $6.5_{-0.4}^{+0.4}$   \\
  2.0   &        $0$        &         $0$        &         $0$        &          $0$          &           $0$           &            $0$           &  $9.2_{-0.5}^{+0.5}$   \\
  4.0   &        $0$        &         $0$        &         $0$        &          $0$          &   $2.0_{-1.0}^{+1.1}$   &  $7.5 _{-1.5 }^{+1.5 }$  &  $1.4_{-0.5}^{+0.5}$   \\
  6.0   &        $0$        &         $0$        &         $0$        &          $0$          &           $0$           &  $30.5_{-1.4 }^{+1.4 }$  &          $0$           \\
  13.0  &        $0$        &         $0$        &         $0$        &  $2.2_{-0.1}^{+0.1}$  &  $37.0_{-1.0}^{+1.0}$   &            $0$           &          $0$           \\
 \hline
 \end{tabular}
\end{table*}

\begin{table*}
 \caption{\label{tab:3}Inference results for grid B, 50 isochrones (abridged; the full table is available as Table\,\ref{tab:d2} in the supplementary online material).}
 \renewcommand{\arraystretch}{1.25} 
 \setlength{\tabcolsep}{12pt} 
 \begin{tabular}{cccccc}
 \hline
 Age   & \mc{4}{c}{$a_i \times 100$} \\
 (Gyr) & \mc{1}{c}{Z=0.008}         & \mc{1}{c}{Z=0.010}          & \mc{1}{c}{Z=0.014}        & \mc{1}{c}{Z=0.017}        & \mc{1}{c}{Z=0.030}  \\
 \hline
  0.2  &            $0$             &            $0$             &            $0$            &            $0$            & $1.3_{-0.2}^{+0.2}$  \\
  0.5  &            $0$             &            $0$             &            $0$            &            $0$            & $4.2_{-0.4}^{+0.4}$  \\
  1.0  &            $0$             &            $0$             &            $0$            &            $0$            & $4.5_{-0.4}^{+0.4}$  \\
  1.6  &            $0$             &            $0$             &            $0$            &            $0$            & $4.3_{-0.6}^{+0.6}$  \\
  2.0  &            $0$             &            $0$             &            $0$            &            $0$            & $4.8_{-0.7}^{+0.7}$  \\
  3.0  &            $0$             &            $0$             &            $0$            &    $9.2_{-1.4}^{+1.2}$    &          $0$         \\
  5.0  &            $0$             &            $0$             &            $0$            &    $15.3_{-1.6}^{+1.6}$   &          $0$         \\
  7.0  &            $0$             &            $0$             &    $8.9_{-2.7}^{+2.7}$    &    $10.3_{-2.8}^{+2.8}$   &          $0$         \\
  10.0 &            $0$             &    $1.7_{-1.0}^{+0.7}$     &    $19.7_{-3.1}^{+3.0}$   &    $9.5_{-2.6}^{+2.6}$    &          $0$         \\
  13.0 &            $0$             &            $0$             &    $1.4_{-1.2}^{+2.1}$    &            $0$            &          $0$         \\
 \hline
\end{tabular}
\end{table*}

\begin{table*}
 \caption{\label{tab:4}Inference results for grid C, 44 isochrones (abridged; the full table is available as Table\,\ref{tab:d3} in the supplementary online material).}
 \renewcommand{\arraystretch}{1.25} 
 \begin{tabular}{cccccccccccc}
 \hline
       &  & \mc{4}{c}{No extinction correction}                                                      &  &  &  \mc{4}{c}{Extinction correction using the Stilism tool$^a$} \\
 Age   &  & \mc{4}{c}{$a_i \times 100$}                                                              &  &  &  \mc{4}{c}{$a_i \times 100$} \\
 (Gyr) &  & \mc{1}{c}{Z=0.010}  &  \mc{1}{c}{Z=0.014}  &  \mc{1}{c}{Z=0.017}  & \mc{1}{l}{Z=0.030}   &  &  &  \mc{1}{c}{Z=0.010} & \mc{1}{c}{Z=0.014}   & \mc{1}{c}{Z=0.017}   & \mc{1}{l}{Z=0.030}   \\
 \hline
  0.1  &  &         $0$         &         $0$          &         $0$          &         $0$          &  &  &         $0$         &         $0$          &         $0$          &         $0$          \\
  0.2  &  &         $0$         &         $0$          &         $0$          &         $0$          &  &  &         $0$         &         $0$          &         $0$          &         $0$          \\
  0.5  &  &         $0$         &         $0$          &         $0$          & $4.4_{-0.4}^{+0.3}$  &  &  &         $0$         &         $0$          &         $0$          & $4.5_{-0.4}^{+0.4}$  \\
  1.0  &  &         $0$         &         $0$          &         $0$          & $5.0_{-0.4}^{+0.4}$  &  &  &         $0$         &         $0$          &         $0$          & $5.3_{-0.5}^{+0.5}$  \\
  1.7  &  &         $0$         &         $0$          &         $0$          & $5.9_{-0.6}^{+0.6}$  &  &  &         $0$         &         $0$          &         $0$          & $5.7_{-0.6}^{+0.6}$  \\
  2.5  &  &         $0$         &         $0$          &  $3.8_{-0.8}^{+0.8}$ & $2.9_{-0.6}^{+0.6}$  &  &  &         $0$         &         $0$          & $5.0_{-0.8}^{+0.8}$  & $2.1_{-0.6}^{+0.6}$  \\
  4.0  &  &         $0$         &         $0$          &  $3.4_{-1.3}^{+1.3}$ &         $0$          &  &  &         $0$         &         $0$          & $3.8_{-1.4}^{+1.3}$  &         $0$          \\
  4.8  &  &         $0$         &         $0$          & $18.7_{-1.7}^{+1.7}$ &         $0$          &  &  &         $0$         &         $0$          & $19.8_{-1.8}^{+1.8}$ &         $0$          \\
  6.5  &  &         $0$         &         $0$          & $13.4_{-2.0}^{+2.0}$ &         $0$          &  &  &         $0$         &         $0$          & $13.8_{-2.0}^{+2.0}$ &         $0$          \\
  8.0  &  &         $0$         &  $1.6_{-1.3}^{+2.1}$ &  $5.3_{-2.3}^{+2.2}$ &         $0$          &  &  &         $0$         & $2.5_{-1.8}^{+2.2}$  & $5.4_{-2.2}^{+2.1}$  &         $0$          \\
  10.0 &  & $6.1_{-0.9}^{+0.6}$ & $19.7_{-2.2}^{+1.9}$ &  $5.8_{-2.2}^{+2.4}$ &         $0$          &  &  & $6.2_{-1.0}^{+0.7}$ & $19.6_{-2.2}^{+1.9}$ & $1.7_{-1.4}^{+2.1}$  &         $0$          \\
 \hline
 \mc{12}{l}{$^a${\citet[][\url{https://stilism.obspm.fr}]{lallement19}}}\\
\end{tabular}
\end{table*}

\begin{figure*}
\begin{center}
    \includegraphics[width=0.8\textwidth]{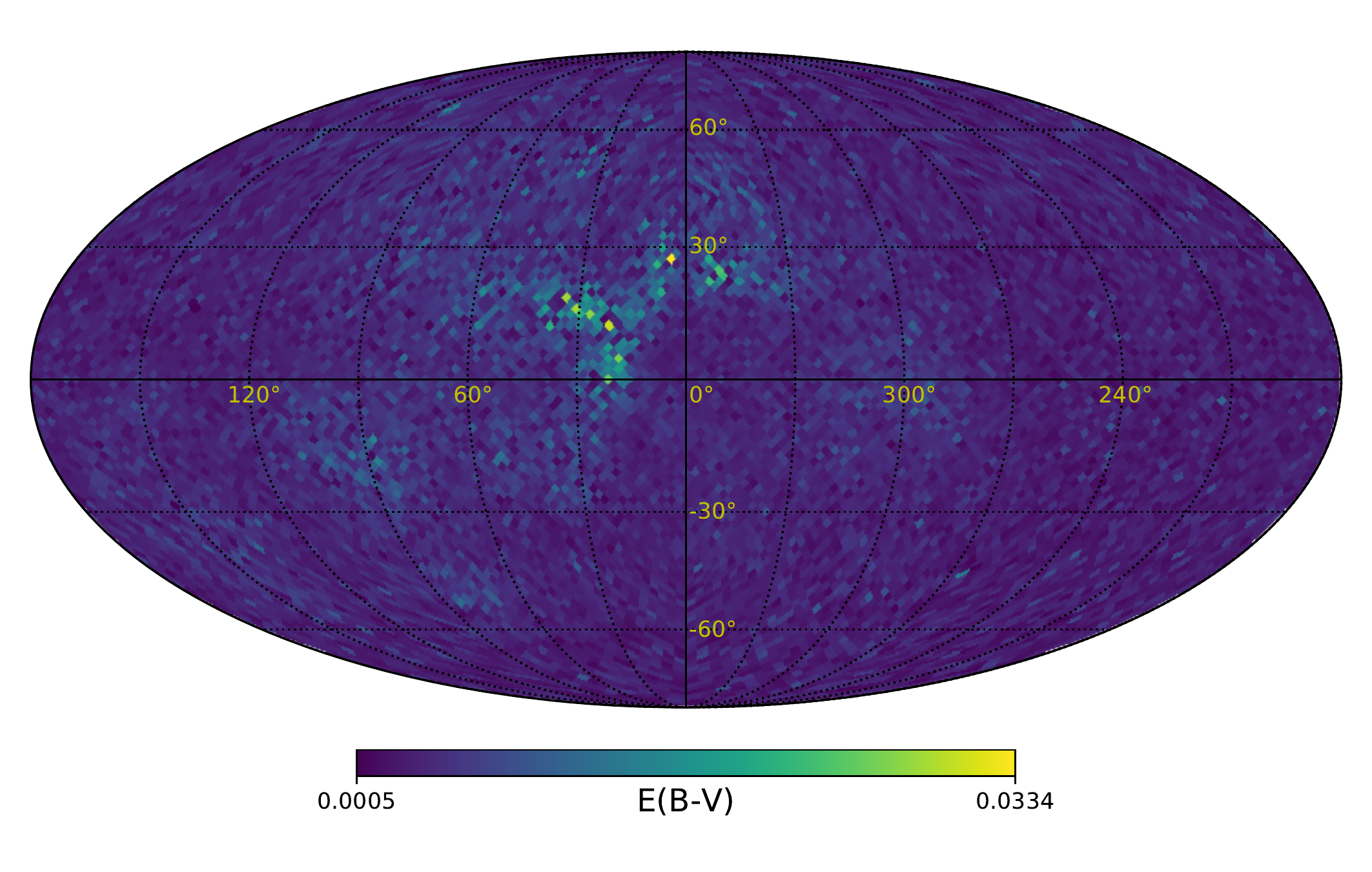}
    \caption{\label{fig:7}Reddening map for $S_{100}^{15}$ derived by the authors using the \citet{lallement19} Stilism tool. The color bar indicates the mean color excess E(B-V) per pixel.}
\end{center}
\end{figure*}

\subsection{Age-Metallicity distribution of the Solar Neighbourhood}\label{amd}

We analyze the \solm sample defined in Section \ref{samp_sel_sol} using the three grids of isochrones defined in Tables \ref{tab:2}, \ref{tab:3}, and \ref{tab:4}. In these tables we list the components of the solution vector $\bm{a}$ grouped as an (age, $Z$) matrix. Values below 1\% are shown as $0$ to highlight the age-metallicity distribution (AMD). The full versions of these tables are available as supplementary online material
(Tables\,\ref{tab:d1}, \ref{tab:d2} and \ref{tab:d3}).

\rfr{Fig. \ref{fig:6}a shows the AMD for the low resolution (in age) grid A, listed in numerical form in Table \ref{tab:2}. It is clear from these plots that the $Z=0.001,\,0.002$ and $0.004$ isochrones do not contribute substantially to the solution and are not needed in our analysis. Stars with probable age $\ge\,6$ Gyr are assigned to the $Z=0.014$ and $0.017$ isochrones.}

\rfr{To break this degeneracy we use grid B. The results are shown in Fig. \ref{fig:6}b and listed in Table \ref{tab:3}. The contribution from the $Z=0.008$ and $0.01$ isochrones is negligible. The maximum contribution from the $Z=0.014$ and $0.017$ isochrones appears now from 7 to 10 Gyr instead of 13 Gyr. The $Z=0.03$ isochrones contribute almost equal amounts to the young 0.5 to 2 Gyr bins.}

\rfr{Next we test grid C, which excludes the $Z=0.008$ isochrones and the 13 Gyr time step due to their minimal contribution to the solution in the previous grids, and varies slightly the time resolution for age $>\,2$ Gyr. In this case we test the effects of interstellar reddening on the solution vector $\bm{a}$. The inferred AMDs for grid C are listed in Table \ref{tab:4} and shown in Fig. \ref{fig:6}c (no extinction correction) and \ref{fig:6}d (extinction corrected). We use the \citet{lallement19} Stilism tool 
(\url{https://stilism.obspm.fr}) to derive the 3D extinction map for all the stars in $S_{100}^{15}$ shown in projection in 
Fig.\,\ref{fig:7}. To work in the Gaia photometric system we use $A_{V}/E(B-V)\,=\,3.16\pm0.15$, and
$A_{G}/A_{V}\,=\,0.789\pm0.005$,
$A_{G_{BP}}/A_{V}\,=\,1.002\pm0.007$, and
$A_{G_{RP}}/A_{V}\,=\,0.589\pm0.004$
\citep[][their table 3]{wang19}.
For simplicity, the extinction correction is applied to the data and not to the statistical model.}

\rfr{From Figs.\,\ref{fig:6}c and \ref{fig:6}d and Table\,\ref{tab:4} we conclude that the correction by extinction does not introduce major differences in the inferred vector $\bm{a}$. Both solutions show two main star formation episodes 10 and 5 Gyr ago, and the maximum of the metallicity distribution occurs at $Z$\,=\,$0.017$.
Nevertheless, some small but noticeable changes are apparent in Table\,\ref{tab:4}.
The extinction correction decreases slightly the contribution of the oldest age bins, increasing the fraction of younger stars of all
metallicities. This means that ignoring extinction biases the AMDs towards older ages.}

\rfr{The results for the three grids shown in Fig. \ref{fig:6} are consistent. 
The AMD in Fig. \ref{fig:6}c shows clearly the presence of three well defined events. 
A maximum in the star formation activity took place $\approx 10$ Gyr ago, forming stars of metallicity slightly below solar
($Z$\,=\,$0.014$). The SFR then decreased, reaching a minimum at $\approx 8$ Gyr ago.
\cite{Sna15} and \cite{Hay16} found evidence of this quenching of star forming activity comparing observed chemical abundances of stars in the solar neighbourhood with the predictions of chemical evolution models. After this minimum, star formation increases again, reaching a local maximum close to 5 Gyr ago at solar metallicity ($Z$\,=\,$0.017$), and then quenches. 
A small amount of residual star formation remains until recent epochs.
The stellar metallicity increases in time from $Z$\,=\,$0.01$ to $0.03$.
Although our sample is very limited compared to {\it all} the stars brighter than $G$\,=\,$12$ in \g DR2, our results are in agreement with the star formation history derived by \cite{Mor19} for the larger sample using a single metallicity model.}

\rfr{The number of stars with $Z$\,=\,$0.03$ for the three grids is higher than expected when compared with chemical distributions reported by, e.g., \cite{Bensby14}. The most likely reason for this effect is that the unresolved binary systems, not included in our statistical treatment, which describe a broad sequence parallel to the MS but just above it are assigned to the $Z$\,=\,$0.03$ isochrones because of their proximity in the CMD (see Section\,\ref{ub}).}

\begin{figure}
\begin{center}
    \includegraphics[width=1.1\columnwidth]{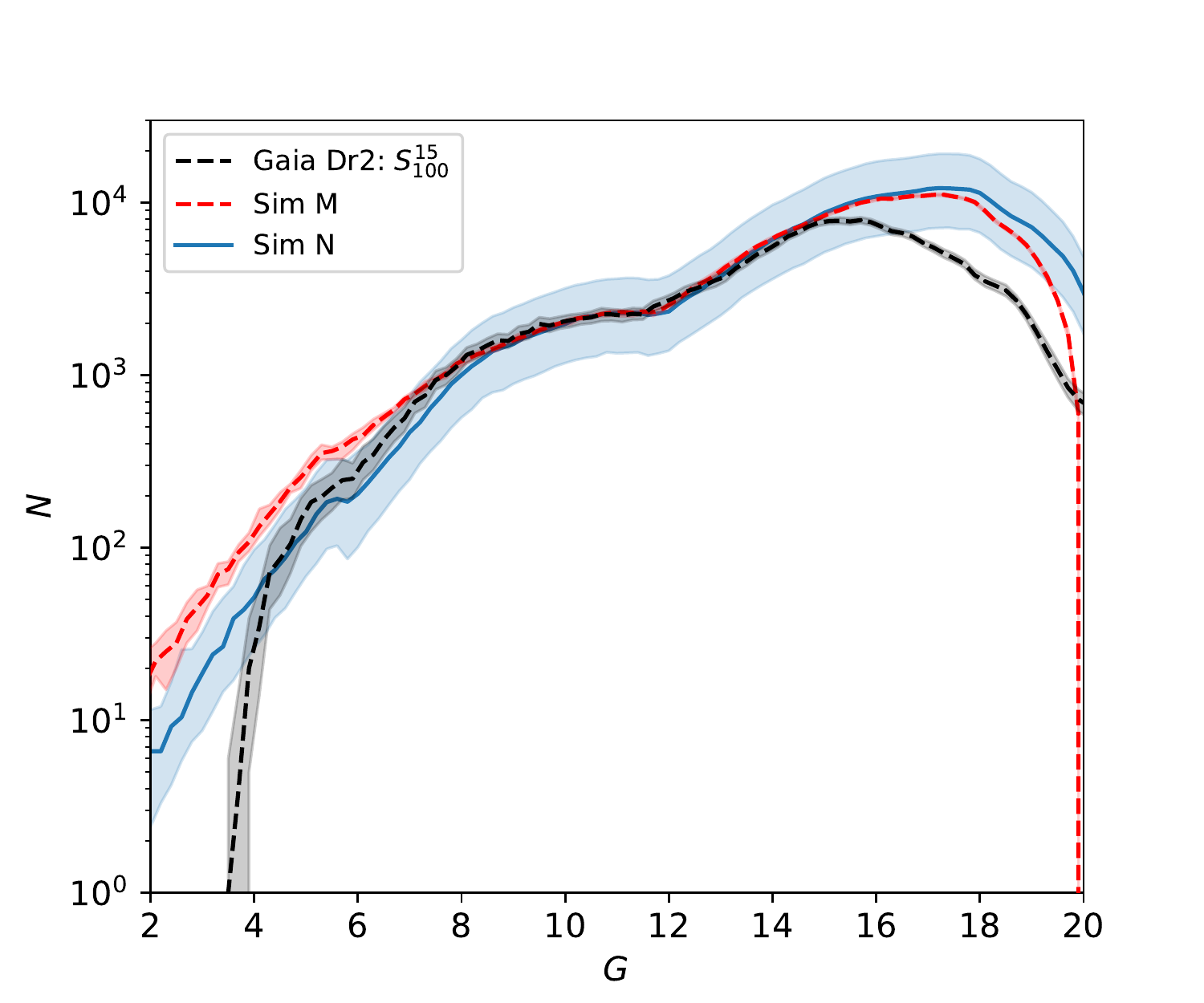}
    \caption{\label{fig:8} {\it Blue line:} Average number counts resulting from 10 \simn\ simulations using the $p50$ values of $a^{*}_i$ in Table\,\ref{tab:C1}. 
    The lower and upper borders of the {\it blue band} correspond to the average counts when we use, respectively, the $p10$ and $p90$ values of $a^{*}_i$.
    The {\it red line} corresponds to the average counts from 10 \simm\ simulations. In this case the {\it red band} indicates the $p10$ and $p90$ percentiles
    determined from the count distribution for the 10 simulations.}
\end{center}
\end{figure}

\begin{figure*}
\begin{center}
    \includegraphics[width=\textwidth,height=0.30\textheight]{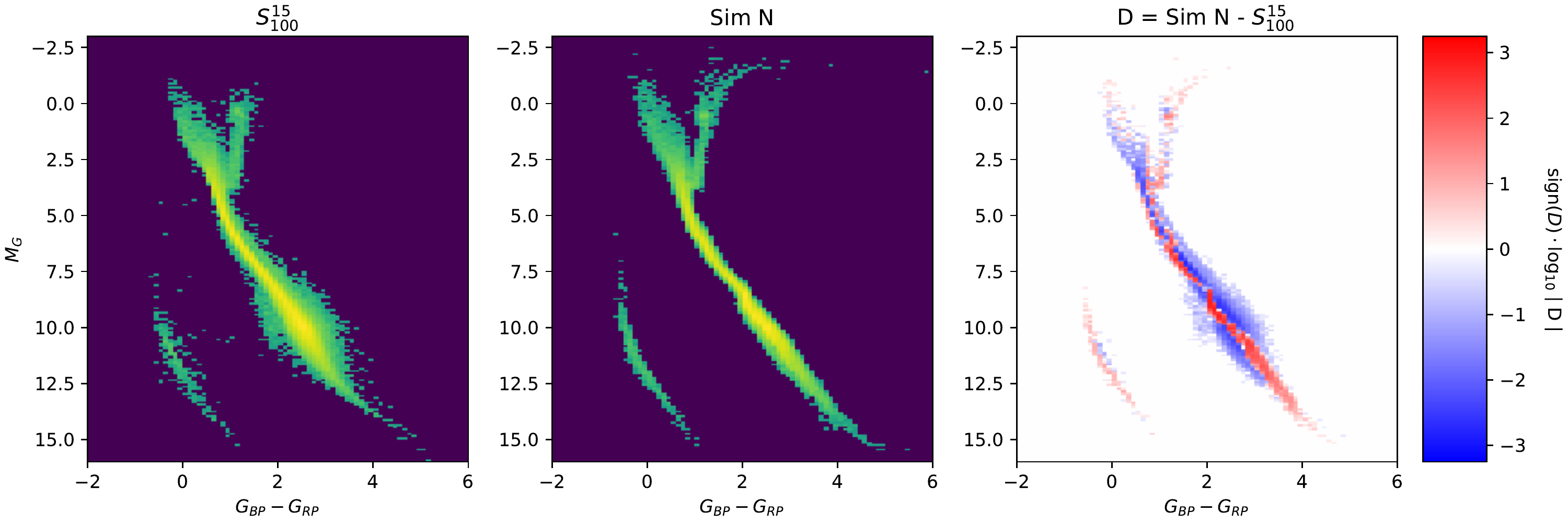}
    \includegraphics[width=\textwidth,height=0.30\textheight]{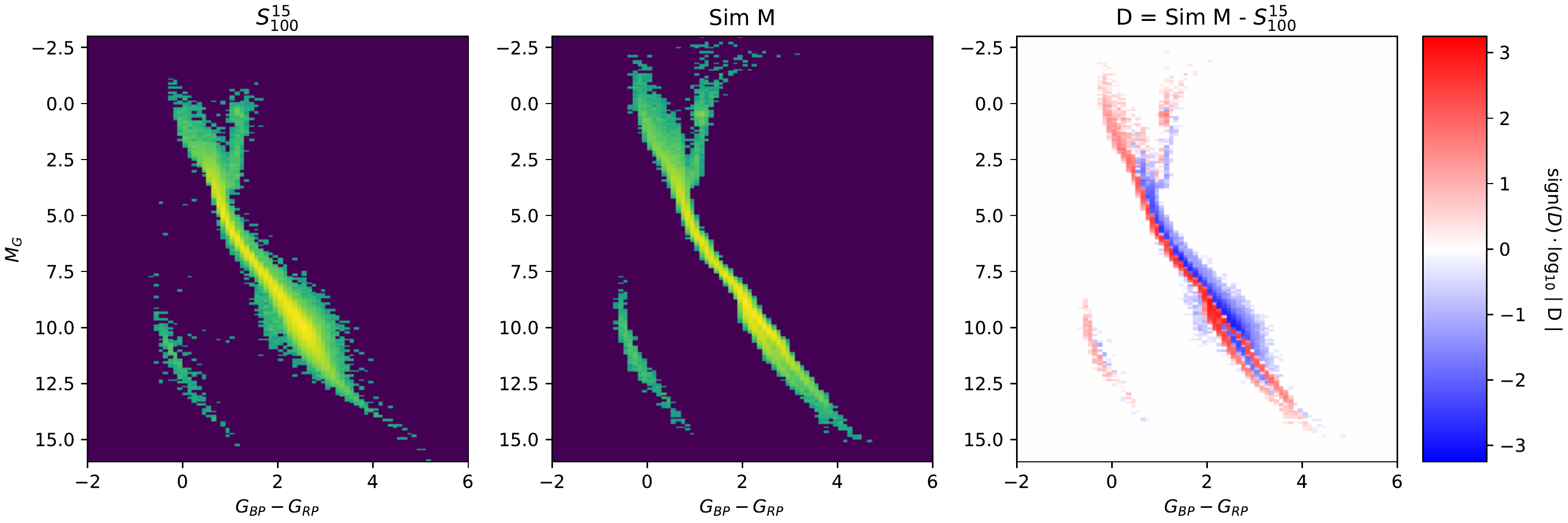}
    \includegraphics[width=\textwidth,height=0.30\textheight]{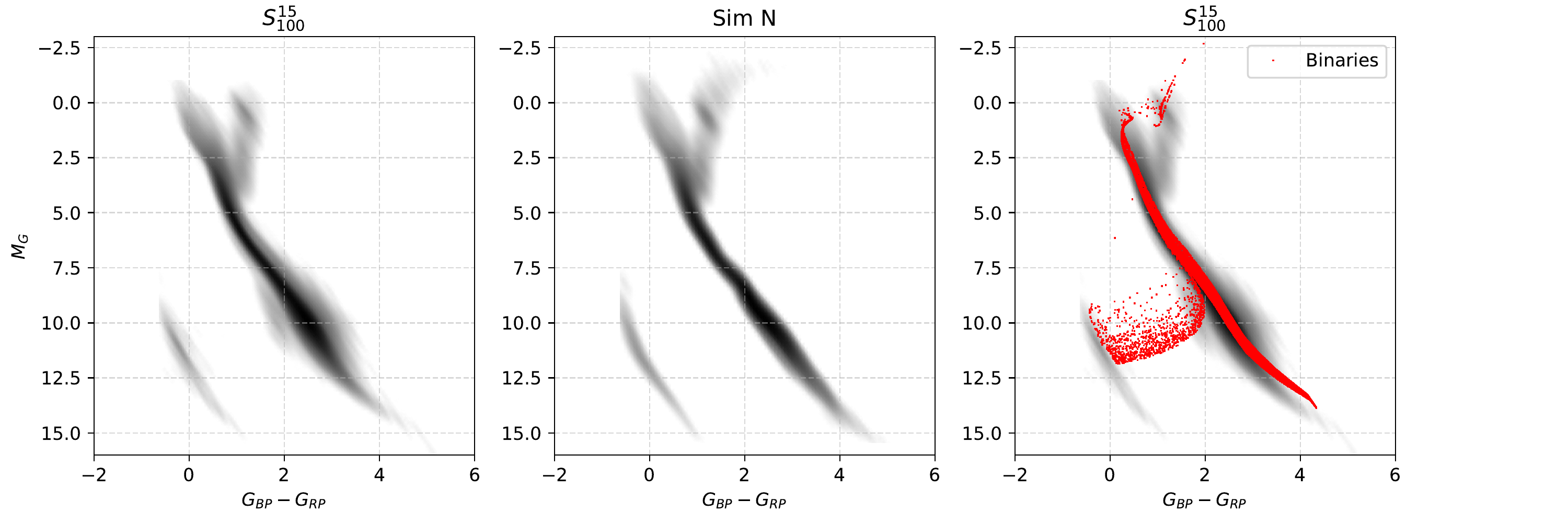}
    \caption{\label{fig:9} ({\it top row}) \gtwo CMD of \solm compared to \simn. The CMDs are binned in 0.1 mag bins.
    The rightmost panel shows the residual \simn\,-\,\solm, colour coded as indicated in the auxiliary axis.
    The residuals show a deficit of stars on the red side of the MS and a excess on the blue side.
    ({\it middle row}) Same as {\it top row} but for \simm. The positive residual is more marked and extends to brighter magnitudes in \simm\ than in \simn.
    ({\it bottom row}) Observed and \simn\ CMDs of \solm. The {\it red dots} in the rightmost panel indicate the expected position
    of unresolved binary systems for a 1 Gyr, $Z=0.014$ population (Section\,\ref{ub}). When the unresolved binary stars are added to all the isochrones
    entering our model, the red strip in this panel becomes broader.
    The \simn\ simulation was computed with the $p50$ values of $a^{*}_i$ in Table\,\ref{tab:C1}. 
    CMDs for $p10$ and $p90$ are available as supplementary online material (Fig.\,\ref{fig:D3}).}
\end{center}
\end{figure*}

\begin{figure*}
\begin{center}
   \includegraphics[width=0.49\textwidth]{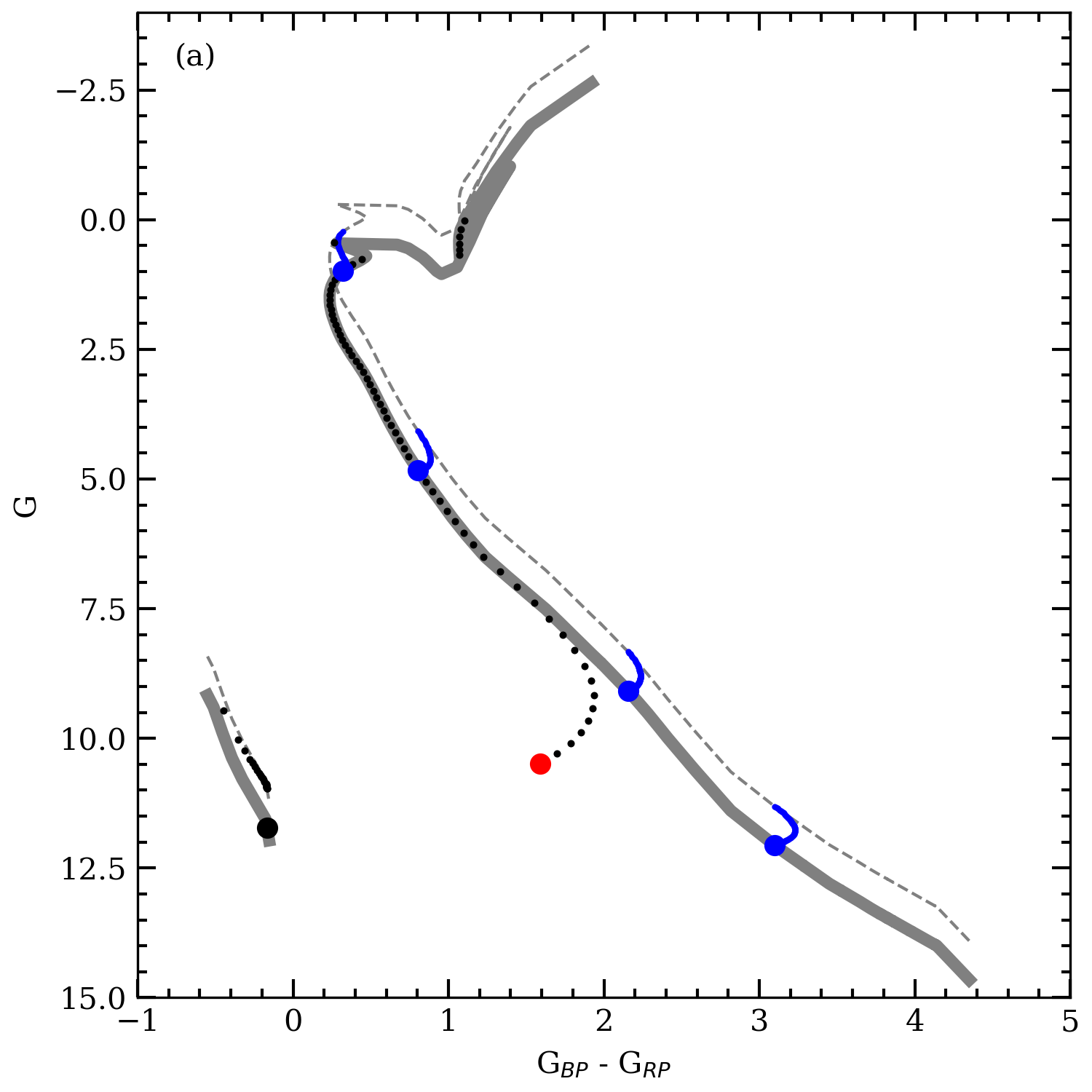}
   \includegraphics[width=0.49\textwidth]{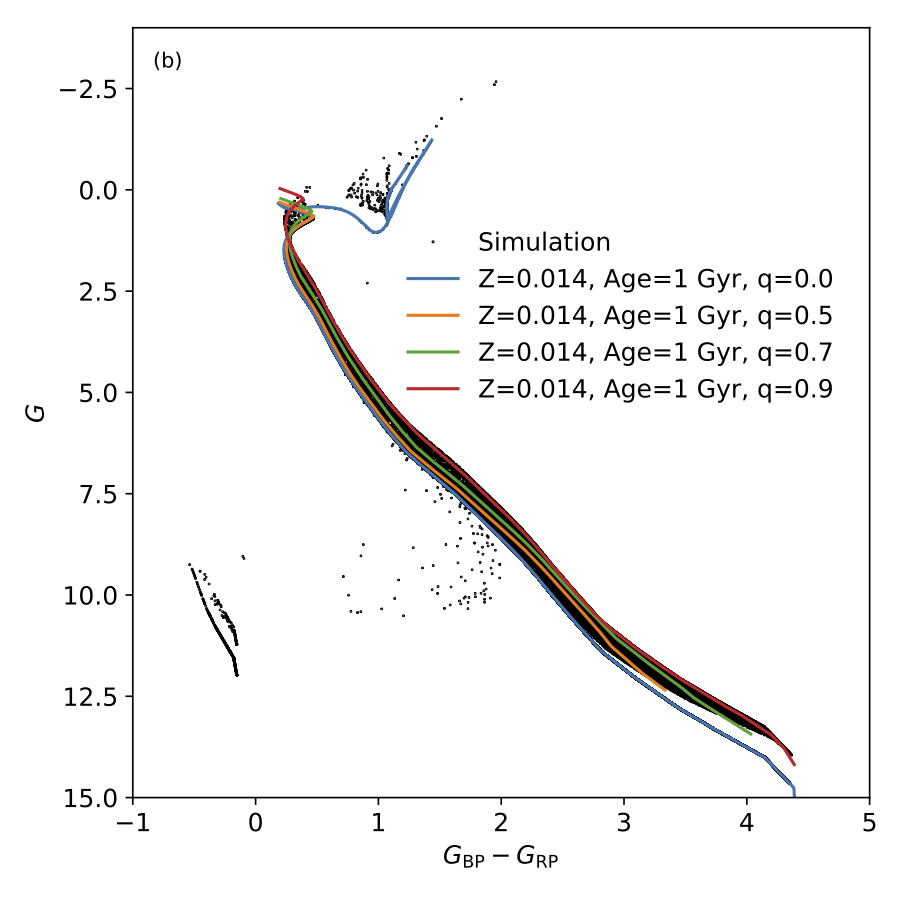}
   \caption{\label{fig:10}{\it (a)} Didactic diagram showing schematically the expected locus defined by unresolved binary stars in the CMD. See Section\,\ref{fp} for details.
   {\it (b)} The {\it black dots} and the {\it black band} show the locus in the CMD defined by our {\it toy-model} binary population assuming that all binary systems are unresolved.
   The {\it coloured} lines show the resulting 1 Gyr $Z$\,=\,$0.014$ isochrones for various values of $q$ built as in Section\,\ref{fp}. The lack of binary stars at the low mass end of the isochrones
   is due to the condition $M_2$\,=\,$qM_1$\,$\ge$\,0.1 M$_\odot$, which translates into $M_1$\,$\ge$\,$0.1/q$ M$_\odot$. See Section\,\ref{mp} for details.}
\end{center}
\end{figure*}

\subsection{Quality control}\label{qc}

\rfr{As a sanity check, we model the \solm\ stellar population using the values of $a_i$ derived above as input to our MW.mx Galaxy model (Appendix\,\ref{mw.mx}). We run two sets of 10 Monte Carlo simulations of \solm, denoted \simn\ and \simm, described in detail in Appendix\,\ref{snmod}. In \simn\ the simulation is stopped when we reach the required number of stars with $G$\,$\leq$\,$15$ for each Galactic component, irrespective of the accumulated mass. In contrast, in \simm\ the simulation is continued until we reach the total mass for each component, obtained by integration of $\rho(R,z)$ over the given volume, irrespective of the number of stars with $G$\,$\leq$\,$15$.}

Fig.\,\ref{fig:8} shows the average number counts in the $G$ band resulting from our simulations. The agreement of the counts for both sets of simulations in the range $G=[8,18]$ is remarkable given the differences in the models and may indicate the lack of sensitivity of the number counts to the model ingredients. The reasons for the excess counts at the bright end in the simulations with respect to \gtwo\ have been analyzed in Section\,\ref{samp_sel_sol}. The \gtwo\ counts are clearly complete to $G=15$. By construction, the \simn\ simulation follows more closely the observed number counts than the \simm\ simulation. Since at the bright end the observed counts are underestimated, it is interesting to explore the nature of the brighter stars in the simulations.

In Fig.\,\ref{fig:9} we compare the \gtwo CMD of \solm\ with \simn\ and \simm. 
The rightmost panel in the {\it top} and {\it middle rows} show the residuals \simn\,-\,\solm\ and \simm\,-\,\solm, respectively, colour coded as indicated in the auxiliary axis.
The residuals show a deficit of stars on the red side of the MS, and an excess on the blue side.
The positive residual is more marked and extends to brighter magnitudes in \simm\ than in \simn.
Similar residual CMD's are presented in \citet[][their figure 1]{Mor19}.

In the {\it bottom row} of Fig.\,\ref{fig:9} we show the observed and simulated (\simn) CMDs of \solm.
The {\it red dots} in the rightmost panel indicate the expected position of unresolved binary systems for a 1 Gyr, $Z$\,=\,$0.014$ population (see Section\,\ref{ub}). When the binary population is added to all the isochrones entering our model, the red strip in this panel becomes broader. The net effect of unresolved binaries is then to widen the single star MS in the CMD towards both {\it brighter-and-redder} and {\it fainter-and-bluer} magnitudes. This may explain the width of the MS in the \gtwo CMD of Figs.\,\ref{fig:1} and \ref{fig:9}. 

\begin{figure*}
\begin{center}
   \includegraphics[width=0.45\textwidth]{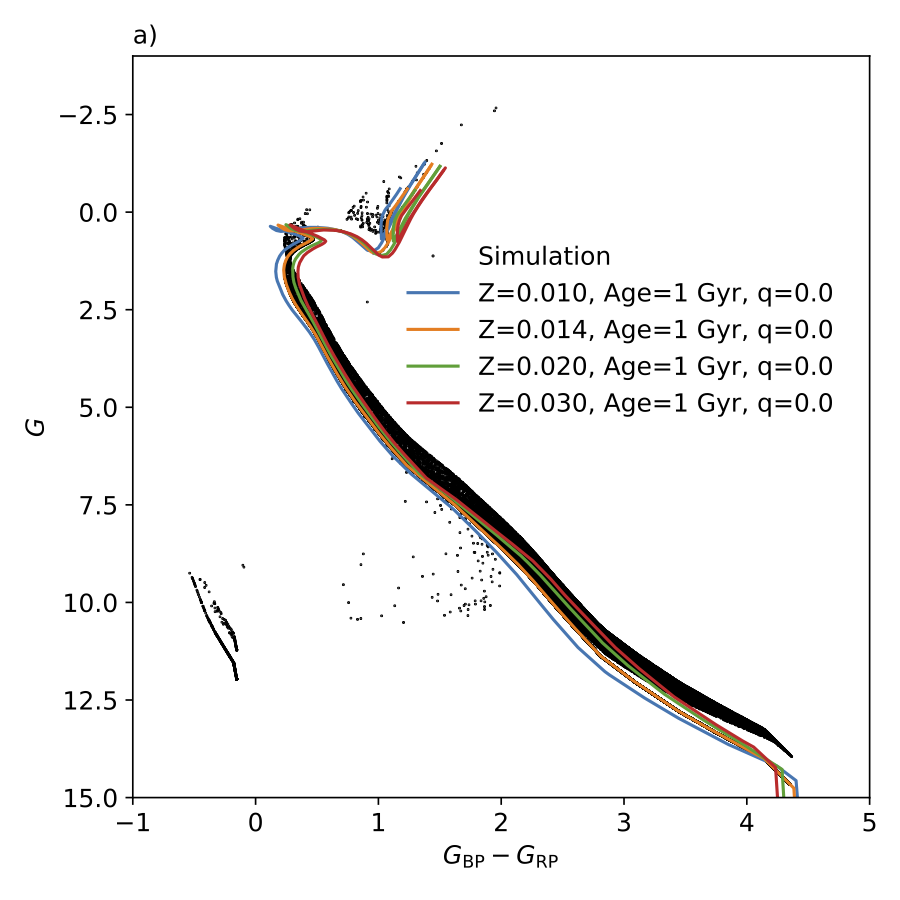}
   \includegraphics[width=0.45\textwidth]{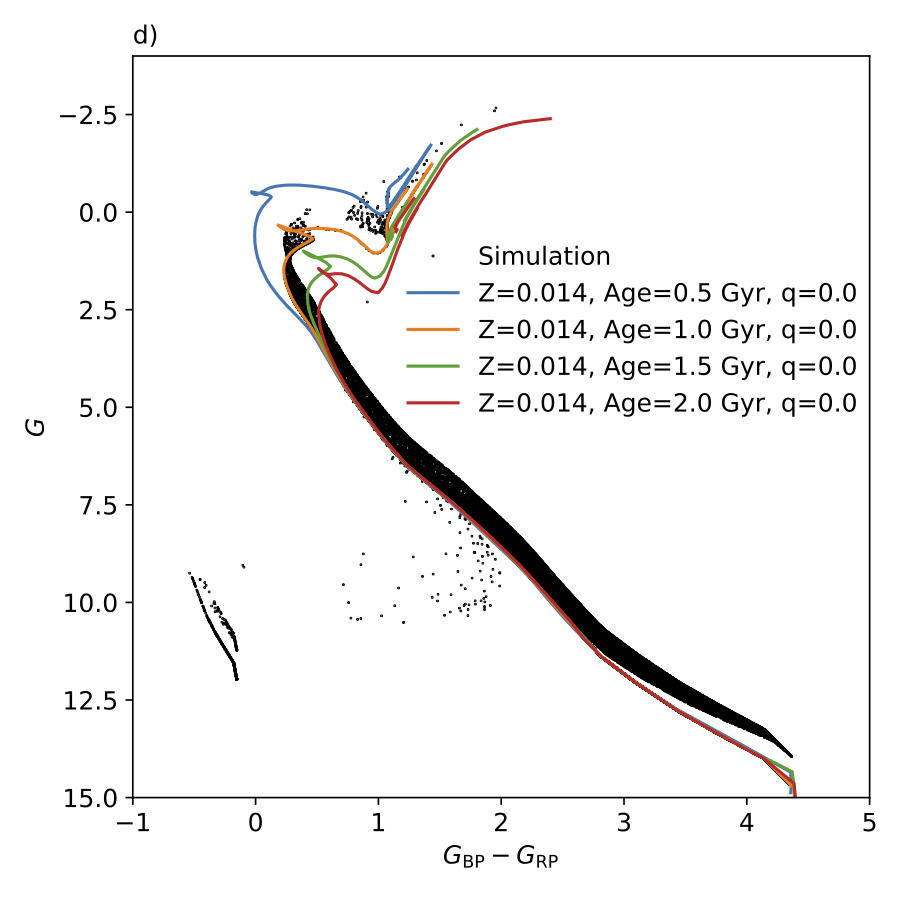}
   \includegraphics[width=0.45\textwidth]{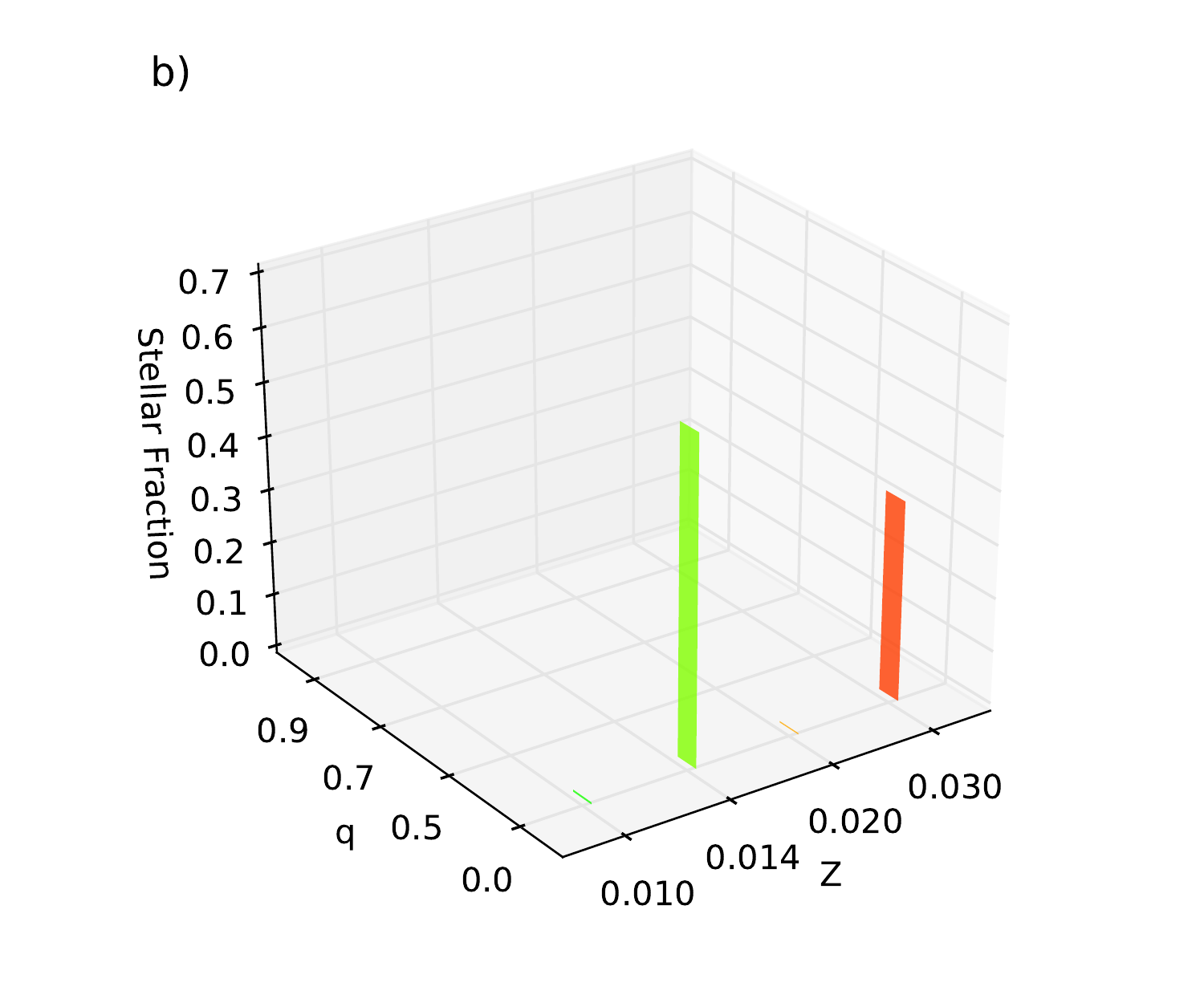}
   \includegraphics[width=0.45\textwidth]{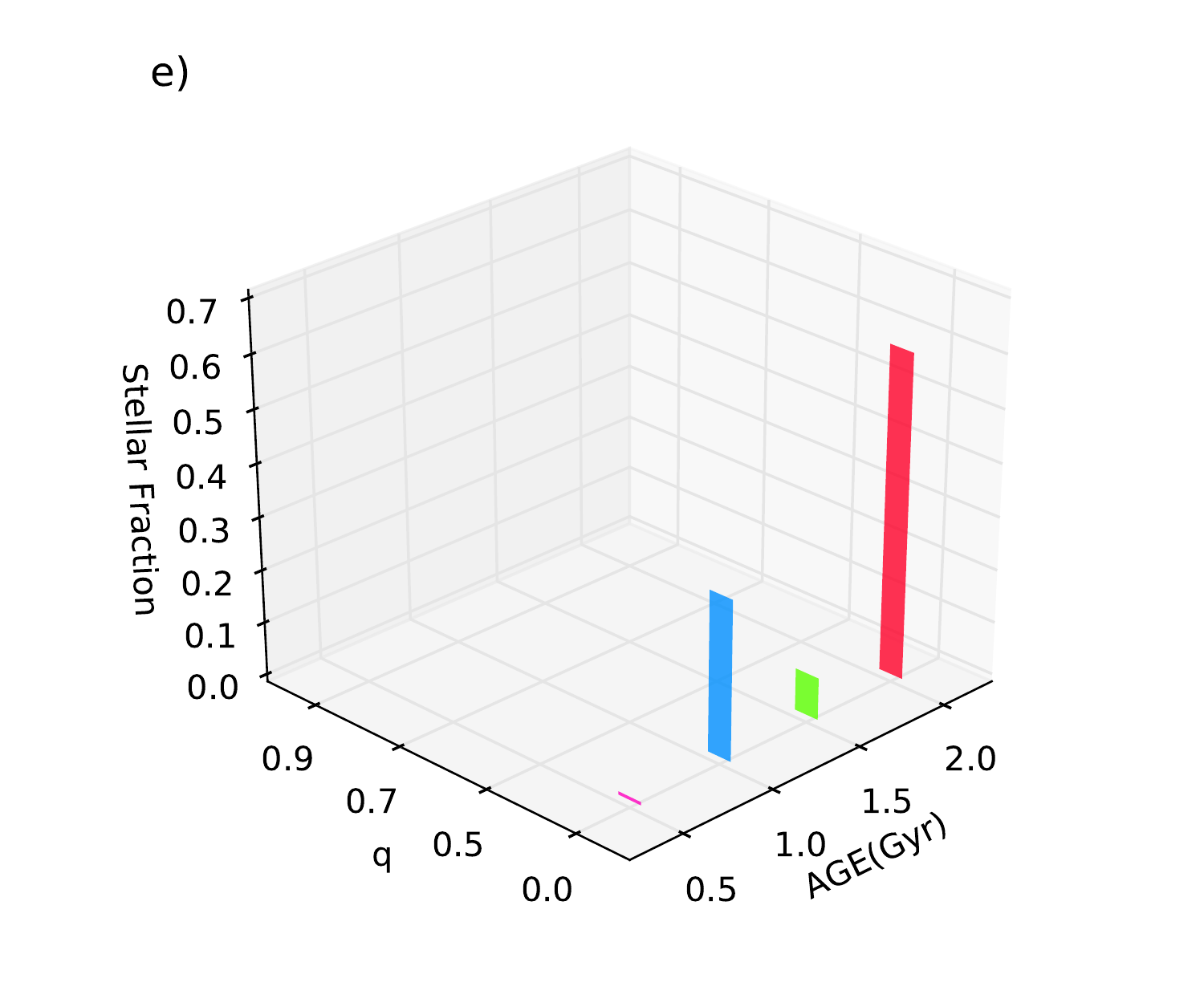}
   \includegraphics[width=0.45\textwidth]{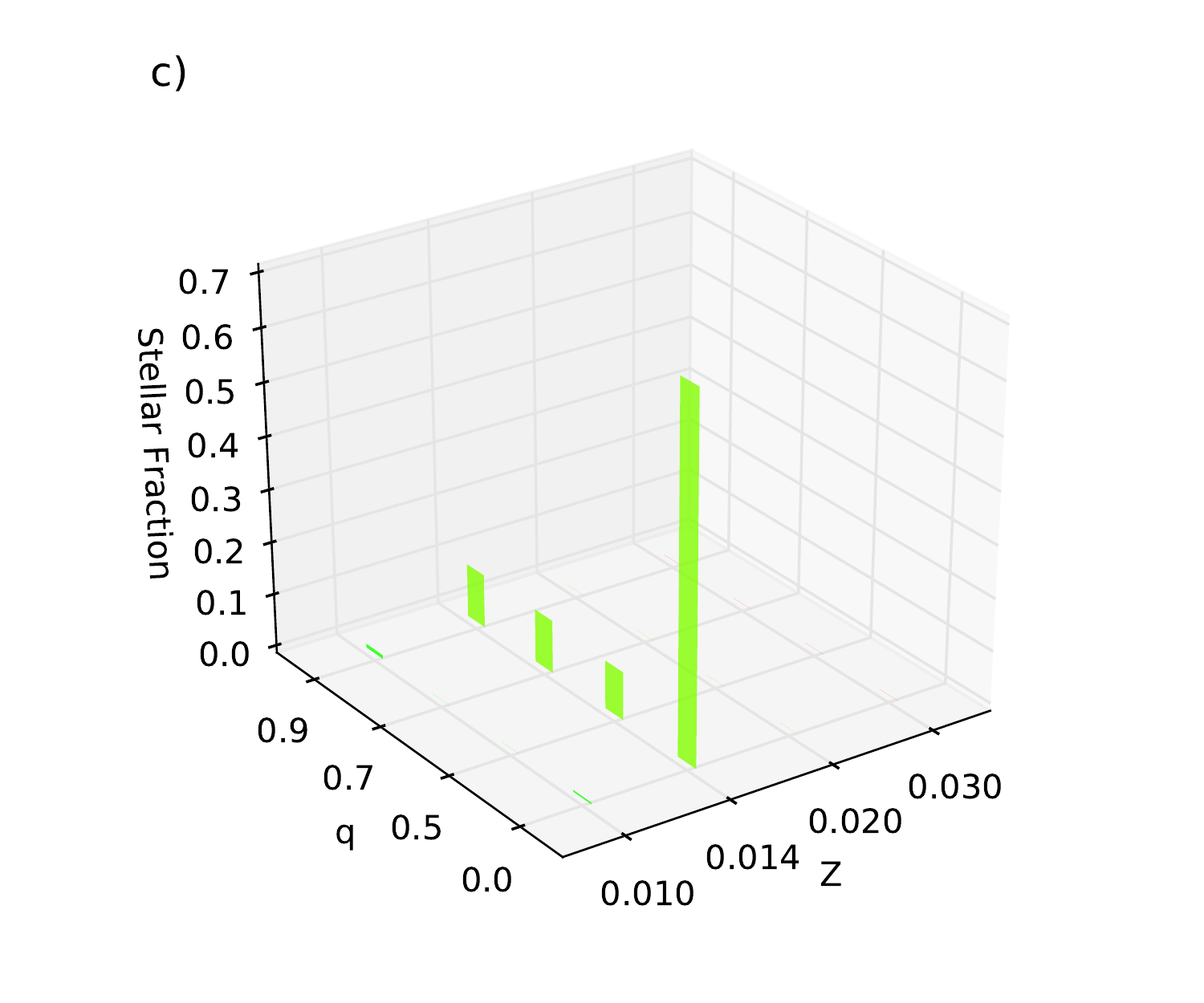}
   \includegraphics[width=0.45\textwidth]{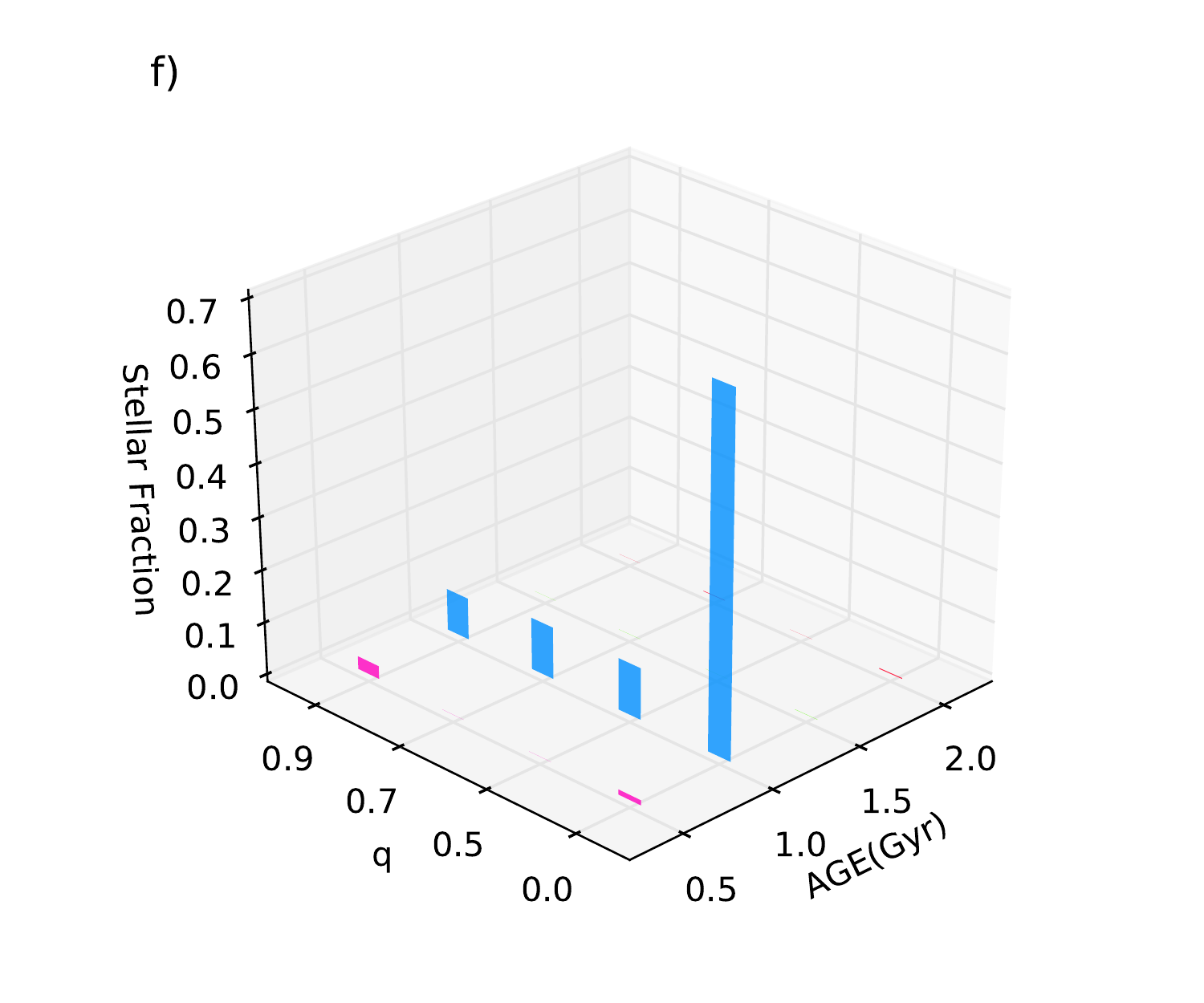}
   \caption{\label{fig:11}CMD of the 1 Gyr $Z$\,=\,$0.014$ mock binary population together with single star isochrones for {\it (a)} 1 Gyr and various
   metallicities, and {\it (d)} $Z$\,=\,$0.014$ and various ages.
   Recovered $Z$ distribution if the presence of binary stars is ignored {\it (b)}, or included in our statistical model {\it (c)}.
   Recovered age distribution if the presence of binary stars is ignored {\it (e)}, or included in our statistical model {\it (f)}.}
\end{center}
\end{figure*}

\section{Unresolved binary stars}\label{ub}

We think that the behavior of the residuals in Fig.\,\ref{fig:9} is due to the fact that in our derivation of the vector $\bm{a}$ we do not correct for the presence of unresolved binary stars in the \solm sample. 
The biases introduced by unresolved binaries in the determination of the SFH using our statistical model will be the subject of a separate paper. Here we explore with a simple {\it toy-model} the expected trends introduced in the derived AMD by ignoring the presence of unresolved binaries in the \solm\ sample. 

\subsection{Footprint of unresolved binaries in the CMD}\label{fp}
For illustration, in this section we show the effects of unresolved binary systems on the position of the stars in the CMD.
The {\it thick gray line} in Fig.\,\ref{fig:10}a is the isochrone described by a 1 Gyr $Z=0.014$ population.
The {\it blue dots} indicate from {\it top to bottom} the position in the MS of single stars of mass $M_1 = 2, 1, 0.5$ and $0.2$ \msun.
At this age stars with $M \geq 2.3$ \msun\ are already in the WD cooling sequence.
The {\it big black dot} signals the position in the WD cooling sequence of a star whose progenitor mass in the MS was 3 \msun. 
The {\it blue arches} describe {\it counter clockwise} the position occupied by unresolved binaries of mass $M = M_1 + M_2$, where the mass $M_2$ of the secondary star obeys $M_2 = qM_1$, and the $q$ parameter is varied in the range $[0.1,1]$ following the \citep{sana12} distribution for $q$. At the bright end of the blue arches $q=1$, $M_2=M_1$, and the unresolved pair is at its brightest, $0.75$ mag brighter than a single star of mass $M_1$.
The {\it gray dashed line} is the same isochrone but displaced by $-0.75$ mag to indicate this absolute limit.

The {\it red dot} in Fig.\,\ref{fig:10}a corresponds to an unresolved pair whose primary star is the WD of progenitor mass $M_1=3$ \msun\ ({\it big black dot}) and $M_2=0.3$ \msun\ ($q=0.1$). The {\it small black dots} indicate {\it counter clock wise} the position of the pair as the mass varies from $M_2=0.3$ to $2.28$ \msun\ ($q=0.76$). For $M_2 \geq 0.6$ \msun\ ($q\geq 0.2$) the pair is dominated by the secondary star and the WD primary goes imperceptible. For $q>0.76$ the secondary is also a WD and the unresolved pair appears in a cooling sequence slightly brighter than the single WD sequence. The last dot corresponds to $q=1$ and the system appears 0.75 mag brighter that any of its members.

Thus, unresolved binary systems broaden the locus defined by single star isochrones in the CMD. An unresolved companion makes a MS primary star look {\it redder, brighter and more massive} in the CMD than it really is. It is then natural that our Bayesian inference lacks stars that are near the MS but redder than in the single star isochrones used as priors. When the primary star is a WD the unresolved system may appear bluer but these systems are less frequent. 

\subsection{Mock population with unresolved binaries}\label{mp}
Following \citet[][]{kouwenhoven09} and \citet{reipurth93} we define the {\it multiplicity or binary fraction} $\mathcal{B}$ of a 
stellar population as
\begin{equation}
\mathcal{B} = \frac{B + T + ...}{S + B + T + ....},
\end{equation}
where $S$ is the number of single stars, $B$ the number of binaries, $T$ the number of triple systems, and so on. The number of systems is $\mathcal{S}$\,=\,$S$\,+\,$B$\,+\,$T$\,+\,$...$\,, and the total number of (individual) stars
is $\mathcal{N}$\,=\,$S$\,+\,$2B$\,+\,$3T$\,+\,$...$\,. Here we consider only single and binary stars, ignoring higher-order systems. Then
\begin{equation}
\mathcal{B} = \frac{B}{S + B},\ \ \ \ \ \ \ \ \ \ \ \mathcal{S}\,=\,S\,+\,B,\ \ \ \ \ \ \ \ \ \ \ \mathcal{N}\,=\,S\,+\,2B. 
\end{equation}
\noindent

We build a mock $Z$\,=\,0.014, 1 Gyr old stellar population of $\mathcal{S}$\,=\,$110,000$ systems which follow the \cite{kr01} IMF distributed with $\rho(r)$\,=\,constant inside a sphere of radius 100 pc using our MW.mx model (Appendix\,\ref{mw.mx}), assuming that $S$\,=\,$55,000$ and $B$\,=\,$55,000$. This translates into a population of $\mathcal{N}$\,=\,$165,000$ individual stars with $\mathcal{B}$\,=\,$0.5$. In the case of the binary stars we interpret the mass of the system as the mass $M_1$ of the primary star.
As above, for each binary pair we assign a mass $M_2$ to the secondary star following $M_2 = qM_1$, where the $q$ parameter is varied in the range $[0.1,1]$ following the \citep{sana12} distribution for $q$. By construction, $M_1$ follows strictly the \citet{kr01} IMF, whereas $M_2$ does not. In our simulations $M_1$\,$\ge$\,$0.1$\,M$_\odot$, which is the lowest stellar mass in our isochrones. We reject all the binary pairs for which the resulting $M_2$\,=\,$qM_1$\,<\,0.1 M$_\odot$ since we cannot follow the evolution of these stars. We end up with a mock population with an effective $\mathcal{B}$\,$\sim$\,$0.4$, which we consider appropriate as a {\it toy-model}. The {\it black dots} and the {\it black band} in Fig.\,\ref{fig:10}b show the locus in the CMD of the binary population assuming that all binary systems are unresolved. The {\it coloured} lines show the resulting 1 Gyr isochrones for various values of $q$ built as in Section\,\ref{fp}.

\subsection{Recovering the AMD}
In Fig.\,\ref{fig:11}a we show in the CMD the mock population of Section\,\ref{mp} together with 1 Gyr single star isochrones of various metallicities. Fig.\,\ref{fig:11}b shows the recovered $Z$ distribution if we ignore the presence of binaries ($q$\,=\,0). Although the simulated stars have $Z$\,=\,$0.014$, a significant contribution appears at $Z$\,=\,$0.030$. If, on the other hand, we allow for the presence of unresolved binaries by using isochrones for various $q$'s (as in Fig. \ref{fig:10}b), we do recover the correct value $Z$\,=\,$0.014$ and a distribution of $q$ values (Fig. \ref{fig:11}c).

Fig.\,\ref{fig:11}d shows the $Z$\,=\,$0.014$ single star isochrones for four different ages. If we ignore unresolved binaries, the recovered age distribution shows besides the 1 Gyr component, an unexpectedly large contribution at 2 Gyr together with minor contributions at 0.5 and 1.5 Gyr (Fig.\,\ref{fig:11}e). Allowing for binaries, i.e., using isochrones for various $q$'s (as in Fig. \ref{fig:10}b), we do recover the correct age of 1 Gyr and a distribution of $q$ values (Fig. \ref{fig:11}f), and the spurious contributions at the wrong age disappear.

In summary, {\it ignoring the presence of unresolved binaries biases the inferred AMD towards older ages and higher $Z$'s than the true values.} It is possible that the $Z$\,=\,$0.03$ contribution to the AMD seen in Fig.\,\ref{fig:6} disappear once unresolved binaries are properly included in the statistical model. This will be explored in a separate paper.

\begin{figure}
\begin{center}
    \includegraphics[width=1.12\columnwidth]{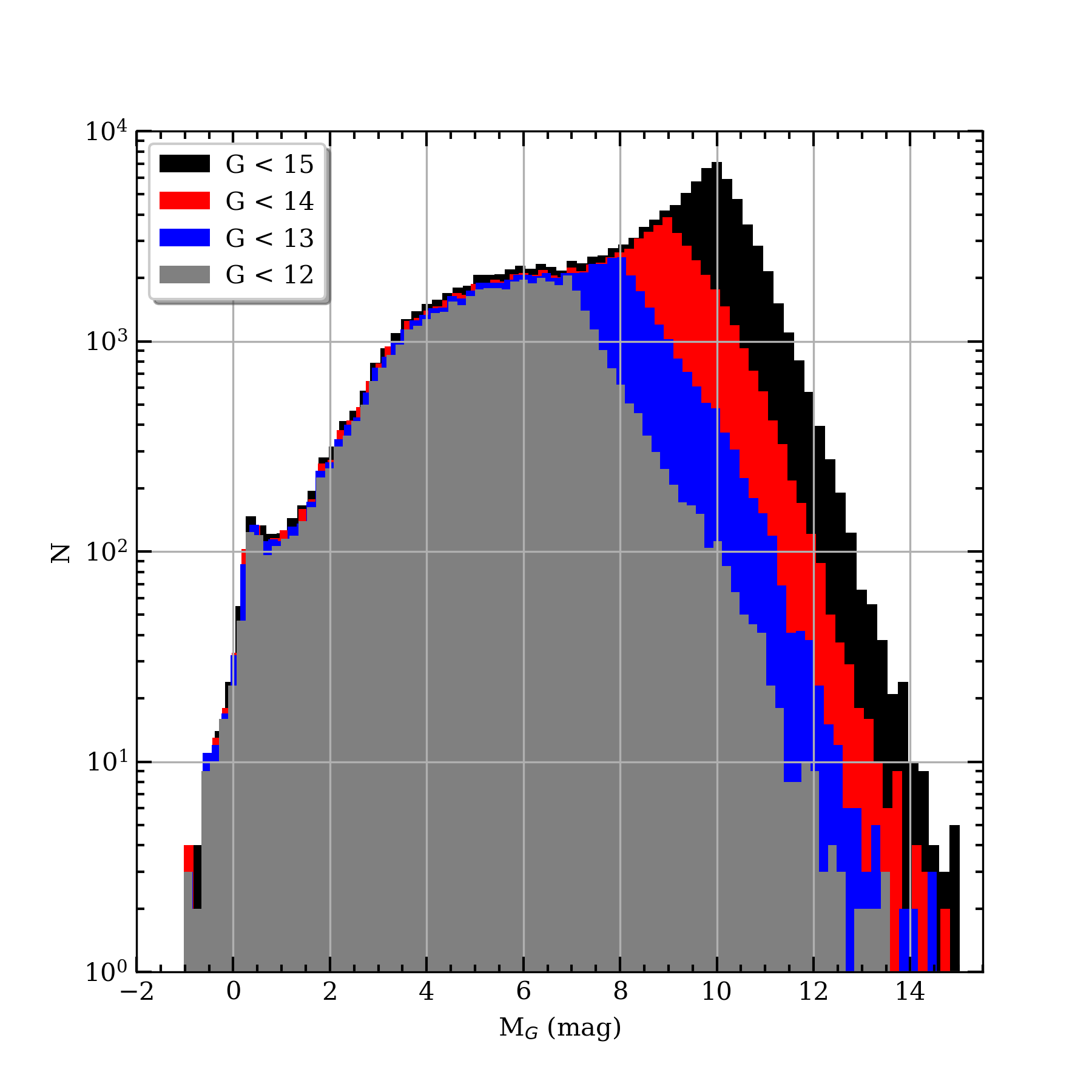}
    \caption{\label{fig:12}Distribution of {\it absolute} $M_G$ magnitude for all the stars in the 
    \sola\ (47,799 stars),
    \solb\ (63,686 stars),
    \solc\ (86,516 stars), and
    \solm\ (120,452 stars) sub-samples.}
\end{center}
\end{figure}

\begin{figure*}
\begin{center}
   \includegraphics[width=\textwidth]{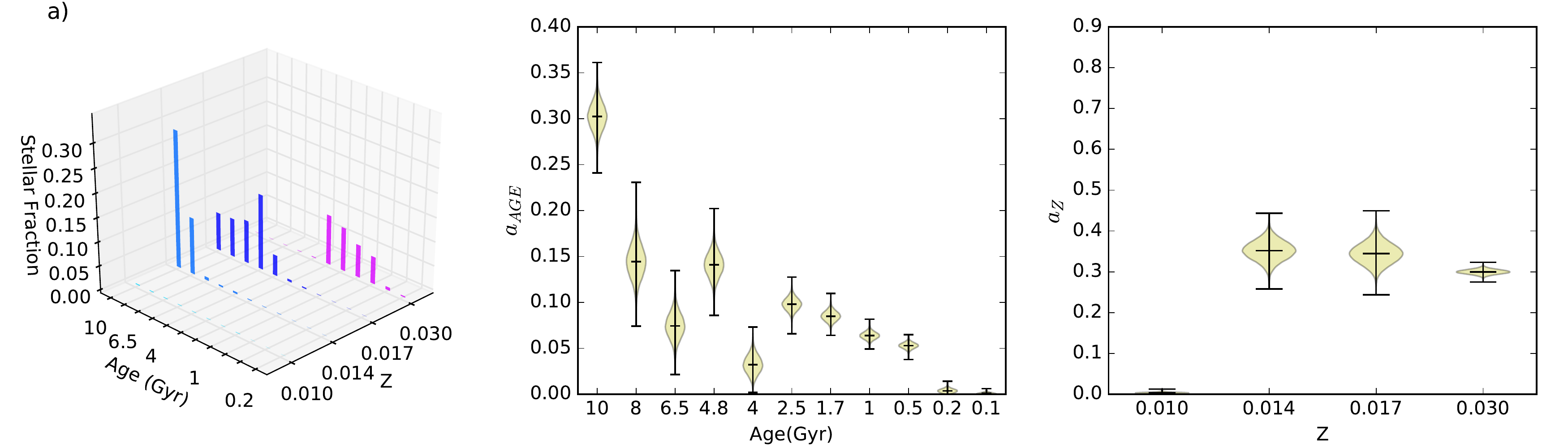}
   \includegraphics[width=\textwidth]{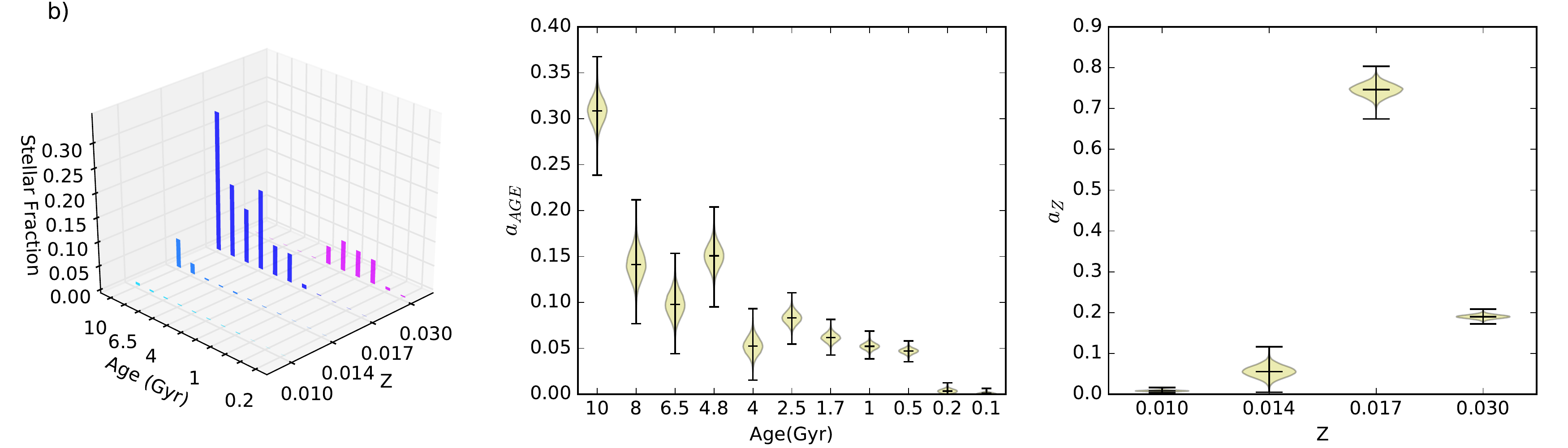}
   \includegraphics[width=\textwidth]{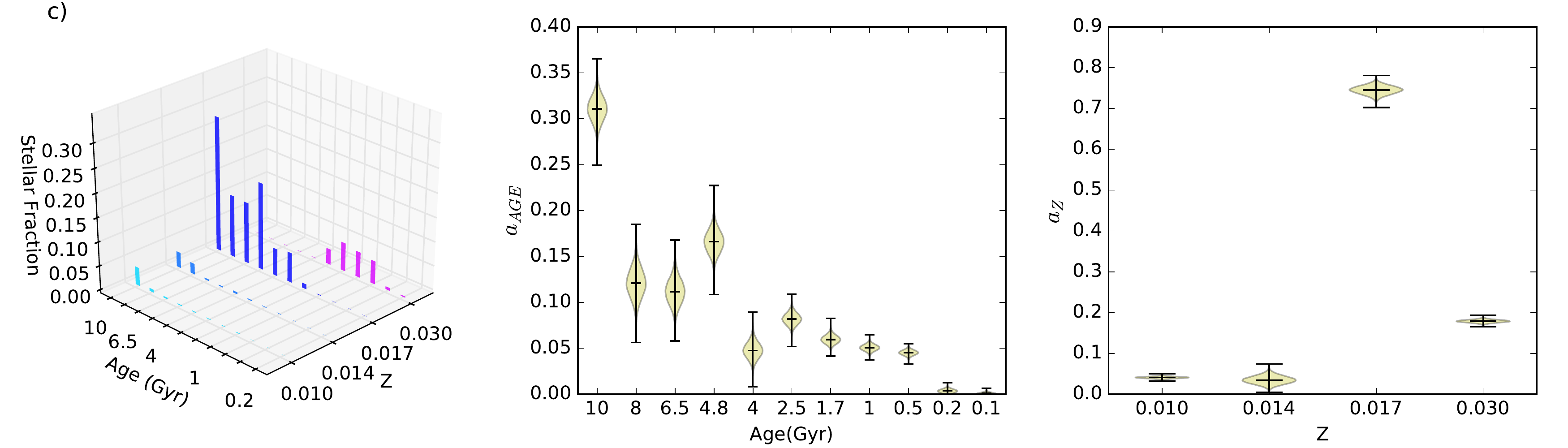}
   \includegraphics[width=\textwidth]{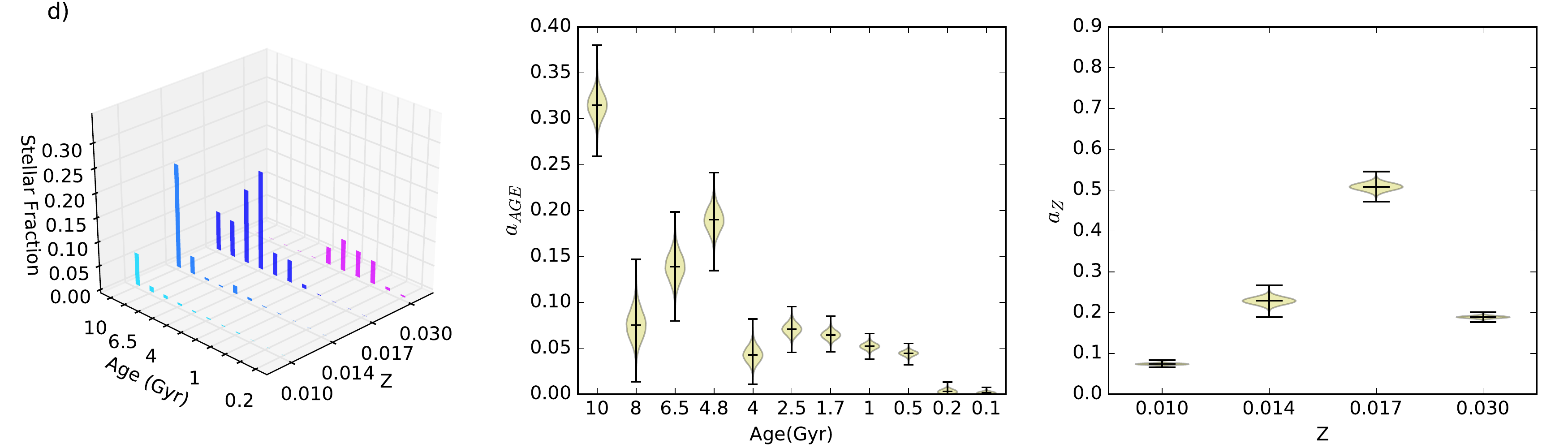}
   \caption{\label{fig:13}AMD for
    {\it (a)} the \sola, 
    {\it (b)} the \solb,
    {\it (c)} the \solc and 
    {\it (d)} the \solm\ samples
   inferred using the grid C of isochrones listed in Table\,\ref{tab:4} with no extinction correction, the \citet{kr01} IMF 
   and $\sigma_{i}=0.075$. The height of the bars in the 3D plots on the left hand side is the median of the distribution
   of $a_i$ for the corresponding isochrone.
   The violin plots summarize the marginalized {\it posterior} PDF for age and $Z$.
   The horizontal lines in each violin represent from bottom to top the 0, 50, and 100 percentiles of the distribution.
   The figure in panel {\it (d)} duplicates Fig.\,\ref{fig:6}c.}
\end{center}
\end{figure*}

\begin{table*}
\begin{center}
 \caption{\label{tab:5}IMF parameters$^a$}
 \begin{tabular}{lcccccccc}
 \hline
\multirow{2}{*}{IMF}& \multirow{2}{*}{$\alpha$} & Mass range  & \multirow{2}{*}{$\mathcal{N}_{*}(m_u)$} & $\mathcal{M}_a$ & \multirow{2}{*}{$f_{\,0.5}$} & \multirow{2}{*}{$f_{\,0.8}$} & \multirow{2}{*}{$f_{\,1.0}$} & \multirow{2}{*}{$f_{\,2.5}$} \\
                    &                           & (M$_\odot$) &                   & (M$_\odot$) \\
 \hline
 \citet{salp55}     &         +2.35             & $m_l \leq m \leq m_u$  & 2.85 & 0.35 & 0.89 & 0.94 & 0.96 & 0.99 \\
\\                                                                                                   
 \citet{kr01}       &         +2.30             & $0.5 \leq m \leq m_u$  & 1.57 & 0.64 & 0.74 & 0.86 & 0.89 & 0.97 \\
                    &         +1.30             & $m_l \leq m  <   0.5$                                            \\
\\                                                                                                   
 \citet{Mor19} - A  &         +1.90             & $1.53 \leq m \leq m_u$ & 0.42 & 2.41 & 0.40 & 0.55 & 0.62 & 0.83 \\
                    &         +1.30             & $0.5  \leq m  <  1.53$                                           \\
                    &         +0.50             & $m_l  \leq m  <  0.5 $                                           \\
\\                                                                                                   
 \citet{Mor19} - B  &         +1.90             & $1.53 \leq m \leq m_u$ & 0.35 & 2.88 & 0.27 & 0.45 & 0.53 & 0.79 \\
                    &         +1.30             & $0.5  \leq m  <  1.53$                                           \\
                    &         -0.50             & $m_l  \leq m  <  0.5 $                                           \\
 \hline
\multicolumn{9}{l}{$^a$For the lowest mass segment of the \cite{Mor19} IMF we use two of their values of $\alpha$: $+0.5$ (case A)}\\
\multicolumn{9}{l}{\ \ \ and $-0.5$ (case B). Recall that the single-star IMF may be steeper than inferred from observations that do}\\
\multicolumn{9}{l}{\ \ \ not resolve binary systems \citep{kr01}.}\\
 \end{tabular}
\end{center}
\end{table*}

\begin{figure*}
\begin{center}
   \includegraphics[width=0.49\textwidth]{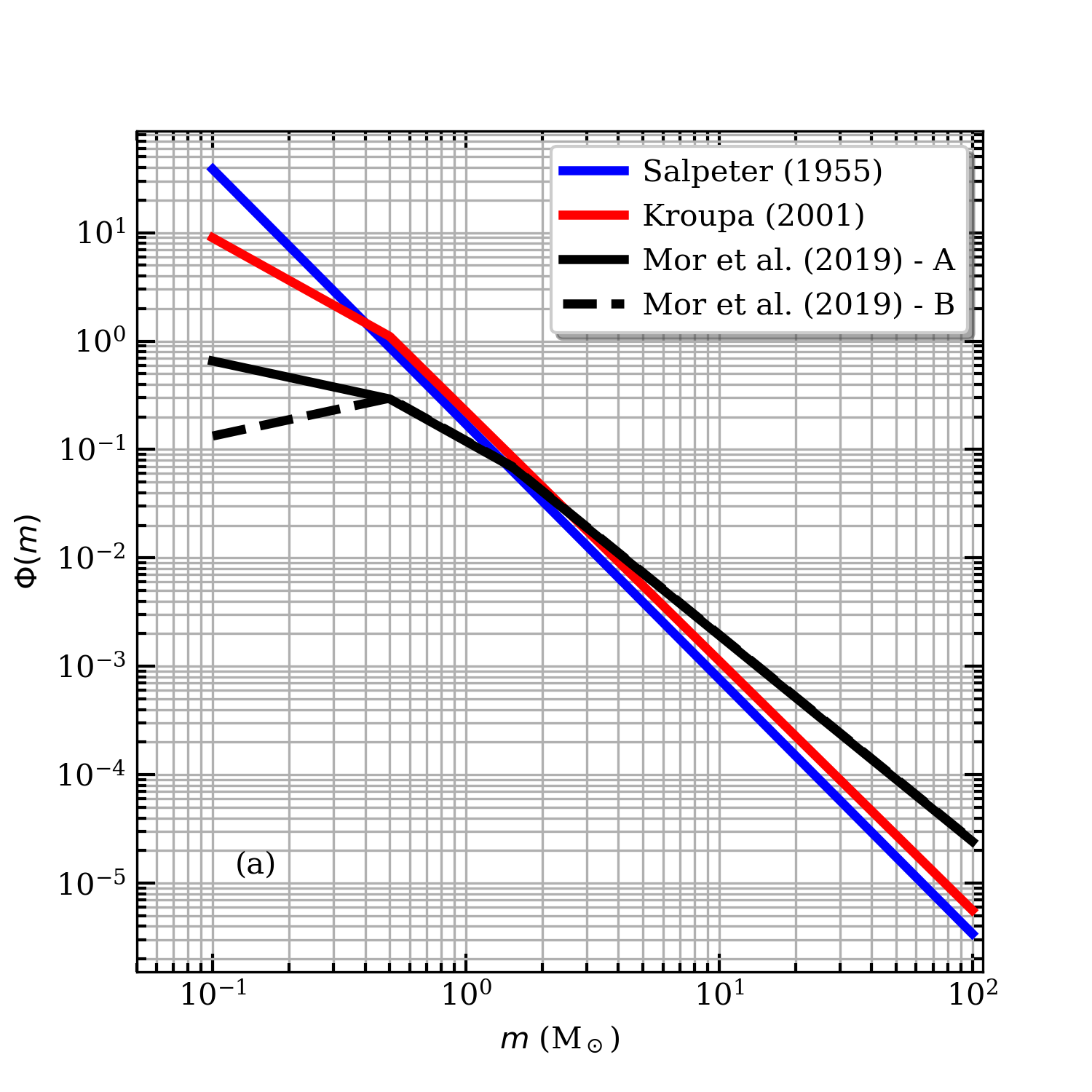}
   \includegraphics[width=0.49\textwidth]{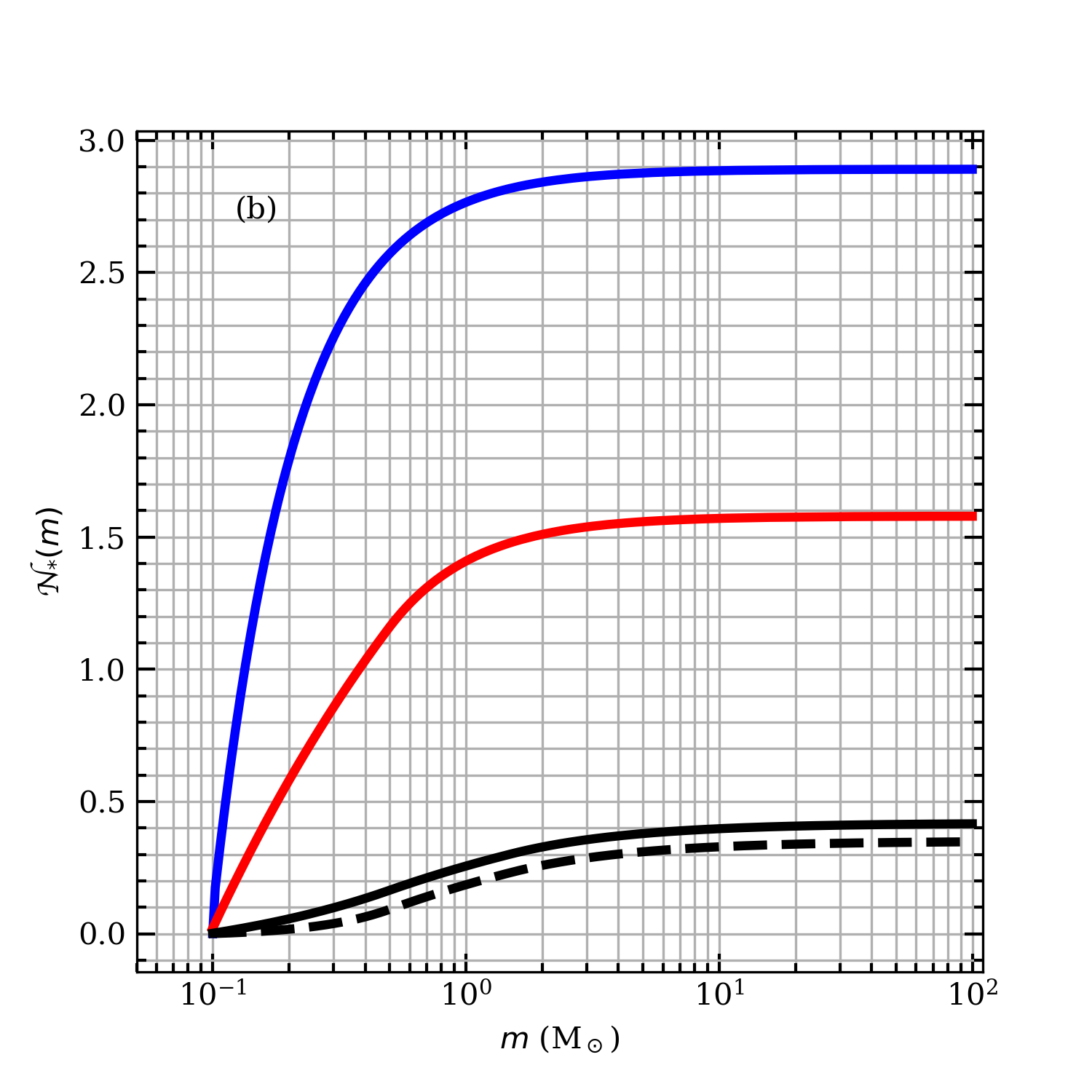}
   \caption{\label{fig:14}{\it (a)} $\Phi(m)$ and {\it (b)} $\mathcal{N}_{*}(m)$ 
   for the \citet{salp55}, \citet{kr01} and \citet{Mor19} IMFs. See Table\,\ref{tab:5} and Eqs.\,(\ref{eq:phi}) to (\ref{eq:frac}).}
\end{center}
\end{figure*}

\begin{figure*}
\begin{center}
   \includegraphics[width=\textwidth]{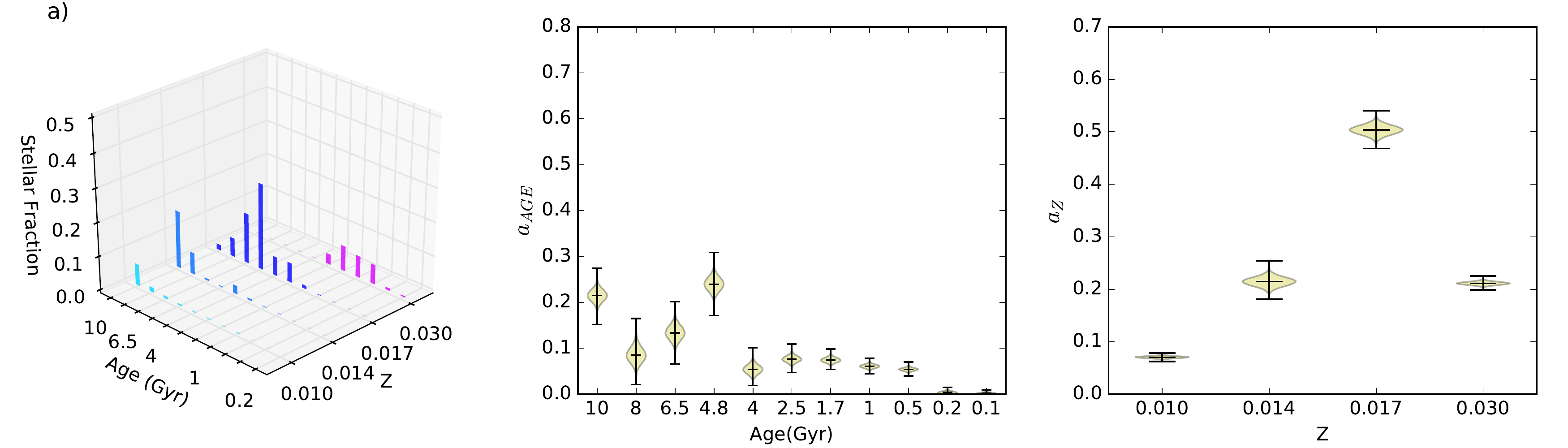}
   \includegraphics[width=\textwidth]{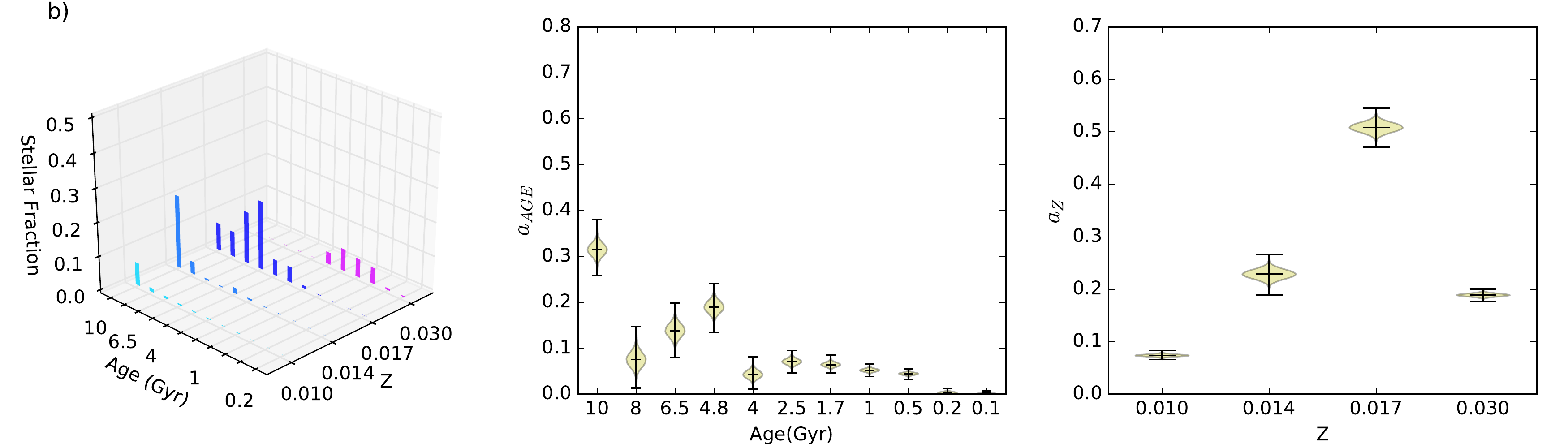}
   \includegraphics[width=\textwidth]{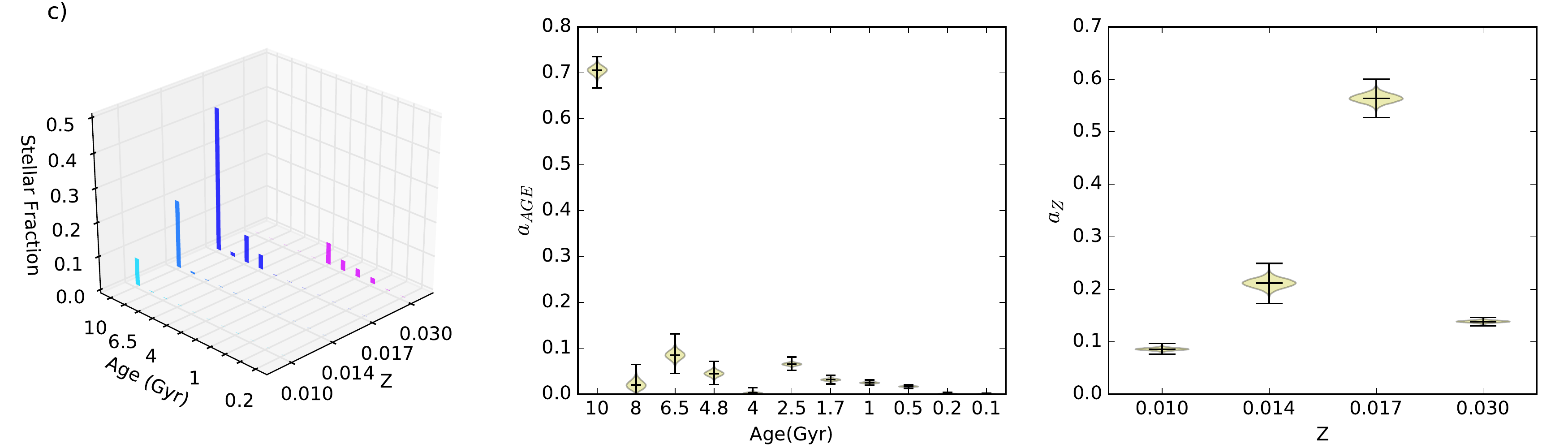}
   \includegraphics[width=\textwidth]{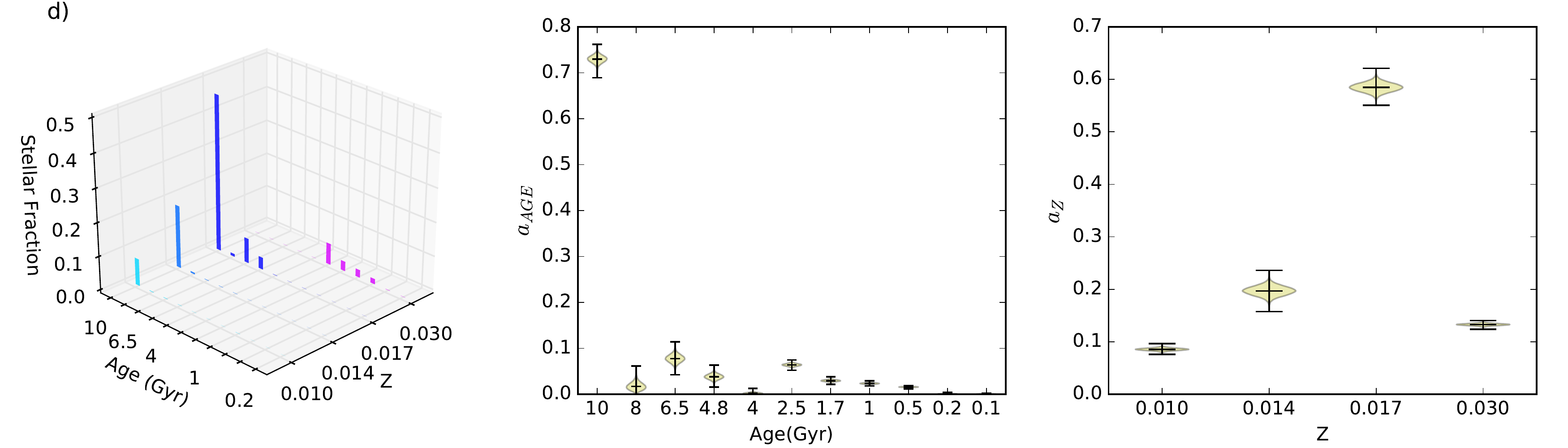}
   \caption{\label{fig:15}AMD for the \solm\ sample inferred assuming
      {\it (a)} the \citet{salp55},
      {\it (b)} the \citet{kr01}, 
      {\it (c)} the case A \citet{Mor19} and
      {\it (d)} the case B \citet{Mor19} IMF, the grid C of isochrones listed in Table\,\ref{tab:4} with no extinction correction, and $\sigma_{i}$\,=\,0.075.
      The height of the bars in the 3D plots on the left hand side is the median of the distribution of $a_i$ for the corresponding isochrone.
      The violin plots summarize the marginalized {\it posterior} PDF for age and $Z$.
      The horizontal lines in each violin represent from bottom to top the 0, 50, and 100 percentiles of the distribution.
      The figure in panel {\it (b)} duplicates Fig.\,\ref{fig:6}c with a different vertical scale.}
\end{center}
\end{figure*}

\section{AMD and sample limiting magnitude}\label{slm}

In this section we explore the dependence of the inferred AMD on the limiting magnitude of the stellar sample. 
We consider the \sola, \solb and \solc\ {\it sub-samples}, defined similarly to \solm\ but using $G$\,=\,12, 13, 14, respectively, as limiting {\it apparent} magnitude (see Section\,\ref{sec_dat_desc} and Fig.\ref{fig:1}).
In Fig.\,\ref{fig:12} we show as histograms the distribution of the {\it absolute} $M_G$ magnitude for all the stars in each sub-sample.

In Fig.\,\ref{fig:13} we compare the AMDs inferred for each sub sample using the grid C of isochrones listed in Table\,\ref{tab:4} with no extinction correction. The marginalized $Z$ distribution varies with the limiting $G$ magnitude, especially at the bright end. 
From Fig.\,\ref{fig:1}b we see that in the MS in the range $5$\,$\le$\,$M_G$\,$\le$\,$7$ the isochrones are degenerate with respect to $Z$. Fig.\ref{fig:12} shows that there are very few stars fainter than $M_G$\,=\,$7.5$ in the \sola\ sample. This explains why the $Z$ distribution in Fig.\,\ref{fig:13}a is flat: it follows the prior since there is no leverage to determine $Z$ in this magnitude range.
As we consider fainter limiting magnitudes the number of stars with $M_G$\,>\,$7.5$ increases considerably; more than half
of the stars in \sols\ are fainter than $M_G$\,=\,$8$. In this regime the isochrones are well separated in $G_{BP}$\,-\,$G_{RP}$ colour and inferring $Z$ becomes possible. 
The broad distribution of MS stars in the CMD in the range $8$\,$\le$\,$M_G$\,$\le$\,$12$ cannot be due only to unresolved binaries or photometric and astrometric errors or to a combination of these factors, and must reflect a true dispersion in the $Z$
value of the stars in \sols.
The contribution of the $Z$\,=\,$0.017$ and $Z$\,=\,$0.030$ components is a robust result for limiting $G$\,$\ge$\,$13$. In the \solm\ sample the number of stars fainter than $M_G$\,=\,$8$ is large enough to allow the inference of a finer $Z$ distribution: the $Z$\,=\,$0.017$ contribution decreases in favour of a relatively large $Z$\,=\,$0.014$ component and a minor one at $Z$\,=\,$0.010$.
The marginalized $Z$ distribution for the stars in \sols\ is then {\it robust with respect to the sample limiting magnitude} 
for $13$\,$\le$\,$G$\,$\le$\,$15$.

The marginalized age distributions in Fig.\,\ref{fig:13} are similar for the four samples.
As it is apparent from Fig.\,\ref{fig:1}b, isochrones of different age are well separated in the CMD for $M_G$\,$\le$\,$5$.
The inferred age distribution is thus determined by the stars in the sample brighter than $M_G$\,=\,$5$.
Fig.\,\ref{fig:12} shows that increasing the limiting apparent magnitude has no effect at the bright $M_G$ end.
Then, the relevant stars for the age inference are the same for the four samples, resulting in a stable age distribution.
The increasing but still low number of WD stars towards fainter limiting magnitudes is not large enough to modify the inferred age distributions.
The marginalized age distribution for the stars in \sols\ is then {\it robust with respect to the sample limiting magnitude} 
for $12$\,$\le$\,$G$\,$\le$\,$15$.

It is possible that some (or all) of the stars in the ($Z$\,=\,$0.014$, 10 Gyr) and ($Z$\,=\,$0.010$, 10 Gyr) bins belong to the
{\it thick} Galactic disk, as opposed to the {\it thin} Galactic disk for the rest of the stars.
This will be explored in detail in a future paper extending our sample to $G$\,>\,$15$.

\section{AMD and the stellar IMF}\label{imf}

\begin{figure}
\begin{center}
   \includegraphics[width=1.1\columnwidth]{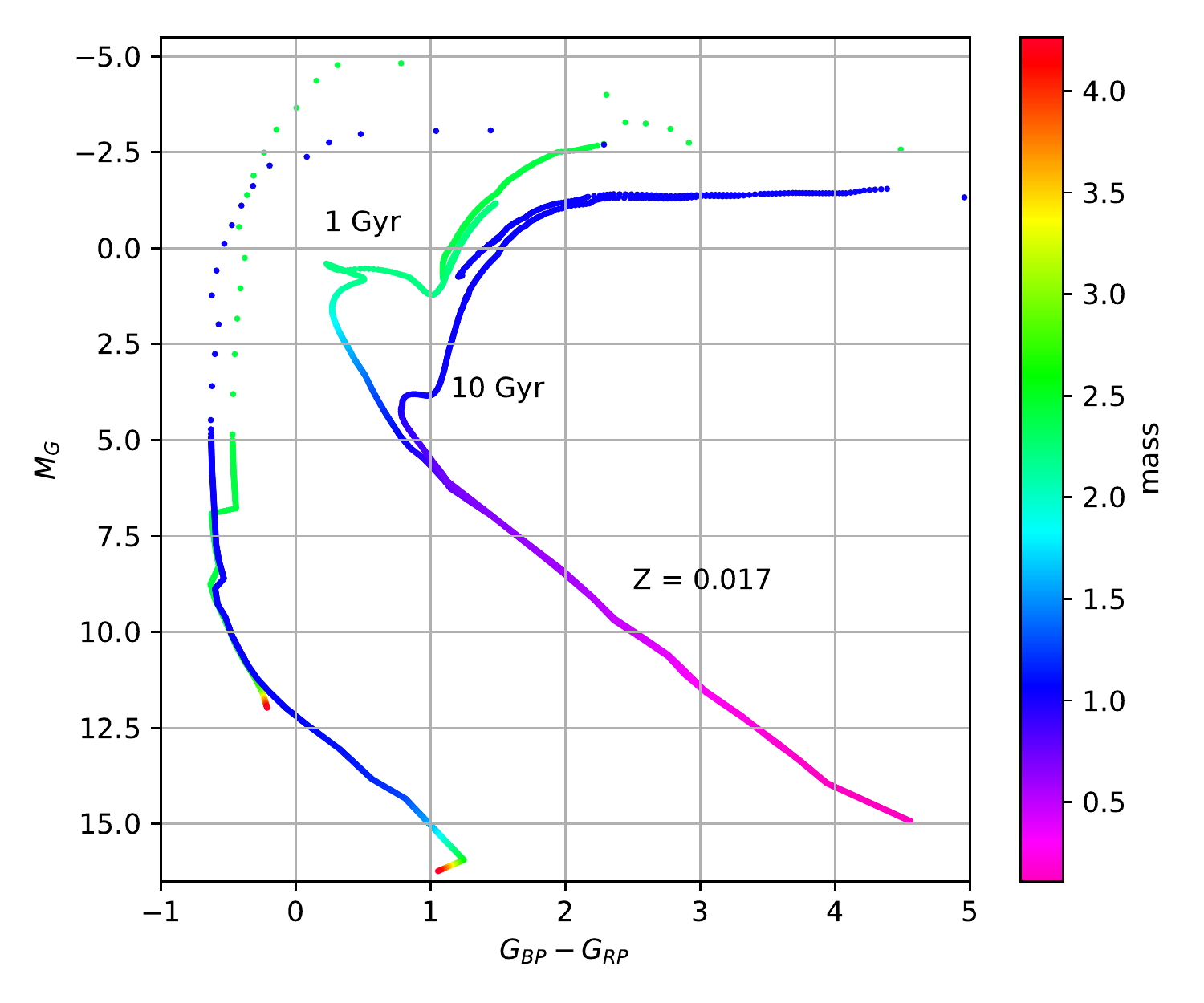}
   \caption{\label{fig:16a}1\,and\,10\,Gyr $Z$\,=\,$0.017$ isochrones, colour coded according to the stellar mass along the isochrone.
   At 1\,Gyr, stars of mass 2.5 M$_\odot$ are found at the MS {\it turn-off} and the PostMS evolutionary phases,
   whereas at\,10\,Gyr these stars are already at the end of the WD cooling sequence, and the MS {\it turn-off} is near 1\,M$_\odot$.}
\end{center}
\end{figure}

\begin{figure}
\begin{center}
   \includegraphics[width=1.1\columnwidth]{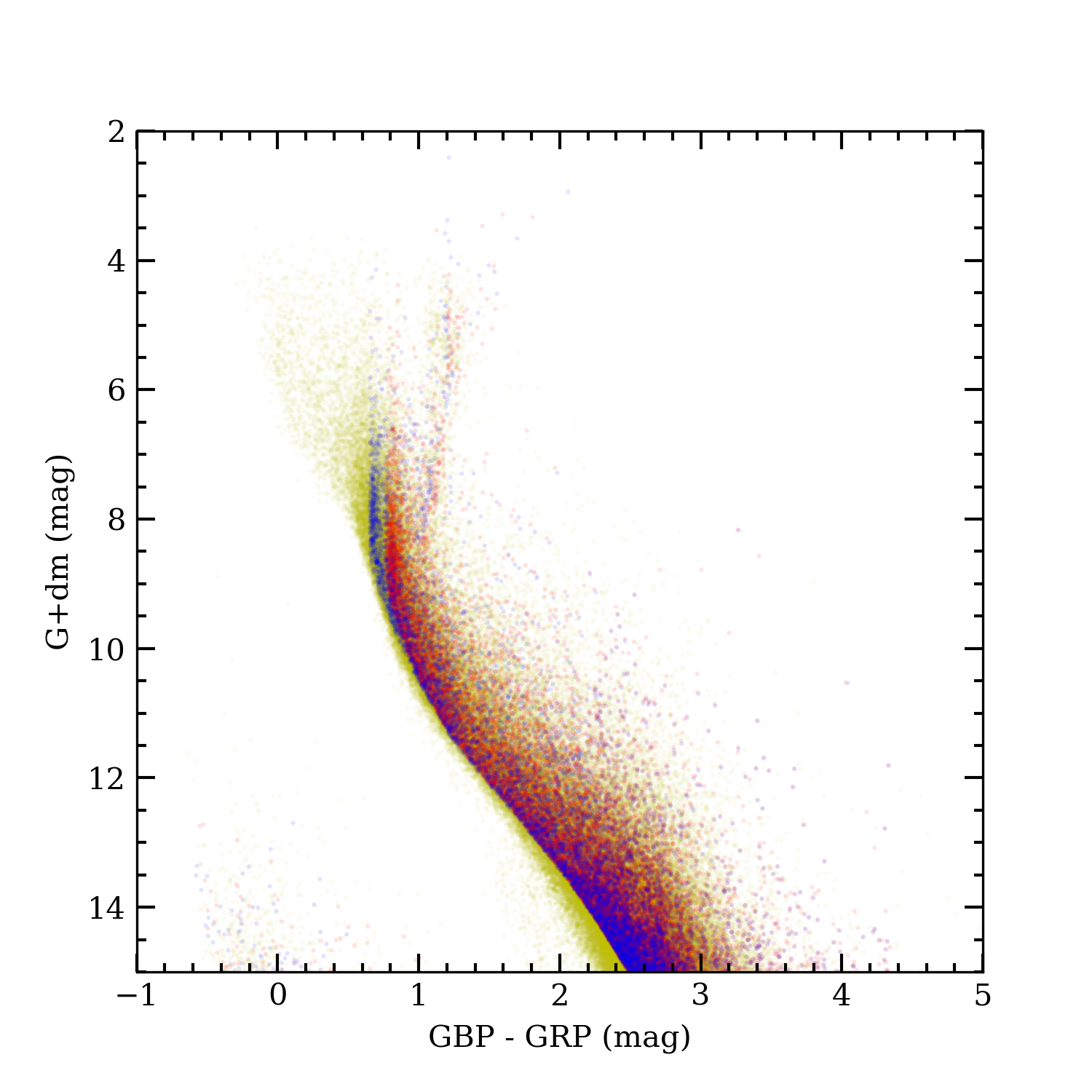}
   \caption{\label{fig:16b}Mock population in the CMD superimposed on the \solm\ CMD. The SFH is shown in Fig.\,\ref{fig:17}a. The oldest population is formed in a
    constant SFR burst lasting from 10 to 7 Gyr ago ({\it red dots}). There is a period with no star formation from 7 to 5 Gyr ago, and
    then a younger population is formed in a second burst lasting from 5 to 3 Gyr ago ({\it blue dots}). Star formation stops again 3
    Gyr ago. 67\% of the stars belong to the old population and 33\% to the younger component. We assume that stars form
    following the \citet{kr01} IMF with constant metallicity $Z$ = 0.017.}
\end{center}
\end{figure}

\begin{figure*}
\begin{center}
   \includegraphics[width=0.33\textwidth]{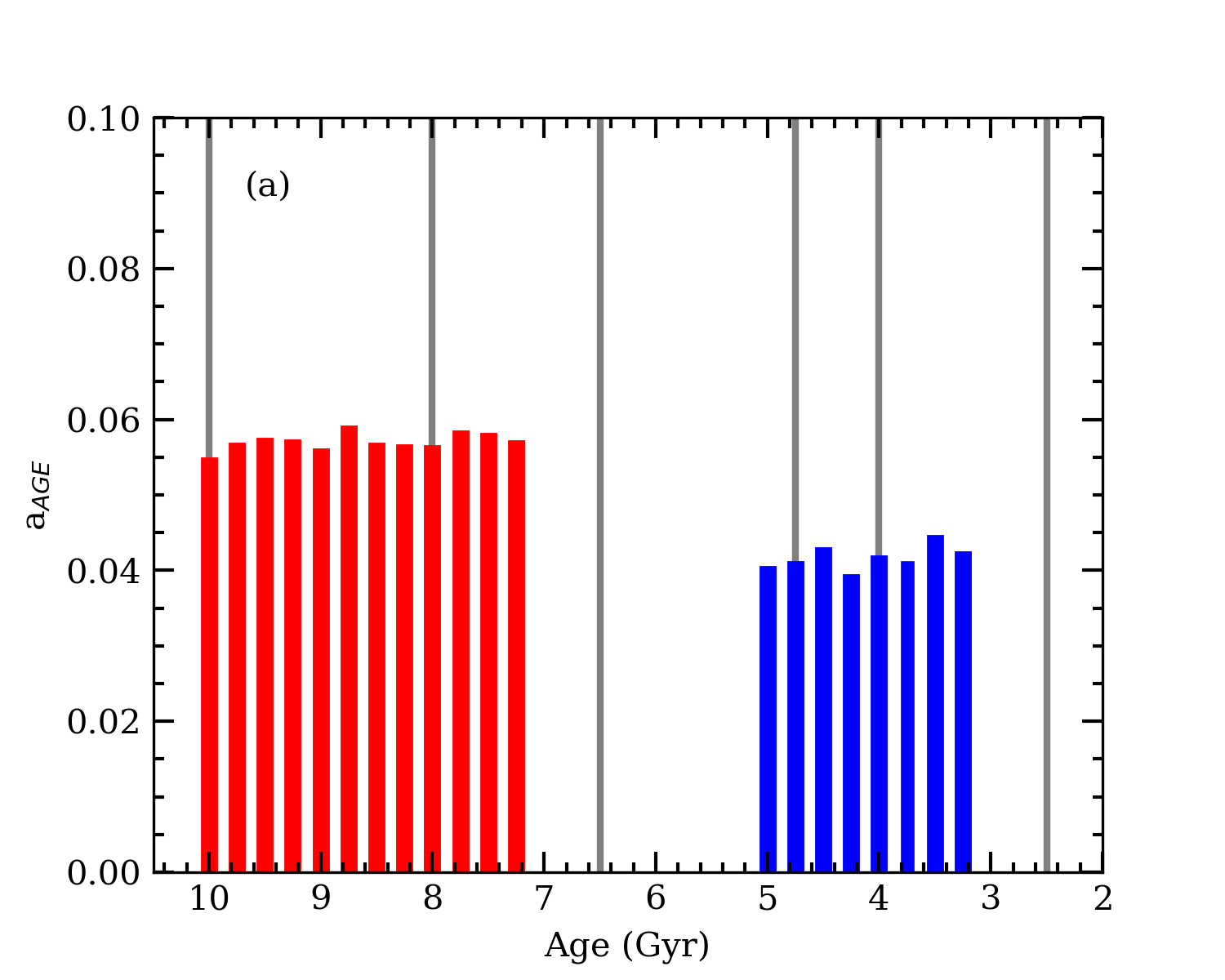}
   \includegraphics[width=0.33\textwidth]{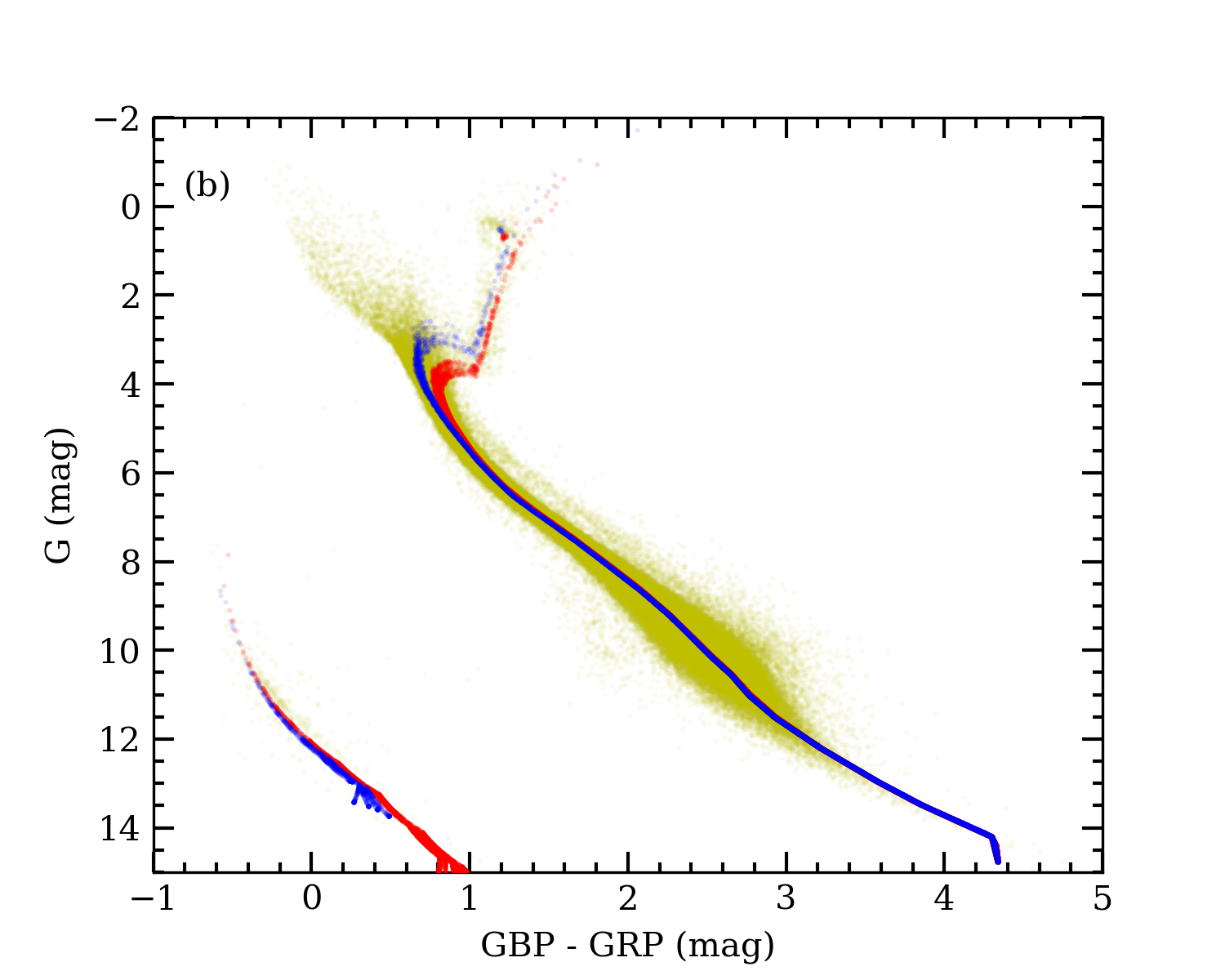}
   \includegraphics[width=0.33\textwidth]{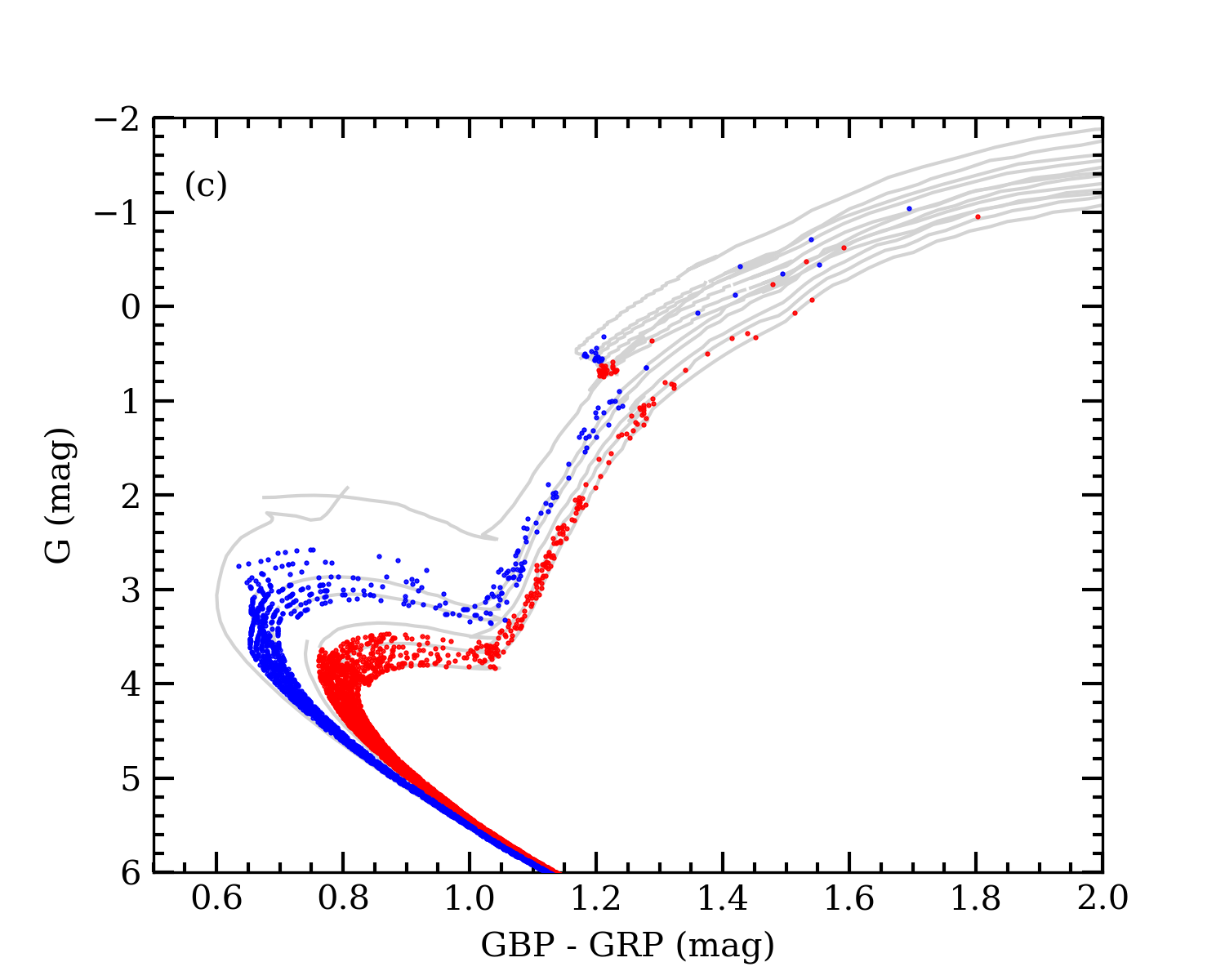}
   \includegraphics[width=\textwidth]{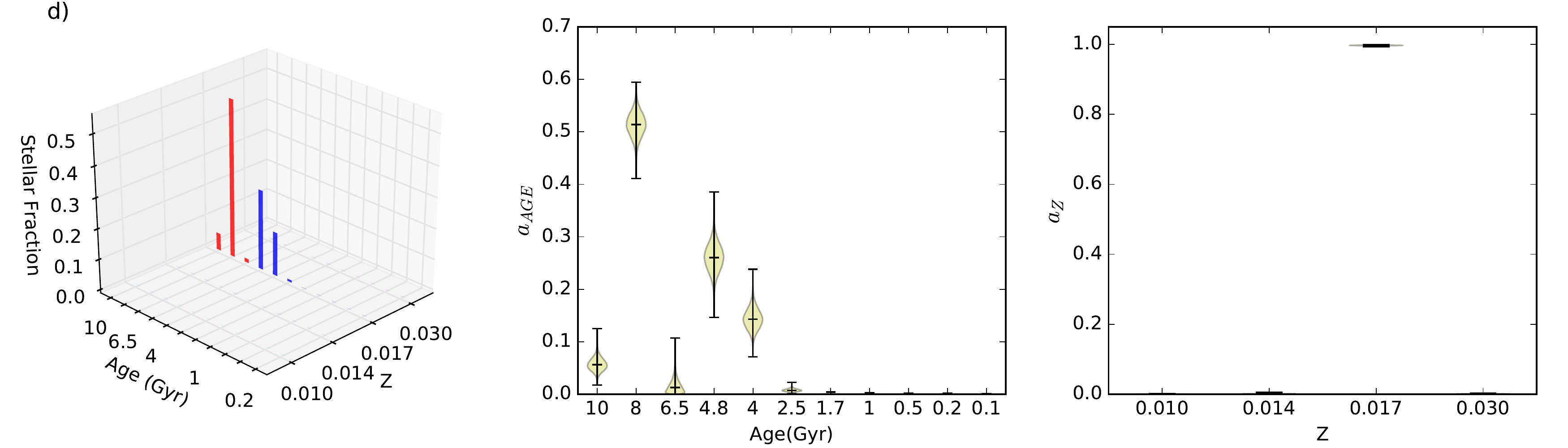}
   \includegraphics[width=\textwidth]{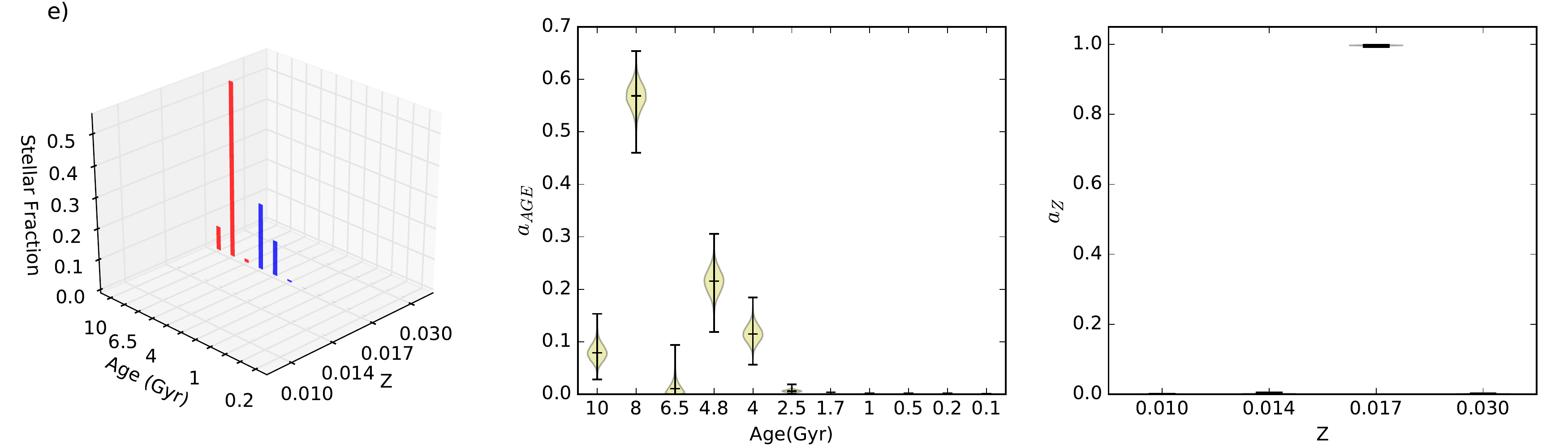}
   \includegraphics[width=\textwidth]{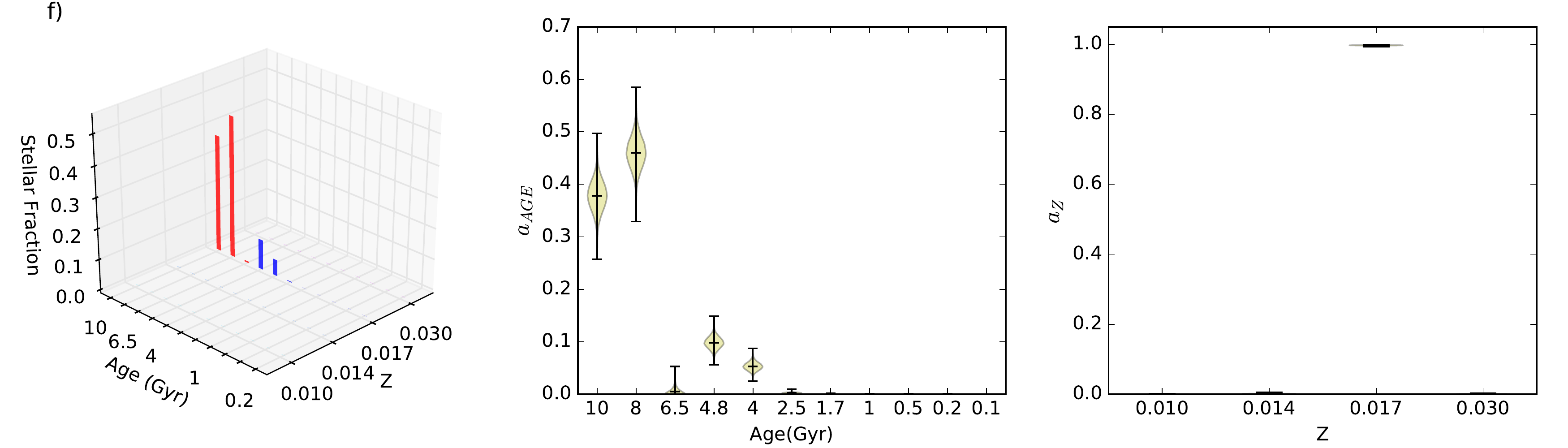}
   \caption{\label{fig:17}{\it (a)} SFH for the mock stellar population. {\it (b)} CMD of the mock population superimposed on the
   \solm\ CMD. {\it (c)} Enlargement of {\it (b)} showing the 10, 8, 6.5, 4.8, 4, and 2.5 Gyr isochrones, corresponding to the
   {\it gray vertical lines} in {\it (a)}. The stars formed in the old (young) burst in {\it (a)} are shown as  {\it red (blue)} dots in {\it (b)}
   and {\it (c)}. AMD for the mock sample inferred assuming the grid C of isochrones listed in Table\,\ref{tab:4} for {\it (d)} the
   \citet{salp55}, {\it (e)} the \citet{kr01} and {\it (f)} the case A \citet{Mor19} IMF. The height of the bars in the 3D plots on 
   the left hand side is the median of the distribution of $a_i$ for the corresponding isochrone. The violin plots summarize the
   marginalized {\it posterior} PDF for age and $Z$. The horizontal lines in each violin represent from bottom to top the 0, 50, and
   100 percentiles of the distribution.}
   \end{center}
\end{figure*}

In the previous section we discussed how the lack of faint stars in the sample affects the inferred AMD.
In this section we explore how the number of stars in specific mass ranges in the statistical model due to the assumed IMF 
modifies the inferred AMD. To this end we compare the AMDs inferred for \solm\ using the \cite{salp55}, the \cite{kr01}, and the \cite{Mor19} IMFs, displayed in Fig.\,\ref{fig:14} and parametrized as indicated in Table\,\ref{tab:5}.
The IMF, written as
\begin{equation}
\Phi(m) = dN/dm = Cm^{-\alpha} = Cm^{-(1+x)},
\label{eq:phi}
\end{equation}
gives the number of stars of mass between $m$ and $m$\,+\,$dm$ born in a star formation event.
The constant $C$ in Eq.\,(\ref{eq:phi}) is determined from the normalization condition
\begin{equation}
\int_{m_l}^{m_u}\Phi(m)mdm = 1\,{\rm M}_\odot,
\label{eq:norm}
\end{equation}
where $m_l$ and $m_u$ are the lower and upper mass limits of star formation, respectively.
The number of stars formed at birth from mass $m_l$ to mass $m$ is
\begin{equation}
\mathcal{N}_{*}(m) = \int_{m_l}^{m}\Phi(m)dm.
\label{eq:nstr}
\end{equation}
The average stellar mass is given by
\begin{equation}
\mathcal{M}_{a} = \frac{\int_{m_l}^{m_u}\Phi(m)mdm}{\int_{m_l}^{m_u}\Phi(m)dm} = \frac{1}{\mathcal{N}_{*}(m_u)},
\label{eq:mav}
\end{equation}
and the fraction of stars formed at birth from mass $m_l$ to mass $m$ is
\begin{equation}
f_m = \frac{\int_{m_l}^{m}\Phi(m)dm}{\int_{m_l}^{m_u}\Phi(m)dm}.
\label{eq:frac}
\end{equation}
We note that $\mathcal{N}_{*}(m)$ is the number of stars formed per unit solar mass, whereas $\mathcal{M}_{a}$ and $f_m$ are independent of the normalization of the IMF. As throughout this paper, we assume $(m_l,m_u)$\,=\,$(0.1,100)$\,M$_\odot$ for all isochrones and all IMFs. The resulting $\Phi(m)$ and $\mathcal{N}_{*}(m)$ are shown in Fig.\,\ref{fig:14} for the various IMF's. $\mathcal{N}_{*}(m_u)$, $\mathcal{M}_{a}$ and $f_m$ are listed in Table\,\ref{tab:5}. $f_{\,0.5}$, $f_{\,0.8}$, $f_{\,1.0}$ and $f_{\,2.5}$ represent the fraction of stars at birth with $m$\,$\leq$\,$0.5$, $0.8$, $1.0$ and $2.5$\,M$_\odot$, respectively. The IMFs in Table\,\ref{tab:5} are sorted in order of increasing fraction of massive stars, i.e., $1\,-\,f_m$. From this table and Fig.\,\ref{fig:14}b we note that 
($f_{\,0.5},\,f_{\,0.8},\,f_{\,1.0},\,f_{\,2.5}$) range from ($89,\,94,\,96,\,99$)\% for the \cite{salp55} IMF to ($27,\,45,\,53,\,79$)\% for the \cite{Mor19} IMF. The number of stars populating an isochrone declines more sharply as a function of age when we use the \cite{Mor19} IMF than for the \cite{kr01} and \cite{salp55} IMFs (cf. Fig.\,\ref{fig:16a}).

In Fig.\,\ref{fig:15} we compare the AMDs inferred for the \solm\ sample using the four IMF's listed in Table\,\ref{tab:5}.
{\it The four AMDs in Fig.\,\ref{fig:15} show the old burst at 10 Gyr, the star formation minimum (quenching) near 8 Gyr, the second
burst from 4 to 6 Gyr, and the residual star formation at recent epochs at $Z$ above solar} discussed in Section\,\ref{amd}.
For the \cite{salp55} and \cite{kr01} solutions, {\it the star formation episodes are bell-shaped or flat as a function of age}.
Instead, for the \cite{Mor19} solutions {\it the bursts peak at the starting age and then decay in time}.
There is a clear anti correlation between $a_i$ and $f_{\,1.0}$ noticeable in the $Z$\,=\,$0.017$ 10 Gyr component. 
From the 3D panels of Fig.\,\ref{fig:15} we see that whereas for the \cite{salp55} IMF (large $f_{\,1.0}$) the 10 Gyr component of $a_i$ is almost entirely due to the $Z$\,=\,$0.014$ population, the contribution of the $Z$\,=\,$0.017$ population at 10 Gyr increases markedly as we switch to the \cite{kr01} IMF (intermediate $f_{\,1.0}$) and the \cite{Mor19} IMFs (low $f_{\,1.0}$). 
This is also clearly noticeable in the marginalized age distribution (10 Gyr bin) and to a lesser extent in the marginalized $Z$
distribution (0.017 bin). 
Even though the contribution of the $Z$\,=\,$0.014$ population at 10 Gyr seems constant, it represents less stars as we 
move to the lower $f_{\,1.0}$ IMFs, in which case the $Z$\,=\,$0.017$ stars are preferred.
As discussed in Section\,\ref{slm}, the inference of the $Z$ distribution depends on the non degeneracy of isochrones for $M_G$\,$\geq$\,$8$. The observed behaviour of the $a_Z$ distribution tells us that the inference of $Z$ is more sensitive to the location of the stars in the CMD than to the number of stars on each isochrone.
The marginalized $Z$ distribution in Fig.\,\ref{fig:15} is less sensitive to variations in the number of stars according to the IMF
than the marginalized age distribution.

To explore further the dependence of the inferred AMD on the assumed IMF we use a mock stellar population whose true parameters we know. 
\begin{figure*}
\begin{center}
   \includegraphics[width=1.02\textwidth]{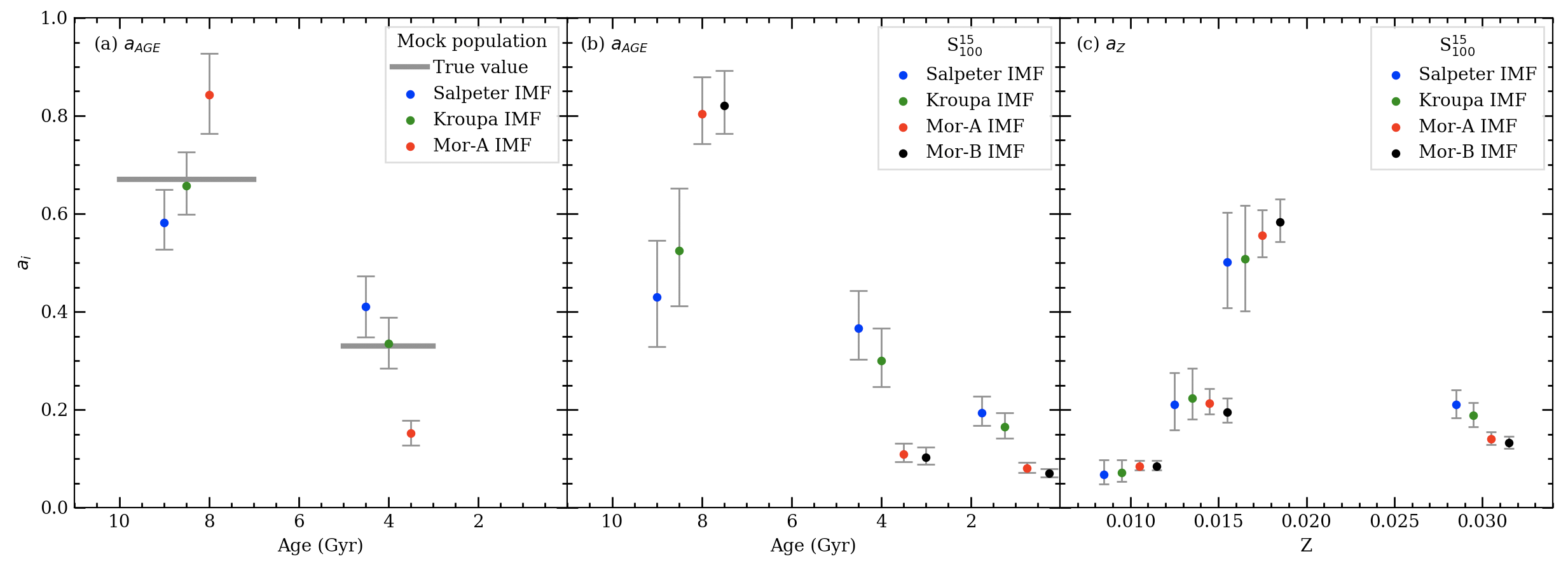}
   \caption{\label{fig:18}{\it (a)} Median values of $a_o$ and $a_y$ for the \citet{salp55} IMF ({\it blue dots}), the \citet{kr01} IMF
   ({\it green dots}) and the case A \citet{Mor19} IMF ({\it red dots}). The height of the error bars correspond to the 10 and 90
   percentiles of the marginalized age PDF shown as violin plots in Fig.\,\ref{fig:17}. The {\it gray horizontal lines} are drawn at 
   the height of the true values $(a_o,\,a_y)\,=\,(0.67,0.33)$. Each {\it group of 3 points} corresponds to the same age but for
   clarity the points are plotted at slightly different ages.
   The {\it left and right groups} include the stars with assigned age in the range $[7,10]$ and $[3,5]$ Gyr, respectively.
   {\it (b)} Same as {\it (a)} but for the \solm\ sample using the marginalized age PDF shown as violin plots in Fig.\,\ref{fig:15}. 
   The ({\it black dots}) correspond to the case B \citet{Mor19} IMF.
   Each {\it group of 4 points} corresponds to the same age but for clarity the points are plotted at slightly different ages.
   The {\it leftmost, middle and rightmost groups} include the stars with assigned age in the range $[6.5,10]$, $[2.5,4.8]$, and
   $[0.1,1.7]$ Gyr, respectively.
   {\it (c)} Same as {\it (b)} but for the marginalized $Z$ PDF shown as violin plots in Fig.\,\ref{fig:15}. 
   Each {\it group of 4 points} corresponds to the same $Z$ but for clarity the points are plotted at slightly different values.
   From {\it left} to {\it right} the groups include the stars with assigned $Z$\,=\,($0.01,0.014,0.017,0.03$).
   }
\end{center}
\end{figure*}

\begin{figure*}
\begin{center}
   \includegraphics[width=1.02\textwidth]{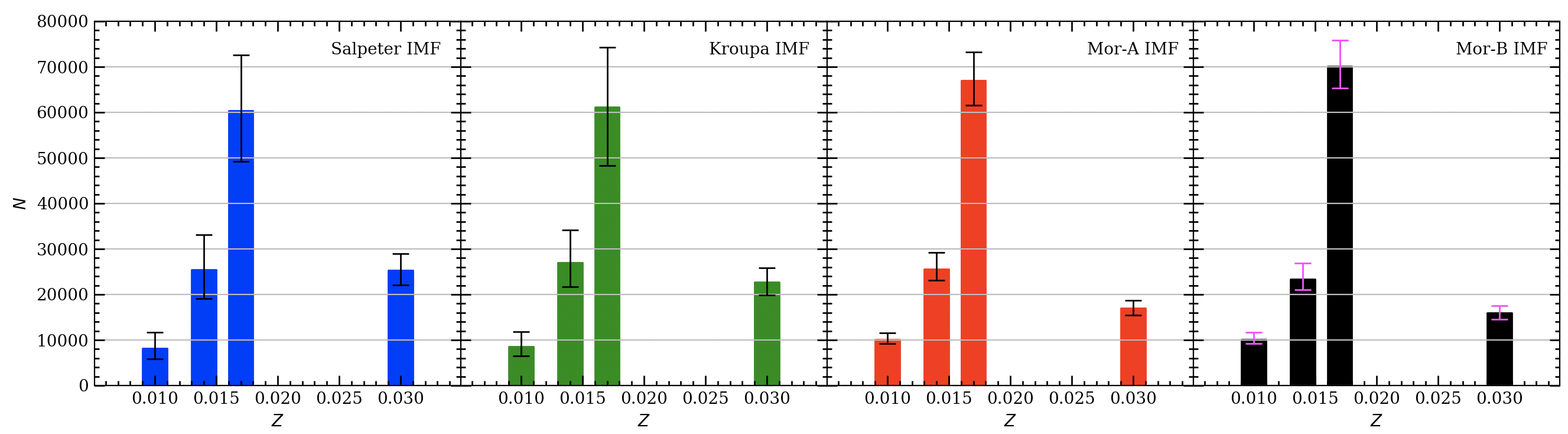}
   \caption{\label{fig:19}{Inferred $Z$ distribution for \solm\ for the different IMFs of Fig.\,\ref{fig:18}c. The height of each bar
   corresponds to $N_{*}$\,$\times$\,$a_{age}(Z)$, where $N_{*}$\,=\,120,452 is the number of stars in the \solm\ sample
   and $a_{age}(Z)$ is the corresponding 50 percentil.}}
\end{center}
\end{figure*}

\subsection{Exploring the IMF dependence of the inferred AMD}

Fig.\,\ref{fig:16b} shows our mock population in the CMD. The SFH shown in Fig.\,\ref{fig:17}a is used as input to our MW.mx model (Appendix\,\ref{mw.mx}). The oldest population is formed in a constant SFR burst lasting from 10 to 7 Gyr ago ({\it red bars} in Fig.\,\ref{fig:17}a). There is a period with no star formation from 7 to 5 Gyr ago, and then a younger population is formed in a second burst lasting from 5 to 3 Gyr ago ({\it blue bars} in Fig.\,\ref{fig:17}a). Star formation stops again 3 Gyr ago. 67\% of the stars belong to the old population and 33\% to the younger component ({\it red and blue dots} in Figs.\,\ref{fig:16b}, \ref{fig:17}a and
\ref{fig:17}b, respectively). We denote these true fractions as ($a_o^t,\,a_y^t$)\,=\,($0.67,\,0.33$). We assume that stars form following the \cite{kr01} IMF with constant metallicity $Z$\,=\,$0.017$.

We use grid C (Table\,\ref{tab:4}) to infer the AMD for the mock population. The {\it gray vertical lines} in Fig.\,\ref{fig:17}a signal the age of the isochrones in grid C in the relevant age range, shown in detail in Fig.\,\ref{fig:17}c.
Figs.\,\ref{fig:17}d,\,\ref{fig:17}e and \ref{fig:17}f show, respectively, the inferred AMDs using the \cite{salp55}, the \cite{kr01} and the \cite{Mor19}-A IMF in our statistical model. The ({\it red, blue}) bars in these figures correspond to the (old, young) population.
From the PDF shown as violin plots in Fig.\,\ref{fig:17} we derive the results shown in Fig.\,\ref{fig:18}a.
Inspection of Fig.\,\ref{fig:17} shows that the periods when the star formation is quenched are well detected by our inference algorithm. Likewise, the metallicity $Z$\,=\,$0.017$ is inferred with no error. 

Fig.\,\ref{fig:18}a shows that $a_o$ is underestimated (overestimated) and $a_y$ is overestimated (underestimated) when using the \cite{salp55} \citep{Mor19} IMF in the statistical model. The true values ($a_o^t,\,a_y^t$) are recovered when using the true \citep{kr01} IMF.
From Table\,\ref{tab:5}, the number of stars populating the isochrones decreases monotonically from the \cite{salp55} to the \cite{kr01} and then to the \cite{Mor19} IMF. In order to reproduce the required number of stars, the 
statistical model assigns a higher weight to the less populated isochrones to compensate for the lack of stars.
This explains the behaviour of $a_o$ with the IMF in Fig.\,\ref{fig:18}a. $a_y$ shows the opposite behaviour
because we impose the condition in Eq.\,(\ref{avec}), $a_o$\,+$\,a_y$\,=\,$1$, in our statistical model.\footnote{$a_o$, and not $a_y$, behaves as described with the IMF because in our mock population twice as many stars are formed in the old as in the young burst. The stochastic sampling in the MCMC process is biased towards the more populated regions in the CMD.} Even though the values of ($a_o,\,a_y$) in Fig.\,\ref{fig:18}a for the \cite{salp55} and the \cite{Mor19}-A IMFs, ($0.58,\,0.41$) and ($0.84,\,0.15$), respectively, differ from the true values ($a_o^t,\,a_y^t$)\,=\,($0.67,\,0.33$), they show the correct trend, $a_o$\,>\,$a_y$. 

Fig.\,\ref{fig:18}b shows $a_{AGE}$ for the four solutions inferred for \solm in the previous section (Fig.\,\ref{fig:15}). We see the same behaviour described above for the mock population (Fig.\,\ref{fig:18}a). The correlation of $\bm{a}$ with the IMF in the older burst indicates that the number of \solm\ stars formed in this burst is substantially larger than the number of stars formed in the younger bursts. It is interesting to note that for a given IMF, $a_{AGE}$ decreases monotonically in time. For all the IMF's considered, there is a period of non detectable (or very low) star formation at the end of the oldest burst, and a residual amount of star formation in recent times.

Fig.\,\ref{fig:18}c shows the equivalent diagram for $a_Z$. The behaviour of $a_Z$ with the IMF for the $Z$\,=\,$0.017$ and $0.03$ bins mimics the behaviour of $a_{Age}$ in Fig.\,\ref{fig:18}b. The positive slope at $Z$\,=\,$0.017$ is compensated by a negative slope at $Z$\,=\,$0.03$. Most of the stars in \solm\ formed in the oldest burst with $Z$\,=\,$0.017$, hence the sensitivity of $a_Z$ at this value of $Z$ to the number of stars predicted by the IMF. For the two lowest metallicity bins $a_Z$ is nearly constant for all 
IMFs. For clarity, we show in Fig.\,\ref{fig:19} the inferred $Z$ distribution of Fig.\,\ref{fig:18}c as bar histograms.

\subsection{Summary of IMF dependence tests}

From this exercise we conclude that our statistical model:
{\it (a)} identifies correctly the periods corresponding to {\it active} and {\it quenched} star formation;
{\it (b)} provides the {\it correct trend} on the number of stars born as a function of time, independently of the assumed IMF; and
{\it (c)} characterizes correctly the {\it prevalent} metallicity in each star formation episode.
The components of the solution vector $\bm {a}$, the number of stars assigned to each population, depend on the assumed IMF.

\section{Conclusions}\label{concl}

We have built a Bayesian hierarchical model designed to infer the age-metallicity distribution or the star formation history of resolved stellar populations.
This model takes into account the possibly non-symmetrical distribution of the inferred quantities, like distance and absolute magnitude, and can handle complete, incomplete, and magnitude limited samples.
We use our model to study the stars within 100 pc of the sun brighter than $G$\,=\,$15$ in the \gtwo catalogue, the \solm sample defined in Section \ref{samp_sel_sol}.
We develop a model of the MW Galaxy tailored after the Besan\c con model \citep{rob03} that we use to text the validity of our inferences and search for biases present in our solutions. 
We describe the bias introduced by the existence of unresolved binaries in \solm\ not included in the model.

Ignoring extinction and unresolved binaries, our results show a maximum of star formation activity about 10 Gyr ago, producing large numbers of stars with slightly sub-solar metallicity ($Z$\,=\,$0.014$), followed by a decrease in star formation up to a minimum level occurring around 8 Gyr ago. After a quiet period, star formation rises to a maximum at about 5 Gyr ago, forming stars of solar metallicity ($Z$\,=\,$0.017$). Finally, star formation has been decreasing until the present, forming stars of $Z$\,=\,$0.03$ at a residual level.

We use the \citet{lallement19} Stilism tool to derive the 3D extinction map for all the stars in $S_{100}^{15}$. The correction by extinction does not introduce major differences in the inferred vector $\bm{a}$, it decreases slightly the contribution of the oldest age bins, increasing the fraction of younger stars of all metallicities. Ignoring extinction biases the AMDs towards older ages.

We build an heuristic model to explore the effects of unresolved binary stars present in the data and ignored in the statistical model and show that ignoring the presence of unresolved binaries biases the inferred AMD towards older ages and higher $Z$'s than the true values. It could happen that the stellar component detected at $Z=0.03$ is an artifact introduced by ignoring unresolved binaries. This will be explored in detail in a separate paper.

We test the sensitivity of the inferred AMD to the apparent limiting magnitude of the sample and conclude that 
the inferred age distribution for the stars in \sols\ is robust with respect to the sample limiting magnitude 
for $12$\,$\le$\,$G$\,$\le$\,$15$. The $Z$ distribution is robust for $13$\,$\le$\,$G$\,$\le$\,$15$.

We show that the components of the solution vector $\bm {a}$ depend on the assumed IMF. The weight $a_i$ assigned to a given isochrone in the inferred AMD anti-correlates with the number of stars populating the isochrone according to the assumed IMF. The smaller the number of stars in the isochrone, the higher the weight $a_i$ required to match highly populated regions in the observed CMD. To fulfill the condition $\sum a_{i}$\,=\,$1$, some $a_i$ show the opposite behaviour. For all the IMFs explored, we obtain a SFH which contains the same basic components. Our model identifies correctly the periods corresponding to {\it active} and {\it quenched} star formation, provides the {\it correct trend} on the number of stars born as a function of time, independently of the assumed IMF, and characterizes correctly the {\it prevalent} metallicity in each star formation episode.

In all the scenarios explored in this paper the resultant AMD shows the behaviour described above: two bursts plus some residual star formation with a similar $Z$ distribution, the $Z=0.017$ (solar) population being the dominant one.
The results for the old population can be improved including fainter stars in the sample, which requires careful modelling of the incompleteness function at the faint end.
The properties of the recent bursts of star formation will become more reliable as the bright end of the CMD becomes
more complete, better calibrated and free of spurious parallaxes in forthcoming \g data releases.

Our most important result is showing that both the {\it star formation} and {\it chemical enrichment} histories of the solar neighbourhood can be derived from the \solm\ sample with the proper statistical treatment.
Our results are consistent with the star formation quenching reported by \cite{Hay16} and \cite{Mor19}.
The enhancement of star formation at later times was also detected by \cite{Mor19}, but they do not allow for metallicity evolution.
They argue that star formation rises too fast and to a level too high to occur in an isolated disc, and suggest that
this event was most likely triggered by an external agent, possibly a merger. 

A natural extension of this work is to explore different scenarios to establish the role of dynamical processes and merger events in determining the SFH of \solm and other population groups in the Galaxy. This will soon be possible once the next \g data releases provide improved positions, velocities and photometry for a large number of Galactic stars.

\section*{Acknowledgements}

We thank the anonymous referee for the careful reading of our manuscript and for very pertinent suggestions that made this
paper more accessible and useful to the interested reader.
We thank Rosa A. Gonz\'alez-L\'opezlira and Bernardo Cervantes Sodi for fruitful discussions at the early stages of this
investigation.
The research in this paper is part of the PhD thesis of J. A. Alzate in the Universidad Nacional Aut\'onoma de M\'exico (UNAM) graduate
program in astrophysics. He thanks the support from the Instituto de Radioastronom\'ia and Astrof\'isica, its staff, and the Consejo
Nacional de Ciencia y Tecnolog\'ia (CONACyT) for the scholarship granted. 
GB and JAA acknowledge financial support from the National Autonomous University of M\'exico (UNAM) through grant DGAPA/PAPIIT IG100319
and from CONACyT through grant CB2015-252364. 

\section*{Data availability}

This work uses data from the European Space Agency (ESA) mission {\it Gaia} (\url{https://www.cosmos.esa.int/gaia}),
processed by the {\it Gaia} Data Processing and Analysis Consortium 
(DPAC,\,\url{https://www.cosmos.esa.int/web/gaia/dpac/consortium}).
Funding for the DPAC has been provided by national institutions, in particular the institutions participating in the {\it Gaia}
Multilateral Agreement.

We use the Stilism tool \citep[][\url{https://stilism.obspm.fr}]{lallement19} and the {\it Stan} MCMC platform (\url{https://mc-stan.org}).

\section*{SUPPORTING INFORMATION}\label{bhmtab}

Supplementary material available at MNRAS online:
\medskip

Tables\,\ref{tab:d1}, \ref{tab:d2} and \ref{tab:d3}.
Figures\,\ref{fig:D3}, \ref{fig:D1} and \ref{fig:D2}.


\bibliographystyle{mnras}
\bibliography{references}

\appendix
\section{Sample completeness}\label{formalism}

In the following subsections we compute the posterior for the cases of $(a)$ a complete and $(b)$ a magnitude-limited sample.
 
\subsubsection{Complete sample}\label{post_comp_samp}

For a complete sample the integral in Eq. (\ref{marginal_1}) can be evaluated for absolute magnitudes in the range from $-\infty$ to $+\infty$. The selection function, being a constant, is irrelevant. After some algebra, Eq. (\ref{marginal_1}) can be written as
\begin{align}\label{post_complete_1}
P(\bm{a}\vert d, \phi) &= \int_{r_o}^{r_{lim}}dr_{j} \prod_{k=1}^{3}\int_{-\infty}^{+\infty} P(\bm{a}, \bm{\beta} \vert \bm{d},\phi) dM_{j}^{k} \nonumber\\
    &\propto P(\bm{a}) \prod_{j=1}^{N_{D}} \sum_{i=1}^{N_{iso}} a_{i} P_{ij},
\end{align}
\noindent
where
\begin{align}\label{post_complete_2}
    P_{ij} &=  \frac{1}{\ell(\textbf{d},S)} \int_{r_o}^{r_{lim}} \int_{m_{l}, i}^{m_{u}, i} \mathcal{N}(\varpi_{j}\vert \varpi_{\rm{true},\textit{j}} \sigma_{\varpi, j})\ P(r_{j})\ \phi(m)\ \times \nonumber\\
    &\times \prod_{k=1}^{3} \mathcal{N}\left(G_{j}^{k}\Big\vert M_{\rm{i}}^{k}+f_{j}\ ,\ \sqrt{\sigma_{j}^{k\ 2}+\sigma_{\rm{i}}^{k\ 2}}\right) \ dm\ dr_{j},
\end{align}
and $f_{j} = 5\log{r_{j}}-5$ is the distance modulus.
The integration of Eq. (\ref{post_complete_2}) with respect to $r_{j}$ and $m$ requieres numerical methods.

In this case the normalization constant, Eq. (\ref{norm_main}), is
\begin{align}\label{norm_complete_1}
    \ell(\bm{a},S) &= \prod_{j=1}^{N_{D}} \int_{r_{0}}^{r_{lim}}dr'_{j} \int_{-\infty}^{+\infty} d\varpi'_{j}\ P(\varpi'_{j}\vert r'_{j})P(r'_{j}) \times \nonumber\\
    &\times \prod_{k=1}^{3} \int_{-\infty}^{+\infty} d{M_{j}^{k}}' \int_{-\infty}^{+\infty} d{G_{j}^{k}}'\ P({G_{j}^{k}}' \vert r'_{j},{M_{j}^{k}}') \ P({M_{j}^{k}}' \vert \bm{a},\phi) \nonumber\\
    &= \left(\sum_{i=1}^{N_{iso}} a_{i} \int_{m_{l}, i}^{m_{u}, i}\phi(m)\ dm\right)^{N_{D}}.
\end{align}

\subsubsection{Magnitude-limited sample}

\begin{figure}
\begin{center}
    \includegraphics[width=1.1\columnwidth]{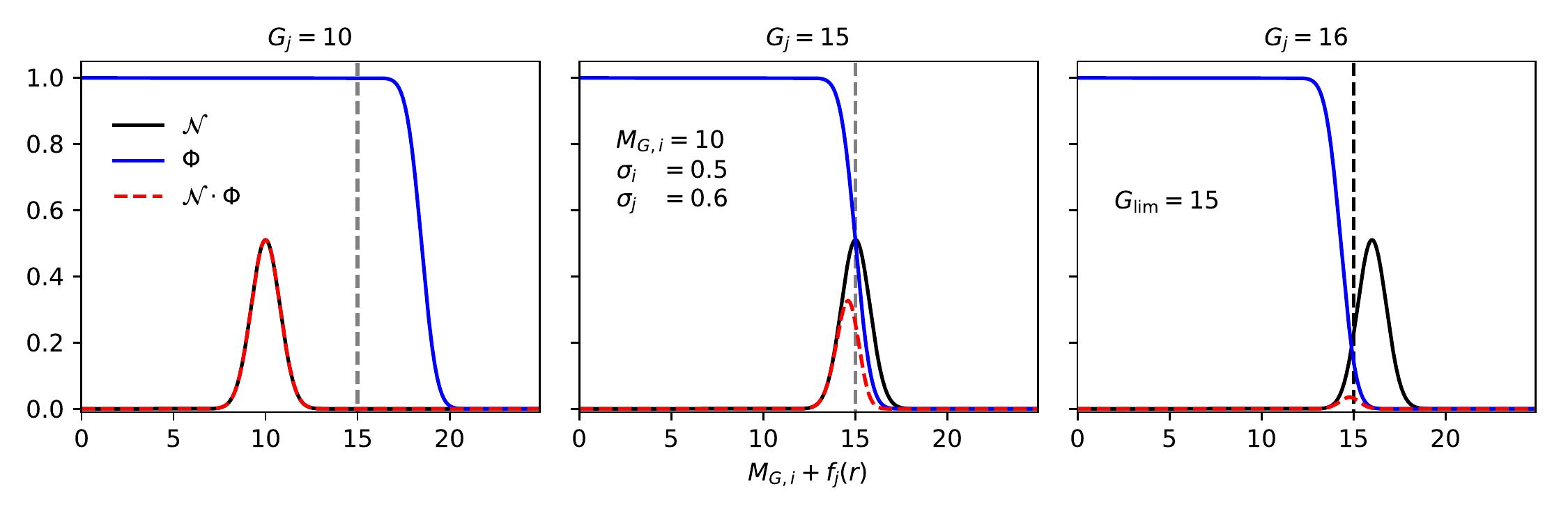}
    \caption{\label{fig:A1}$N_{ij}^{k}$ ({\it black line}), $\Phi_{ij}^{k}$ ({\it blue line}), and
    $N_{ij}^{k}\dot\Phi_{ij}^{k}$  ({\it red dashed line}) for $k=1$
    vs. apparent magnitude $M_{i}^{k}+f_{j}$, see Eqs. (\ref{post_trunc_4}-\ref{post_trunc_3}).
    $N_{ij}^{k}$ is shown for three hypothetical values of $G_j = 10, 15$, and $16$ ({\it left, center, right}), and
    $\Phi_{ij}^{k}$ for $G_{\rm lim}=15$, $M_{G,i} = 10$, $\sigma_{j} = 0.6$, $\sigma_{i}=0.5$, and the same three values of $G_{j}$.
    $\Phi_{ij}^{k}$ acts like a filter that limits the visibility of stars approaching apparent magnitude 15.
    }
\end{center}
\end{figure}

If the sample is complete up to apparent magnitude $G_{\rm lim}^{k}$, Eq. (\ref{post_main}) can be written as
\begin{align}\label{post_trunc_1}
    P(\bm{a}\vert d, \phi) &= \int_{r_o}^{r_{lim}}dr_{j} \prod_{k=1}^{3}\int_{-\infty}^{G_{lim}^{k}-f_{j}} P(\bm{a}, \bm{\beta} \vert \bm{d},\phi) dM_{j}^{k}\propto  \nonumber\\
    &\propto P(\bm{a}) \prod_{j=1}^{N_{D}} \sum_{i=1}^{N_{iso}} a_{i} P_{ij},
\end{align}
\noindent 
where
\begin{align}\label{post_trunc_2}
    P_{ij} & = \frac{1}{\ell(\textbf{d},S)} \int_{r_o}^{r_{lim}} \int_{m_{l}, i}^{m_{u}, i} \mathcal{N}(\varpi_{j}\vert \varpi_{\rm{true},\textit{j}} \sigma_{\varpi, j})\ P(r_{j})\ \phi(m)\times \nonumber\\
    &\times \prod_{k=1}^{3}N_{ij}^{k}(r_{j},m)\ \Phi_{ij}^{k}(r_{j},m)\ dm\ dr_{j},
\end{align}
\noindent 
\begin{equation}\label{post_trunc_4}
    N_{ij}^{k}=\mathcal{N}\left(G_{j}^{k}\Big\vert M_{\rm{i}}^{k}+f_{j}\ ,\ \sqrt{\sigma_{j}^{k\ 2}+\sigma_{\rm{i}}^{k\ 2}}\right),\end{equation}
\begin{equation}\label{post_trunc_3}
    \Phi_{ij}^{k}=\Phi\left(\frac{G_{lim}^{k}-\frac{\sigma_{i}^{k\ 2}}{(\sigma_{i}^{k\ 2}+\sigma_{j}^{k\ 2})}G_{j}^{k}-\frac{\sigma_{j}^{k\ 2}}{(\sigma_{i}^{k\ 2}+\sigma_{j}^{k\ 2})}\left(M_{i}^{k}+f_{j}\right)}{\frac{\sigma_{i}^{k}\sigma_{j}^{k}}{\sqrt{\sigma_{i}^{k\ 2}+\sigma_{j}^{k\ 2}}}}\right),
\end{equation}
\noindent
and $\Phi$ is the cumulative distribution function for a normal PDF.
If $G_{\rm lim}^{k}\rightarrow +\infty$, $\Phi\rightarrow 1$. 
$N_{ij}^{k}$ and $\Phi_{ij}^{k}$ depend on $m$ and $r_{j}$ through $M_{i}^{k}$ and $f_{j}.$ 
In this case the normalization constant, Eq. (\ref{norm_main}), is
\begin{align}\label{norm_trunc_1}
    \ell(\bm{a},S) &= \prod_{j=1}^{N_{D}} \int_{r_{0}}^{r_{lim}}dr'_{j} \int_{-\infty}^{+\infty} d\varpi'_{j}\ P(\varpi'_{j}\vert r'_{j})P(r'_{j}) \times \nonumber\\
    &\times \prod_{k=1}^{3} \int_{-\infty}^{G_{lim}^{k}} d{M_{j}^{k}}' \int_{-\infty}^{+\infty} d{G_{j}^{k}}'\ P({G_{j}^{k}}' \vert r'_{j},{M_{j}^{k}}') \ P({M_{j}^{k}}' \vert \bm{a},\phi) \nonumber\\
    &= \prod_{j=1}^{N_{D}} \sum_{i=1}^{N_{iso}} a_{i} C_{ij},
\end{align}
where
\begin{align}\label{norm_trunc_2}
    C_{ij} &=  \frac{1}{\ell(\textbf{d},S)} \int_{r_o}^{r_{lim}} \int_{m_{l}, i}^{m_{u}, i} P(r_{j})\ \times \nonumber\\
    &\times \phi(m_i) \prod_{k=1}^{3}\Phi\left(\frac{G^{k}_{lim}-\left(M^{k}_{i}+f_{j}\right)}{\sigma^{k}_{i}}\right)\ dm_{i}\ dr_{j}.
\end{align}
Fig. \ref{fig:A1} serves to clarify the role played by Eqs. (\ref{post_trunc_4}) and (\ref{post_trunc_3}).

\section{The MW.\lowercase{mx} model}\label{mw.mx}

As part of this investigation we have built a model of the Galaxy to compute the star counts in any direction of the sky in {\it any photometric system}.
Since results of this model are used in the main body of this paper, in this Appendix we summarize its basic ingredients and how they are assembled to model the MW. 
We adapt the recipes used in the Besan\c con model \citep{rob03} to our needs to describe the structure of the Galaxy.
To differentiate our model from the Besan\c con model and in lack of a better name, we call our model the MW.mx model.
In these models the Galaxy is divided into 4 structurally independent components: {\it Bulge, Halo, Thick disc} and {\it Thin disc}.
The mass density laws $\rho(R,z)$ for the 4 components are given in Tables\,\ref{tab:B1} and \ref{tab:B2}, which are based on tables 1, 2, and 3 of \cite{rob03}.
The thin disc population is divided into 7 different age-metallicity groups which follow the same functional form of $\rho(R,z)$ but with different parameters.
We note that whereas the SFR in the thin disc is assumed constant from 0 to 10 Gyr, the thick disc, halo, and bulge populations form in instantaneous bursts older than the thin disc.
The sixth and seventh columns of Table\,\ref{tab:B2} list the age and metallicity for each stellar group used in the Besan\c con model.
In the three rightmost columns we list the corresponding values used in MW.mx.
The values of $\left[\frac{Fe}{H}\right]$ for the PARSEC tracks used in our model (Section\,\ref{methods}) differ slightly from the Besan\c con values but agree within the errors.
In the Besan\c con model they use one IMF for the thin disc, another for the thick disk and halo, and a third one for the bulge \citep[][table 1]{rob03}. In our case we use the \citet{kr01} IMF for all the components. We performed detailed comparisons of our model predictions with the Besan\c con model in several directions in the sky.
We adjusted the age of the thick disc and halo in our model to get the closest match between the two models.

Modelling the stellar populations in the Galaxy is a special case of stellar population synthesis (SPS) in which we must account for the exact position of each star in the sky.
Standard SPS \citep[e.g.,][]{bc03} is a four parameter problem. Each star is characterized by its mass $m$, age $t$, metallicity $Z$, and absolute photometry/spectrum $F$.
Stars with the same values of $m,t$ and $Z$ have identical $F$, i.e., all the stars are assumed to be at a distance of 10 pc.
In SPS models for resolved clusters or galaxies \citep[e.g.,][]{gb10}, the IMF is sampled stochastically to account for fluctuations in the number of stars of a given mass in
sparsely populated systems, but again all the stars are placed at the same distance. The Galaxy is a special case of resolved population because we are inside it and it appears different in every direction. In this case we need to add three more parameters to the SPS problem, the distance $r$ to the star as measured from the sun and its Galactic coordinates $(l,b)$.
The Galaxy model is then built by sampling stochastically the IMF, the SFH, and the mass distribution function $M(r)$ for as many stars as required for each component of the Galaxy.
In Section\,\ref{s:rdist} we explain how this sampling is performed.

\begin{table*}
\begin{center}
\caption{\label{tab:B1}Density laws for the different Galactic components \citep[after][table 3]{rob03}.
$R$ is the galactocentric distance, $z$ the height above the Galactic plane, $a^{2}=R^{2}+\frac{z^{2}}{\epsilon^2}$, $\epsilon$ the axis ratio,
$\rho_0$ the density in the solar vicinity and $d_0$ a normalization constant. $\rho_{0}$ and $\epsilon$ are listed in Table\,\ref{tab:B2}.
}
\renewcommand{\arraystretch}{1.70} 
\setlength{\tabcolsep}{12pt}       
\begin{tabular}{llll}
\hline
\multicolumn{4}{c}{Density Laws} \\
\hline
\multirow{4}{0.5cm}{Thin disc}  & $\rho_{0}/d_{0}\times \lbrace\exp(-(a/h_{R_+})^2)-\exp(-(a/h_{R_-})^2)\rbrace$ & $h_{R_+}$=5000 pc & \multirow{2}{*}{age$\ \leq0.15$ Gyr} \\
                                &                                                                                & $h_{R_-}$=3000 pc &  \\ \cline{2-4}
& $\rho_{0}/d_{0}\times \lbrace\exp(-(0.5^{2}+a^2/h_{R_+}^2)^{1/2})-\exp(-(0.5^{2}+a^2/h_{R_-}^2)^{1/2})\rbrace$ & $h_{R_+}$=2530 pc & \multirow{2}{*}{age$\ >0.15$ Gyr} \\
&                                                                                                                & $h_{R_-}$=1320 pc & \\
\hline
\multirow{2}{0.5cm}{Thick disc} & $\rho_{0}/d_{0}\times \exp{-(\frac{R-R_{\odot}}{h_{R}})}\times (1-\frac{1/h_{z}}{x_{l}\times(2+x_{l}/h_{z})}\times z^2)$       & $h_{R}$=2500 pc & $\vert z\vert\leq x_{l}=400$ pc \\ 
                           & $\rho_{0}/d_{0}\times \exp{-(\frac{R-R_{\odot}}{h_{R}})}\times \frac{\exp(x_{l}/h_{z})}{1+(x_{l}/2h_{z})}\exp(-\frac{|z|}{h_{z}})$  & $h_{z}$=800 pc  & $\vert z\vert>x_{l}$  \\
\hline
\multirow{2}{0.5cm}{Halo}  & $\rho_{0}/d_{0}\times(\frac{a_c}{R_{\odot}})^{-2.44}$ & \multirow{2}{*}{$a_{c}$=500 pc} & $a\leq a_{c}$  \\
                           & $\rho_{0}/d_{0}\times(\frac{a}{R_{\odot}})^{-2.44}$   &                                 & $a>a_{c}$    \\
\hline
                           & $N\times\exp(-0.5\times r_S^2)$                                                                                             &    $x_0=1.59$ kpc     & $\sqrt{x^2+y^2}<R_c$ \\ 
 Bulge                 & $N\times\exp(-0.5\times r_S^2) \times \exp\left[-0.5\left(\frac{\sqrt{x^2+y^2}-R_c}{0.5}\right)^2\right]$ &    $y_0=0.424$ kpc  & $\sqrt{x^2+y^2}>R_c$ \\ 
                           &\quad with $r_s^2=\sqrt{\left[(\frac{x}{x_0})^2 + (\frac{y}{y_0})^2\right]^2 + (\frac{z}{z_0})^4}$, \quad\quad\quad\quad $R_c=2.54$ kpc             &    $z_0=0.424$ kpc  & $N=13.70$ stars pc$^{-3}$  \\
\hline
\end{tabular}
\end{center}
\end{table*}

\begin{table*}
\begin{center}
\caption{\label{tab:B2}Parameters defining different Galactic components \citep[after][tables 1 and 2]{rob03}.}
\begin{tabular}{clcclccccccccr}
\hline
\mr{3}{*}{$i$} & \mr{3}{*}{Component} &                          &                       & \mr{3}{*}{SFR}       & & & \mc{2}{c}{ Besan\c con Model}                     & & & \mc{3}{c}{ MW.mx Model} \\
               &                      & $\rho_0$                 & \mr{2}{*}{$\epsilon$} &                      & & & Age      & \mr{2}{*}{$\left[\frac{Fe}{H}\right]$} & & & Age      & \mr{2}{*}{Z} & \mr{2}{*}{$\left[\frac{Fe}{H}\right]$}  \\
               &                      & (M$_{\odot}$ pc$^{-3}$)  &                       &                      & & & (Gyr)    &                                        & & & (Gyr)    &              &                                         \\
\hline                                                                                                                            
1              & \mr{7}{*}{Thin disc} & 4.0$\times10^{-3}$       & 0.0140                &                      & & & 0-0.15   & +0.01$\pm$0.12                         & & & 0-0.15   & 0.017        &   +0.020                                \\
2              &                      & 7.9$\times10^{-3}$       & 0.0268                &                      & & & 0.15-1   & +0.03$\pm$0.12                         & & & 0.15-1   & 0.017        &   +0.020                                \\
3              &                      & 6.2$\times10^{-3}$       & 0.0375                &                      & & & 1-2      & +0.03$\pm$0.10                         & & & 1-2      & 0.017        &   +0.020                                \\
4              &                      & 4.0$\times10^{-3}$       & 0.0551                & Constant             & & & 2-3      & +0.01$\pm$0.11                         & & & 2-3      & 0.017        &   +0.020                                \\
5              &                      & 5.8$\times10^{-3}$       & 0.0696                &                      & & & 3-5      & -0.07$\pm$0.12                         & & & 3-5      & 0.014        &   -0.070                                \\
6              &                      & 4.9$\times10^{-3}$       & 0.0785                &                      & & & 5-7      & -0.14$\pm$0.17                         & & & 5-7      & 0.010        &   -0.222                                \\
7              &                      & 6.6$\times10^{-3}$       & 0.0791                &                      & & & 7-10     & -0.37$\pm$0.20                         & & & 7-10     & 0.008        &   -0.322                                \\ \\
8              & Thick disc           & 1.34$\times10^{-3}$      &                       & $\delta (t-t_{0})$   & & & 11       & -0.78$\pm$0.30                         & & & 10       & 0.008        &   -0.322                                \\ \\
9              & Halo                 & 9.32$\times10^{-3}$      & 0.7600                & $\delta (t-t_{0})$   & & & 14       & -1.78$\pm$0.5                          & & & 11       & 0.0002       &   -1.937                                \\ \\
10             & Bulge                &                          &                       & $\delta (t-t_{0})$   & & & 10       & +0.00$\pm$0.40                         & & & 10       & 0.017        &   +0.020                                \\
\hline
\end{tabular}
\end{center} 
\end{table*}

\begin{figure}
\begin{center}
    \includegraphics[width=\columnwidth]{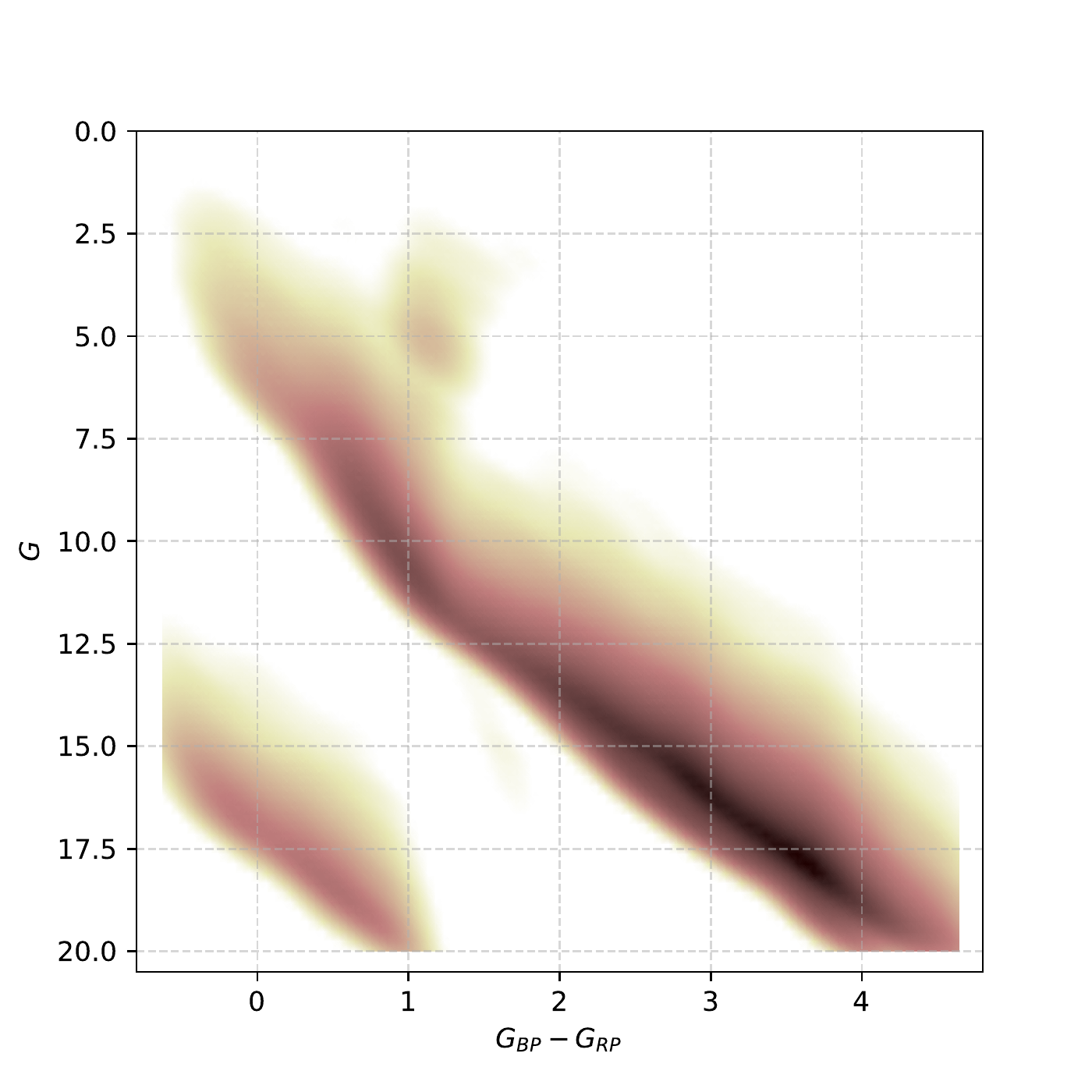}
    \caption{\label{fig:B1}
    CMD in the \g bands resulting from a MW.mx simulation of \sols using the parameters in Tables\,\ref{tab:B1} and \ref{tab:B2}.}
    \end{center}
\end{figure}

\begin{figure*}
\begin{center}
    \includegraphics[width=0.95\textwidth]{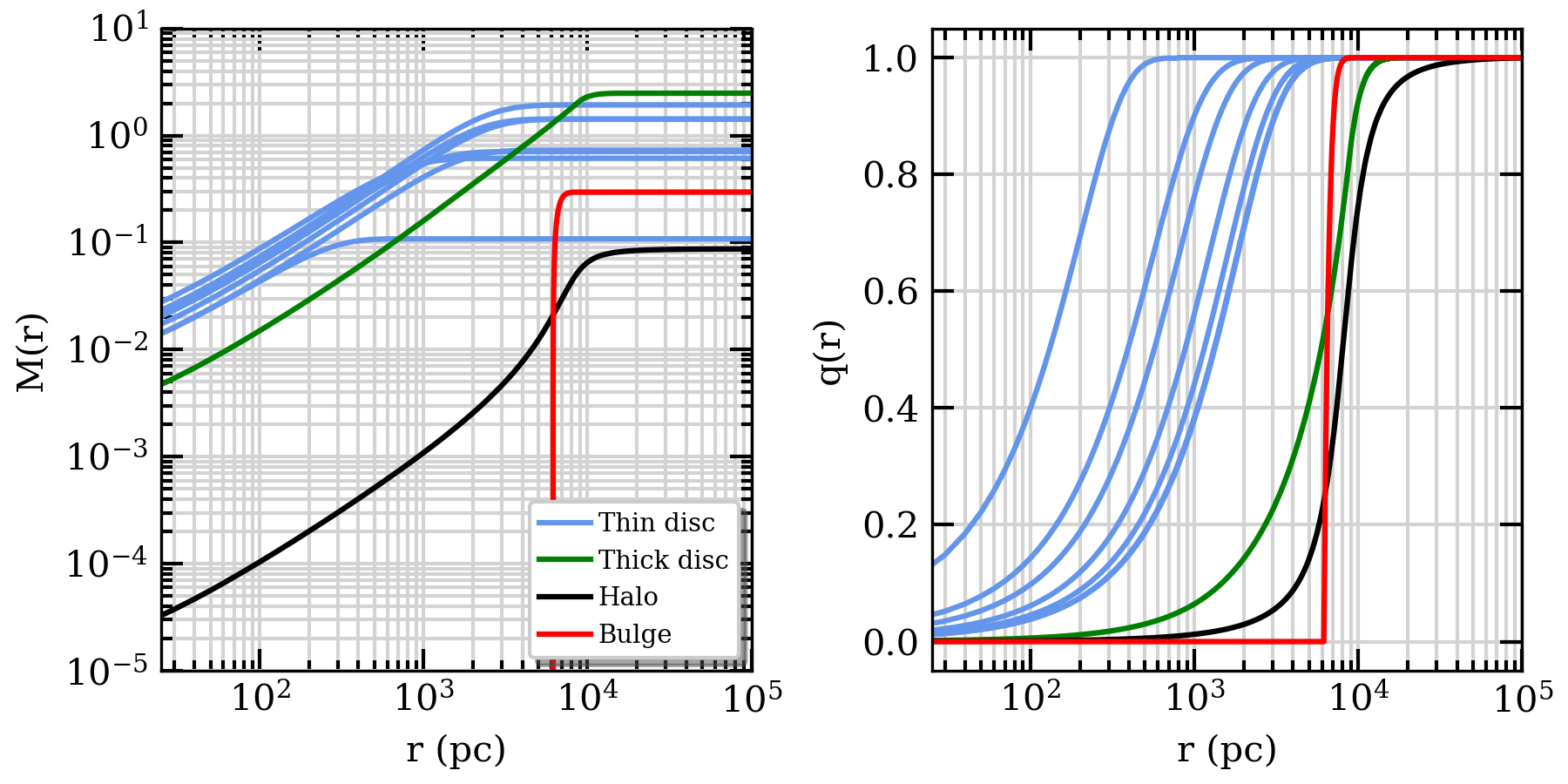}
    \caption{\label{fig:B2}
    {\it (left)} $M(r)$, cumulative mass vs. $r$ inside solid angle $\Omega = 0.15$ square degrees in the direction $l = 0\degree, b = 15\degree$ for the ten Galaxy populations
     listed in Table\,\ref{tab:B2}.
    {\it (right)} $q(r)$, normalized cumulative mass distribution from {\it left}.}
    \end{center}
\end{figure*}


$M_i(r)$, the mass in stars of the $i^{th}$ population group per unit solid angle in direction $(l,b)$, is obtained by integrating $\rho(R,z)$ along the radial 
distance $r$ measured from the sun, where
\begin{equation}
R = \sqrt{x^2 + y^2}
\end{equation}
is the galactocentric distance in the galactic plane, and
\begin{equation}
    x = R_\odot - r\ cos\,b\ cos\,l,\\
    y = r\ cos\,b\ sin\,l,\\
    z = r\ sin\,b.
\label{lbc}
\end{equation}

To construct a photometrically useful 3D model, we must build up the mass $M_i$ out of individual stars sampled stochastically from the IMF and placed at a random distance $r$ such that the grow of the integrated mass with $r$ obeys $M_i(r)$.
If we are modelling the full sky or a large solid angle we must also assign to each star a random angular position $(l,b)$ in the celestial sphere.
Finally, an age must be assigned to each star according to the SFR. 
In practical terms, we generate series of five independent random numbers $(N,Q,L,B,T)$ in the range $[0,1]$ that we use to assign a mass $m$, a distance $r$, an angular 
position $(l,b)$ and an age $t$ to each star, as described in Section\,\ref{s:rdist}.
The absolute magnitude of the star is obtained by interpolating in mass the corresponding $(t,Z)$ isochrone (Section\,\ref{methods}).
From $r$ we get the distance modulus and hence the star apparent magnitude.
For magnitude limited samples we must account for the contribution to $M_i$ from stars fainter than the limit.
This process is repeated until the mass $M_i$ is reached for the 10 components. The final catalogue is the union of the 10 partial catalogues.
For more realistic results, we use the trends of the errors on $\varpi$ \citep[][Eqs. A1 \& A2]{luri18} and $G,\,G_{BP}$ and $G_{RP}$ determined by us from \gtwo,
to add Gaussian noise to the parallax $\varpi^*$ and apparent $G^*,\,G^*_{BP}$ and $G^*_{RP}$ magnitudes of each star in the simulation.

Fig.\,\ref{fig:B1} shows the CMD in the \g bands obtained from a MW.mx simulation of \sols using the parameters in Tables\,\ref{tab:B1} and \ref{tab:B2}.
In our simulation there are $\approx 132,000$ stars with $G<15$ (compared with $\approx 120,500$ stars in \g DR2) and $\approx 343,000$ stars with $G<20$.
We ran 10 similar simulations of \sols to study the completeness of \solm in Section\,\ref{sec_dat_desc}, where we show that {\it (a)} the simulated counts match well the observed counts in the range $G=[7.5,12.5]$, {\it (b)} there is a small and constant excess in the simulated counts in the range $G=[12.5,15]$, and {\it (c)} the observed underestimate the expected counts at the bright and faint ends, the differences being much larger than the statistical fluctuations in our model counts due to the likely reasons discussed in Section\,\ref{sec_dat_desc}.

\subsection{Building the Galaxy model}\label{s:rdist}

Here we outline the procedure followed to assign random values to the quantities $(m,t,r,l,b)$ in the MW.mx model.

\subsubsection{Random placement of stars in $(r,l,b)$}\label{s:q}

As indicated above, integrating $\rho(R,z)$ along the radial distance $r$ measured from the sun in direction $(l,b)$ we obtain $M_i(r)$, the mass in stars per unit solid angle up to distance $r$ for the $i^{th}$ galactic component.
For illustration purposes we show in Fig.\,\ref{fig:B2} the case of a solid angle $\Omega = 0.15$ square degrees in the direction $l = 0\degree, b = 15\degree$.
The lines in the {\it left} and {\it right} hand side panels of Fig.\,\ref{fig:B2} show, respectively, $M_i(r)$ and its normalized counterpart $q_i(r)$, obtained by dividing 
$M_(r)$ by the total mass $M_i$ in the $i^{th}$ component. By definition, $0 \le q \le 1$. 
To assign a distance $r$ to a star of population $i$, we generate a random number $Q$ and use the function $r(q)$ in Fig.\,\ref{fig:B2} 
for the $i^{th}$ component to obtain $r(Q)$. This guarantees that the stars will be distributed following $q(r)$ along the direction $(l,b)$.

To model the full sky or a large solid angle we divide the celestial sphere into one square degree solid angle patches.
We then compute $M_i(r,l,b)$, the mass inside each solid angle up to distance $r$ for the $i^{th}$ galactic component in direction $(l,b)$.
Integrating over the other two variables we obtain the marginalized distributions $M_i(r), M_i(l)$, and $M_i(b)$, which we normalize to 1.
We use random numbers $(Q,L,B)$ to assign a distance $r(Q)$, a longitude $l(L)$, and a latitude $b(B)$ to each star as explained above (Fig.\,\ref{fig:B2}).

\subsubsection{Random selection of isochrones}\label{s:sfr}

In the case of an instantaneous burst, the selection of the isochrone is unique and corresponds to the current age of the burst. 
This is the case for Galactic thick disc, halo and bulge (Table\,\ref{tab:B2}).

For a time dependent SFR, $\Psi(t)$, the mass formed in stars from time 0 to $t$ is
\begin{equation}
M(t) = \int_{0}^{t}\Psi(t')dt'.
\label{pl1}
\end{equation}
If star formation ends at $t = \tau$, the total mass formed in stars is $M_\tau$. We can then write Eq.\,(\ref{pl1}) as
\begin{equation}
T(t) = M_\tau^{-1}  \int_{0}^{t}\Psi(t')dt',
\label{tcum}
\end{equation}
which is a cumulative function that takes values between 0 and 1. 
Using a random number generator to sample $T$, we use Eq.\,(\ref{tcum}) 
to select the age $t$ corresponding to $T$.

For the simple case of a SFR which is constant from $t = t_1$ to $t = t_2$, as assumed for the thin disk (Table\,\ref{tab:B2}), we can write
\begin{equation}
T(t) = \frac{t_2-t}{t_2-t_1}\ \ \ \ \ \ \ \ \ \ \ \ \ \ \ t_1 \le t \le t_2,
\label{csfr}
\end{equation}
and the age $t$ of the isochrone corresponding to $T$ is
\begin{equation}
t(T)= t_2 - T(t_2 - t_1)\ \ \ \ \ \ \ \ \ \ \ \ \ \ \ 0 \le T \le 1.
\label{csfr2}
\end{equation}
$T = 1$ signals the beginning of star formation. 
For $T = 0$, all the mass has formed into stars and star formation ends.

\subsubsection{Stochastic Sampling of the IMF}\label{s:imf}

We sample stochastically the IMF following the procedure outlined by \citet{sf97} for the case of a 
single power law IMF. Here we generalize this procedure for the case of a double power law
\citep{kr01} and a lognormal \citep{ch03} IMF.

\begin{flushleft}
{\it (a) Single power law IMF}
\end{flushleft}
\noindent
The IMF, written as
\begin{equation}
\Phi(m) = dN/dm = Cm^{-(1+x)},
\label{phi}
\end{equation}
gives the number of stars of mass between $m$ and $m+dm$ born in a star formation event. 
$\Phi(m)$, normalised as usual,
\begin{equation}
C = \frac{x}{m_l^{-x} - m_u^{-x}},
\label{cons}
\end{equation}
where $m_l$ and $m_u$ are the lower and upper mass limits of star formation, respectively,
obeys $\Phi(m) \ge 0$, and
\begin{equation}
\int_{m_l}^{m_u}\Phi(m')dm' = 1.
\label{norm}
\end{equation}
$\Phi(m)$ can be thought as a PDF giving the probability that a random mass $m$ is in the range between $m$ and $m+dm$.
$\Phi(m)$ can be transformed into another PDF $g(N)$, such that the probability of occurrence of the random variable
$N$ within $dN$ and the probability of occurrence of the random variable $m$ within $dm$ are the same,
\begin{equation}
|\Phi(m)dm| = |g(N)dN|,
\label{pg}
\end{equation}
where $N$ is a single-valued function of $m$. From Eq.\,(\ref{phi}),
\begin{equation}
N(m) = \int_{m_l}^{m}\Phi(m')dm',
\label{nm}
\end{equation}
is a cumulative distribution function which gives the probability that the mass is $\leq m$.
Using Eq.\,(\ref{pg}), it follows that
\begin{equation}
g(N) = 1, ~~~~ 0 \le N \le 1.
\label{gm}
\end{equation}
$g(N)$ is thus a uniform distribution for which any value is equally likely in
the interval $ 0 \le N \le 1$, and from Eq.\,(\ref{nm}) we can write $m$ as a function of $N$,
\begin{equation}
m = [(1-N)m_l^{-x} + Nm_u^{-x}]^{-\frac{1}{x}}.
\label{nc}
\end{equation}
If we sample $N$ using a random number generator, the values of $m$ from Eq.\,(\ref{nc}) will follow the IMF in Eq.\,(\ref{phi}).

\begin{flushleft}
{\it (b) Double power law IMF}
\end{flushleft}
\noindent

In the case of a two-segment power law IMF, e.g., the thin disk IMF used by \citet{rob03} or the \citet{kr01} universal IMF, written in general as,
\begin{equation}
\Phi(m) = 
\begin{cases}
    C_1\ m^{-(1+x_1)}\ \ \ {\rm if}\ \ \ m_l \le m \le m_c\cr
    C_2\ m^{-(1+x_2)}\ \ \ {\rm if}\ \ \ m_c \le m \le m_u,
\end{cases}
\label{phi2}
\end{equation}
we follow a similar procedure. The normalisation of the IMF is derived from
\begin{equation}
C_1 H(m_l,m_c,x_1) + C_2 H(m_c,m_u,x_2) = 1,
\label{norm1}
\end{equation}
where
\begin{equation}
H(m_1,m_2,x)  = \int_{m_1}^{m_2}m^{-x}dm.
\label{pl2}
\end{equation}
Using the continuity condition of the IMF at $m = m_c$,
\begin{equation}
C_2  = C_1\ m_c^{(x_2 - x_1)},
\label{pl3}
\end{equation}
we obtain $C_1$ and then $C_2$ from
\begin{equation}
C_1  = [H(m_l,m_c,x_1)\ +\  m_c^{(x_2 - x_1)} H(m_c,m_u,x_2)]^{-1}.
\label{pl4}
\end{equation}
The fraction of stars (by number) formed from $m_l$ to $m_c$, is
\begin{equation}
N_c  = \frac{C_1 H(m_l,m_c,1+x_1)}{C_1 H(m_l,m_c,1+x_1)\ +\ C_2 H(m_c,m_u,1+x_2)},
\label{nc2}
\end{equation}
and the equation for the randomly selected mass is written as
\begin{equation}
m = \begin{cases} 
       [ \frac{(N_c-N)m_l^{-x_1} + Nm_c^{-x_1}}{N_c} ]  ^{-\frac{1}{x_1} }\ \ \ \ \ \ \ \ {\rm if}\ \ \ N \le N_c \cr
       [ \frac{(1 - N)m_c^{-x_2} + (N-N_c)m_u^{-x_2} }{1 - N_c } ]  ^{-\frac{1}{x_2} } \ \ \ {\rm if}\ \ \ N > N_c,
       \end{cases}
\label{norm2}
\end{equation}
where again $0 \le N \le 1$.
Sampling $N$ with a random number generator we obtain from Eq.\,(\ref{norm2}) values of $m$ that follow the IMF in Eq.\,(\ref{phi2}).

In the case of the \citet{kr01} universal IMF, the power law segments are defined by
$(m_1,m_2,x) = (0.1,0.5,0.3)$ and $(0.5,100,1.3)$, respectively.
We obtain from (\ref{nc2}) $N_c = 0.72916$, indicating that 72.9\% of the stars are born with $m \le 0.5~M_\odot$.
For the double power law thin disk IMF of \citet{rob03}, defined by
$(m_1,m_2,x) = (0.1,1,0.6)$ and $(1,100,2)$, respectively,
$N_c = 0.90857$, indicating that 90.8\% of the stars are born with $m \le 1~M_\odot$.

\begin{flushleft}
{\it (c) Chabrier IMF}
\end{flushleft}
\noindent

The case of the \citet{ch03} IMF is more complicated. The Chabrier IMF is written as follows,
\begin{eqnarray}
\phi(\log m)
\propto\
\begin{cases}
    \exp\left[ -{\frac{(\log m - \log m_0)^2}{2\sigma^2}}\right]\,,&{\rm for\ } m\leq 1M_\odot\,,\cr
    m^{-1.3}\,, & {\rm for\ } m >   1M_\odot\,,\cr
\end{cases}
\label{cimf}
\end{eqnarray}
with $m_0=0.08M_\odot$ and $\sigma=0.69$. The two expressions in
Eq.\,(\ref{cimf}) are forced to coincide at $1M_\odot$. It can be shown numerically that 
for this IMF, $N_c = 0.8774$, i.e., 87.7\% of the stars are formed below $m_c = 1 M_\odot$.
For $N \le N_c$ we determine $m$ by interpolation of $m(N)$.
For $N > N_c$  we use the second equation in (\ref{norm2}) with $m_c = 1$, $m_u = 100 M_\odot$,
$x_2 = 1.30$, and $N_c = 0.8774$.
\section{Modelling \solm\ with our AMD}\label{snmod}

\begin{table}
\caption{\label{tab:C1}Weights adopted to model \solm\ (based on the full version of Table\,\ref{tab:4} available as supplementary online material).}
\begin{tabular}{ccrrrr}
\hline
Thin   & Age      & \mc{4}{c}{$a^{*}_i\ (p10)$} \\
disc   & (Gyr)    & \mc{1}{c}{$Z$=0.01} & \mc{1}{c}{$Z$=0.014} & \mc{1}{c}{$Z$=0.017} & \mc{1}{c}{$Z$=0.03} \\
\hline
1      & 0-0.15   &  0.00001 & 0.00001 & 0.00001 & 0.00033 \\
2      & 0.15-1   &  0.00040 & 0.00007 & 0.00009 & 0.08706 \\
3      & 1-2      &  0.00006 & 0.00007 & 0.00071 & 0.05336 \\
4      & 2-3      &  0.00012 & 0.00023 & 0.03044 & 0.02382 \\
5      & 3-5      &  0.00032 & 0.00114 & 0.19084 & 0.00005 \\
6      & 5-7      &  0.00034 & 0.00021 & 0.11390 & 0.00002 \\
7      & 7-10     &  0.05250 & 0.17884 & 0.06521 & 0.00003 \\
\hline
Thin   & Age      & \mc{4}{c}{$a^{*}_i\ (p50)$} \\
disc   & (Gyr)    & \mc{1}{c}{$Z$=0.01} & \mc{1}{c}{$Z$=0.014} & \mc{1}{c}{$Z$=0.017} & \mc{1}{c}{$Z$=0.03} \\
\hline
1      & 0-0.15   &  0.00005 & 0.00005 & 0.00007 & 0.00153 \\
2      & 0.15-1   &  0.00108 & 0.00047 & 0.00061 & 0.09727 \\
3      & 1-2      &  0.00042 & 0.00045 & 0.00350 & 0.05929 \\
4      & 2-3      &  0.00073 & 0.00150 & 0.03832 & 0.02949 \\
5      & 3-5      &  0.00206 & 0.00692 & 0.22095 & 0.00037 \\
6      & 5-7      &  0.00218 & 0.00137 & 0.13358 & 0.00013 \\
7      & 7-10     &  0.06453 & 0.21300 & 0.11034 & 0.00018 \\
\hline
Thin   & Age      & \mc{4}{c}{$a^{*}_i\ (p90)$} \\
disc   & (Gyr)    & \mc{1}{c}{$Z$=0.01} & \mc{1}{c}{$Z$=0.014} & \mc{1}{c}{$Z$=0.017} & \mc{1}{c}{$Z$=0.03} \\
\hline
1      & 0-0.15   &  0.00016 & 0.00017 & 0.00022 & 0.00321 \\
2      & 0.15-1   &  0.00220 & 0.00152 & 0.00200 & 0.10816 \\
3      & 1-2      &  0.00137 & 0.00143 & 0.00794 & 0.06520 \\
4      & 2-3      &  0.00233 & 0.00496 & 0.04597 & 0.03544 \\
5      & 3-5      &  0.00669 & 0.01918 & 0.25073 & 0.00121 \\
6      & 5-7      &  0.00688 & 0.00446 & 0.15330 & 0.00042 \\
7      & 7-10     &  0.07830 & 0.25207 & 0.15615 & 0.00062 \\
\hline
\end{tabular}
\end{table}

\begin{table*}
\caption{\label{tab:C2}Global properties of the solar neighbourhood stellar population.
$N_{*}$, $M_{*}$ and {\bf G}$_{*}$ are, respectively, the number, total mass and integrated absolute $G$ magnitude of the stars
in the simulation brighter than the limiting magnitude. $M/L_G$ was computed using $G_\odot=5$.}
\begin{tabular}{lrrrrrrrrrrrr}
\hline
Simulation:        & \mc{3}{c}{Sim N} &&& \mc{3}{c}{Sim N} &&& \mc{2}{c}{Sim M} \\
Limiting magnitude & \mc{3}{c}{$G \leq 15$} &&& \mc{3}{c}{$G \leq 20$} &&& \mc{1}{c}{$G \leq 15$} & \mc{1}{c}{$G \leq 20$} \\
\hline
Percentile & \mc{1}{c}{p50} & \mc{1}{c}{p10} & \mc{1}{c}{p90} &&& \mc{1}{c}{p50} & \mc{1}{c}{p10} & \mc{1}{c}{p90} \\
$N_{*}$                           & 120551   &  70633 & 191900 &&&   358037 & 210225 &   568974 &&&  126405 &    341223 \\
$M_{*}$ (M$_\odot$)               &  76443   &  44111 & 122933 &&&   159810 &  93553 &   255398 &&&   84597 &    151613 \\
{\bf G}$_{*}$                     &  -7.97   &  -7.22 &  -8.59 &&&    -7.97 &  -7.22 &    -8.59 &&&   -8.77 &     -8.77 \\
$M/L_{G}$    (M$_\odot/L_{\odot}$)&   0.49   &   0.57 &   0.45 &&&     1.03 &   1.22 &     0.94 &&&    0.26 &      0.47 \\
\hline
\end{tabular}
\end{table*}




\begin{figure*}
\begin{center}
    \includegraphics[width=\textwidth]{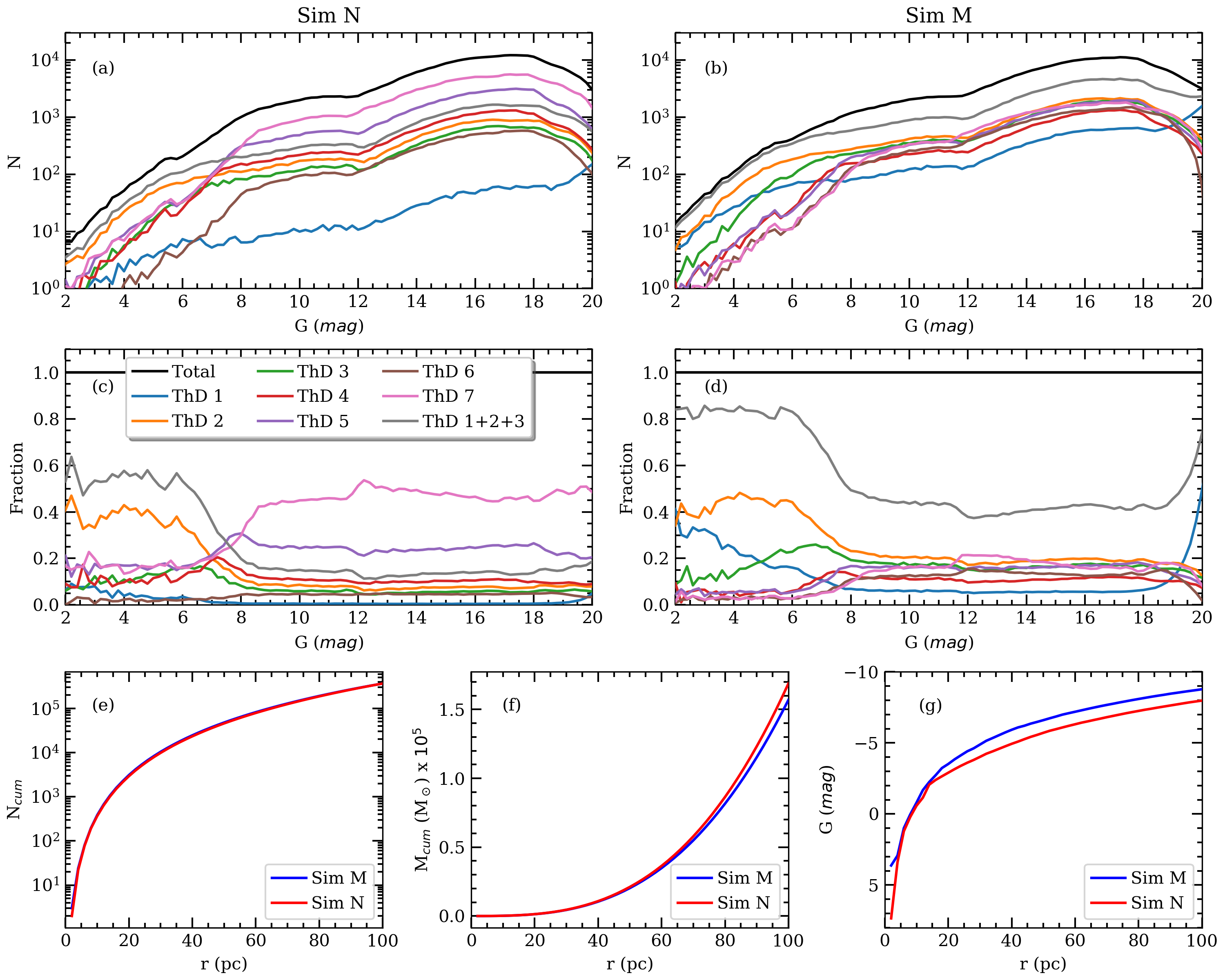}
    \caption{\label{fig:C1} {\it (a,b)} Contribution to the number counts from each of the thin disc age groups listed in Table\,\ref{tab:C1} for the \simn\ and \simm\ simulations,
    respectively. {\it (c,d)} Same as {\it (a,b)} but in fractional form. {\it (e)} Cumulative number of thin disk stars of all age groups vs. $r$ for the two simulations.
    {\it (f)} Same as {\it (e)} but for the stellar mass. {\it (g)} Same as {\it (e)} but for the luminosity in the $G$ band. }
\end{center}
\end{figure*}

\begin{figure*}
\begin{center}
    \includegraphics[width=\textwidth,height=0.30\textheight]{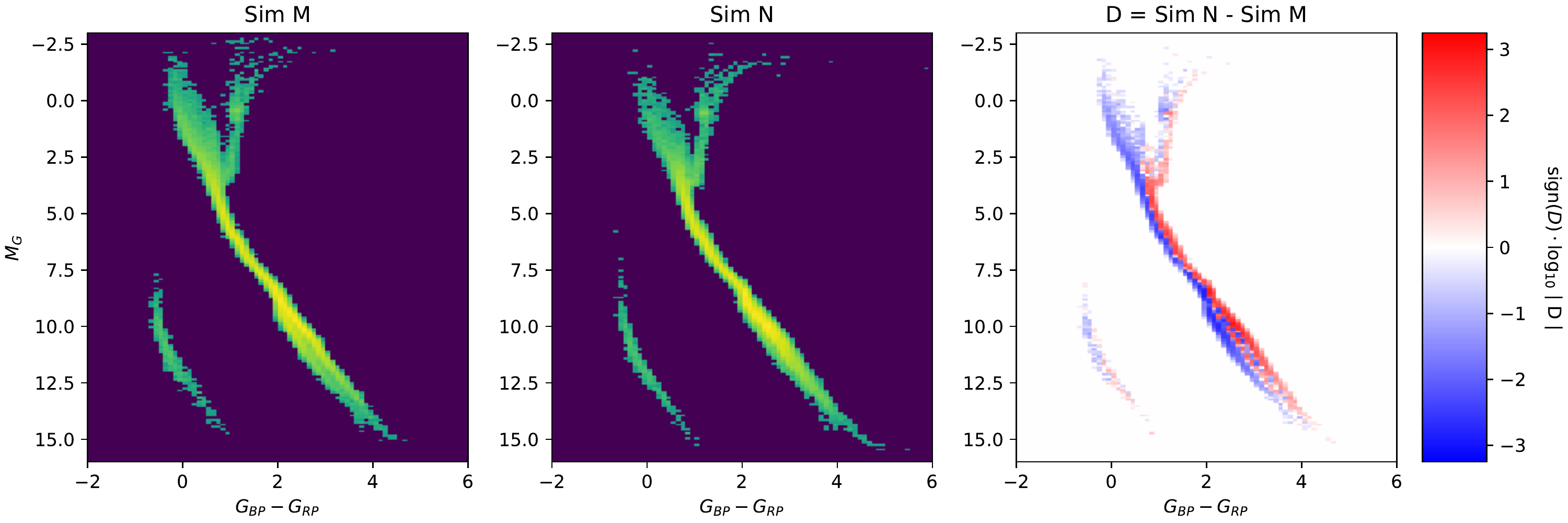}
    \includegraphics[width=\textwidth,height=0.30\textheight]{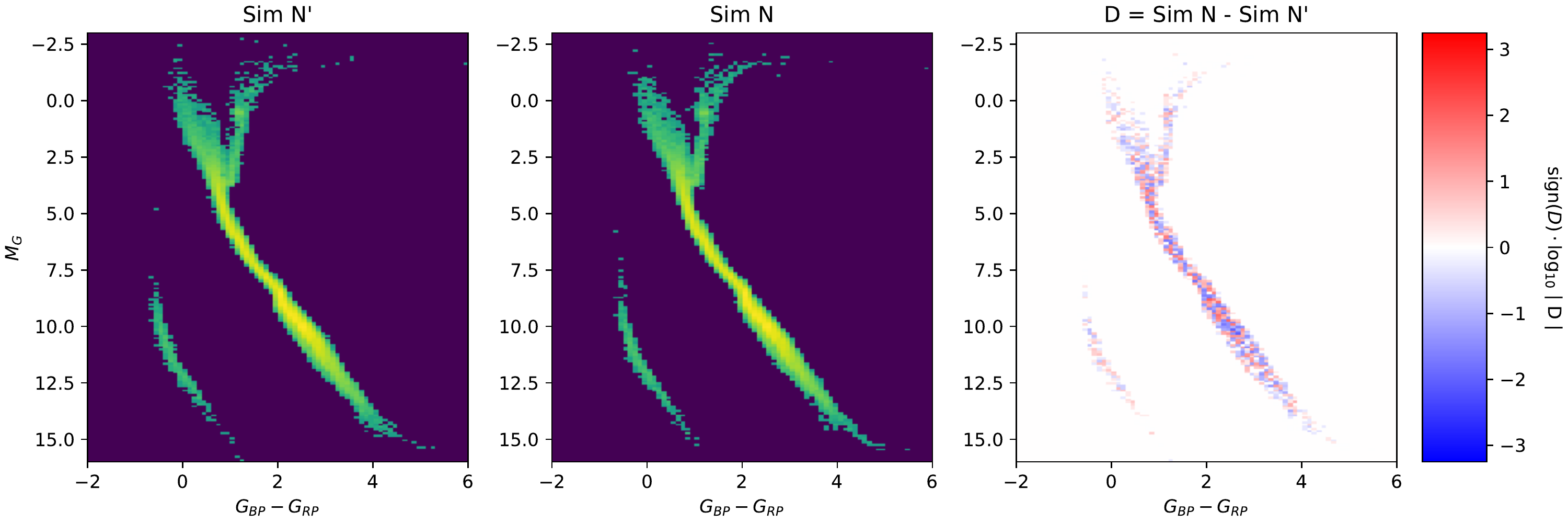}
    \caption{\label{fig:C2} ({\it top row}) CMDs resulting from the \simn\ and \simm\ simulations binned in 0.1 mag bins
    The rightmost panel shows the residual \simn\,-\,\simm, colour coded as indicated in the auxiliary axis.
    The residuals show an excess of stars on the red side of the MS, and a deficit on the blue side. 
    ({\it bottom row}) Same as {\it top row} but for two different simulations, \simn\ and {\it Sim N'}.
    The residuals show the flat behaviour characteristic of unbiased samples.
    The \simn\ and {\it Sim N'} simulations were computed with the $p50$ values of $a^{*}_i$ in Table\,\ref{tab:C1}.
    CMDs for $p10$ and $p90$ are shown in Figs.\,\ref{fig:D1} and \ref{fig:D2}.}
\end{center}
\end{figure*}

To test our results, we model the solar neighbourhood population using the values of $a_i$ derived in Section\,\ref{amd} as input to the MW.mx Galaxy model. 
We interpret the populations detected in Section\,\ref{amd} as belonging to the {\it thin disc}.
We map each of the eleven components per $Z$ for the no extinction solution listed in the full version of Table\,\ref{tab:4} (see Table\,\ref{tab:d3}) to one of the seven thin disc age groups used in MW.mx, defined in Table\,\ref{tab:B2}.
Table\,\ref{tab:C1} lists the resulting fractional weights $a^{*}_i$ for the percentiles $p10$, $p50$ (median) and $p90$. 
The number of stars with $G\leq15$ belonging to each group is then $a^{*}_i \times 120,452$ (cf. Section\,\ref{samp_sel_sol}).
We perform 10 Monte Carlo simulations of \solm\ for each set of $a^{*}_i$ in Table\,\ref{tab:C1} (\simn, hereafter) and another 10 simulations using the age groups as defined in the three rightmost columns of Table\,\ref{tab:B2} (\simm, hereafter).
In \simn\ the simulation is stopped when we reach the required number of stars with $G\leq15$ for each component, irrespective of the accumulated mass.
In contrast, in \simm\ the simulation is continued until we reach the total mass for each population, obtained by integration of $\rho(R,z)$ over the given volume, irrespective of the number of stars with $G\leq15$.

Fig.\,\ref{fig:8}, discussed in Section\,\ref{qc}, shows the average number counts in the $G$ band resulting from our simulations. 
Fig.\,\ref{fig:C1} shows the contribution to the number counts from each of the thin disk age groups. Stars younger than 2 Gyr (groups 1, 2, 3) account
for 85\% of the stars brighter than $G=6$ in \simm\ but only for 55\% of these stars in \simn. Group 7 contributes about 50\% of the stars fainter than $G=8$ in \simn,
but no group contributes more than 20\% in \simm. In Table\,\ref{tab:C2} we list relevant astrophysical properties of the solar neighbourhood derived from these simulations.
The stellar population of \sols\ is more numerous, less massive, and brighter according to \simm\ than to \simn, resulting in a lower $M/L$ ratio in the \simm\ simulation.
This is consistent with the brightness profile in Fig.\,\ref{fig:C1} which shows that the integrated magnitude of \sols\ is 1 magnitude brighter in \simm\ than in \simn.
Observations at the bright end of the same quality as {\it Gaia}'s are needed to judge in favour or against one of these scenarios.

\begin{figure*}
\begin{center}
    \includegraphics[width=\textwidth,height=0.30\textheight]{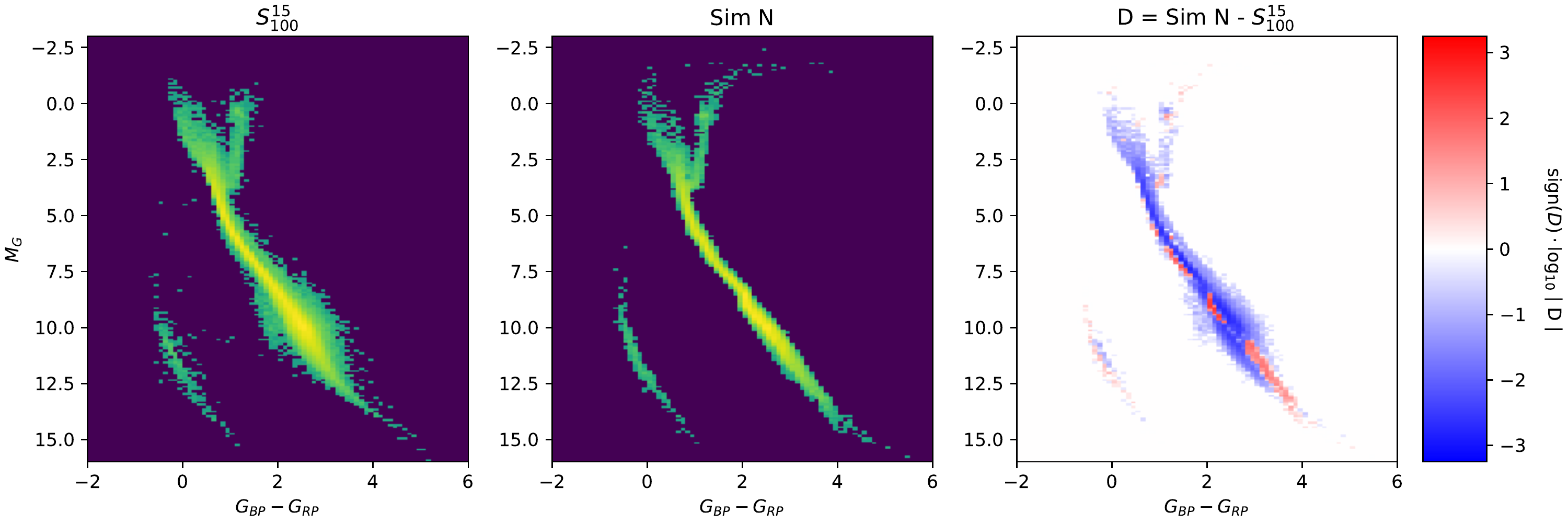}
    \includegraphics[width=\textwidth,height=0.30\textheight]{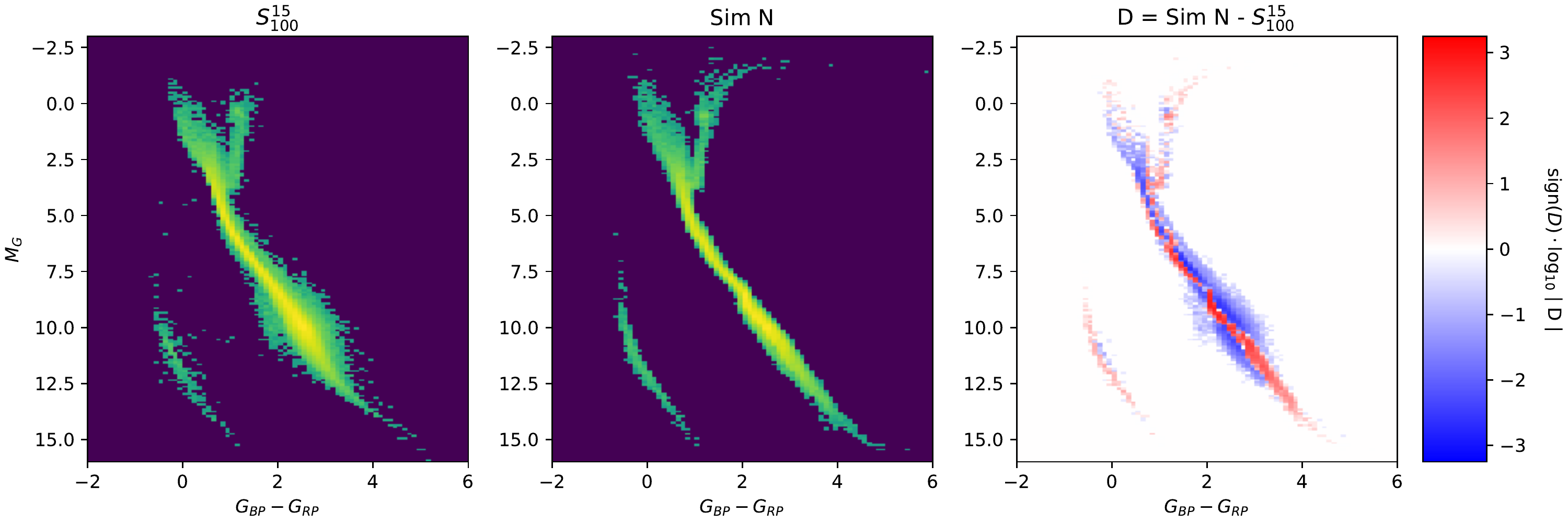}
    \includegraphics[width=\textwidth,height=0.30\textheight]{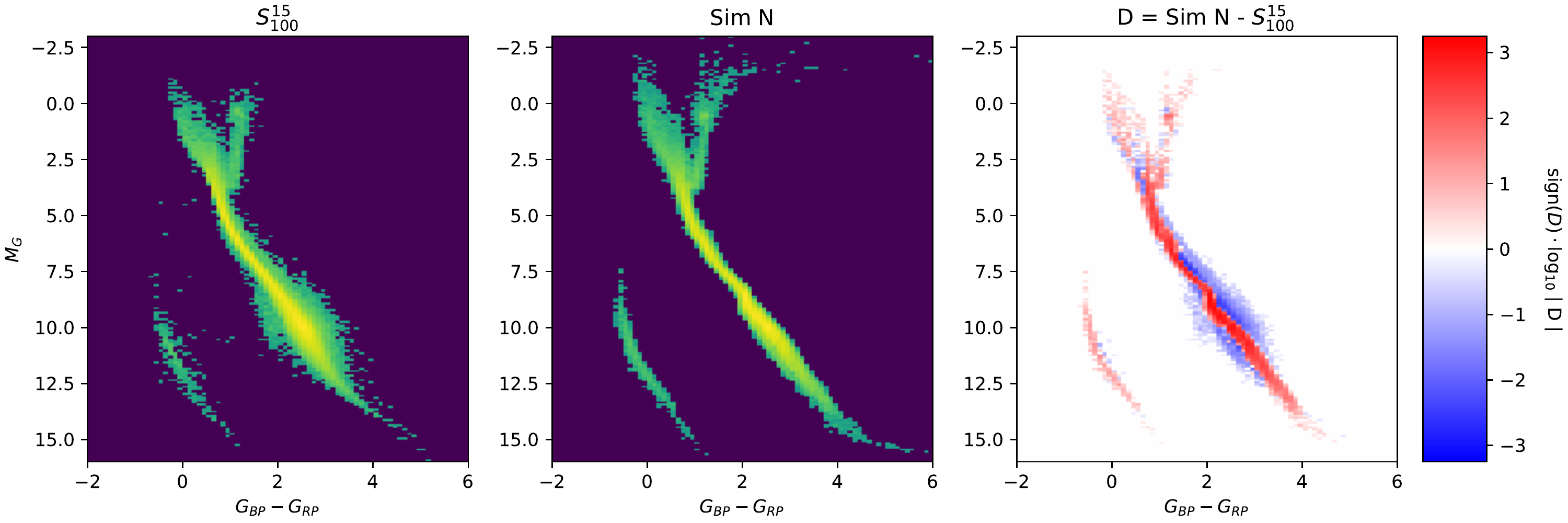}
    \caption{\label{fig:D3} \gtwo CMD of \solm compared to \simn. The CMDs are binned in 0.1 mag bins.
    The rightmost panel shows the residual \simn\,-\,\solm, colour coded as indicated in the auxiliary axis.
    The residuals show a deficit of stars on the red side of the MS and an excess on the blue side.
    ({\it Top, middle} and {\it bottom row}) \simn\ simulation computed with the $p10, p50$ and $p90$ values of $a^{*}_i$ in Table\,\ref{tab:C1}, respectively.}
\end{center}
\end{figure*}

\begin{figure*}
\begin{center}
    \includegraphics[width=\textwidth,height=0.30\textheight]{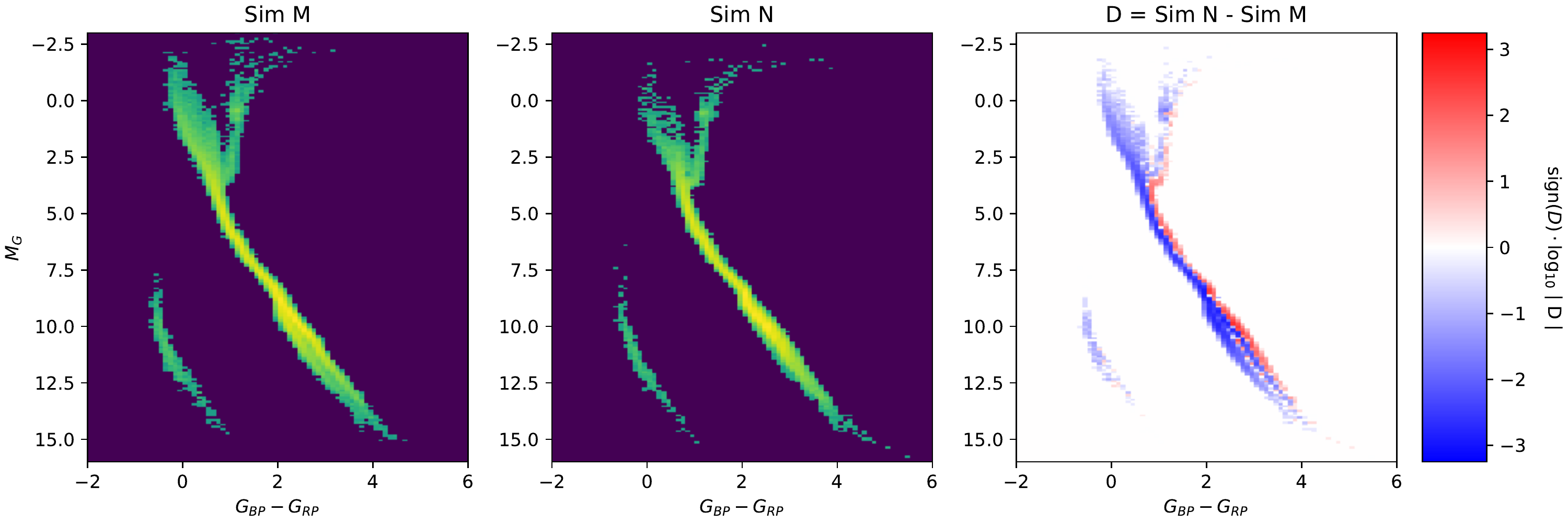}
    \includegraphics[width=\textwidth,height=0.30\textheight]{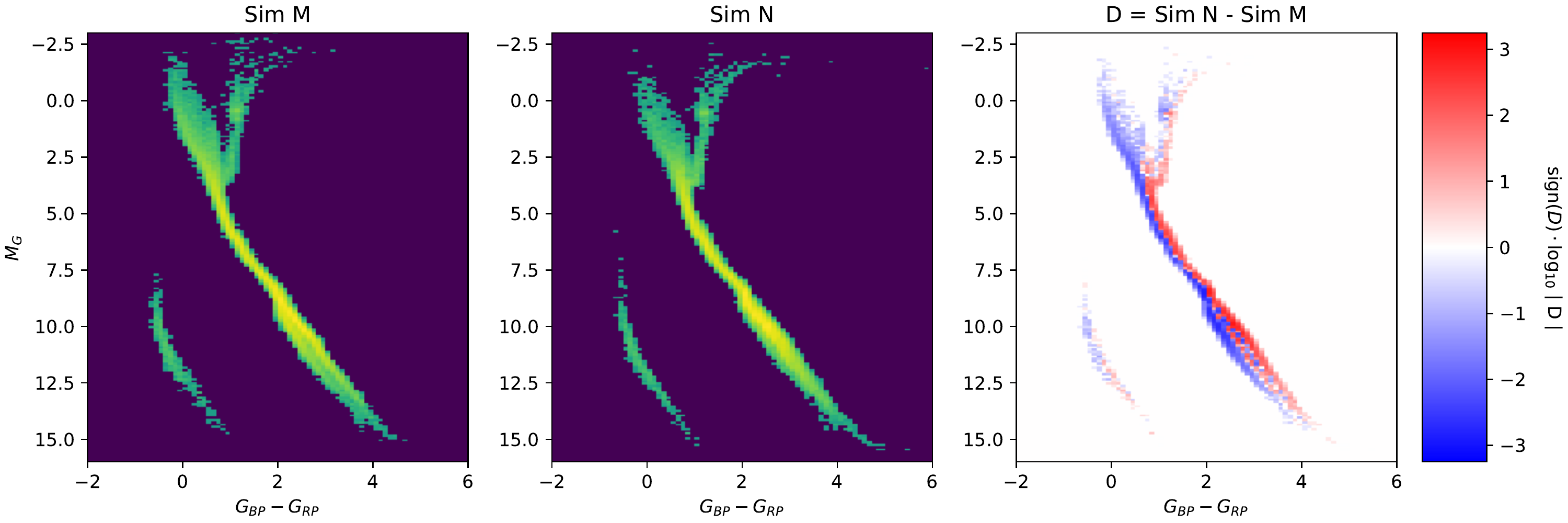}
    \includegraphics[width=\textwidth,height=0.30\textheight]{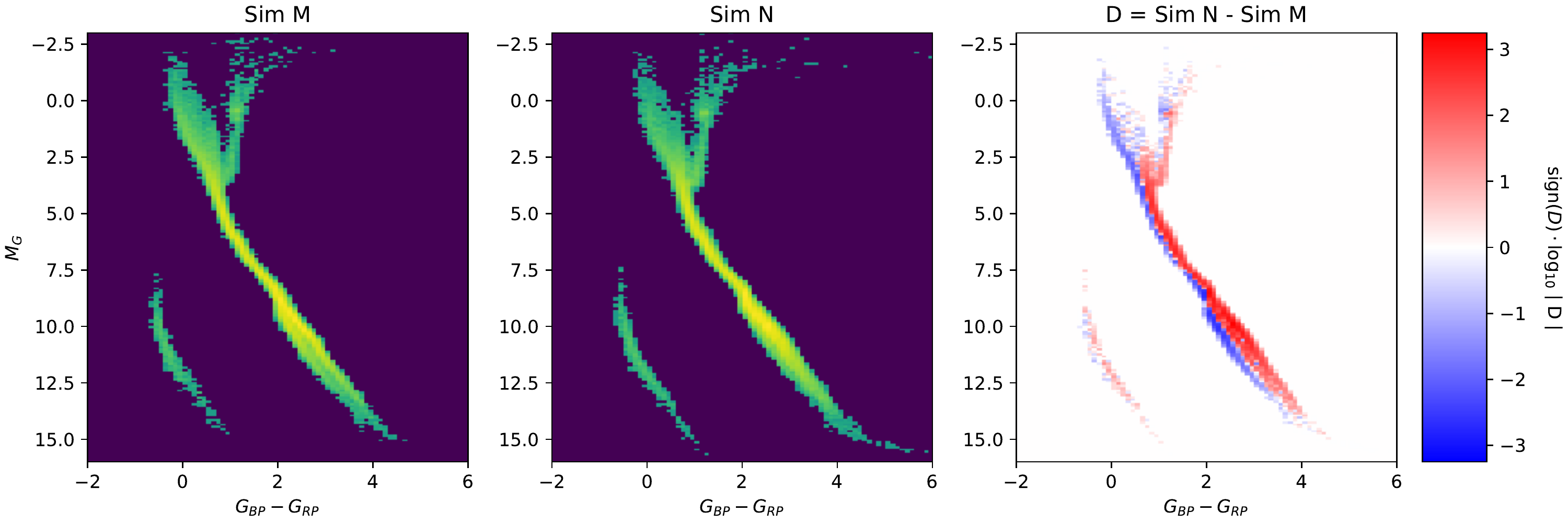}
    \caption{\label{fig:D1}CMDs from simulations \simm\ and \simn. The CMDs are binned in 0.1 mag bins.
    The rightmost panel shows the residual \simn\,-\,\simm, colour coded as indicated in the auxiliary axis.
    The residuals show an excess of stars on the red side of the MS and a deficit on the blue side.
    ({\it Top, middle} and {\it bottom row}) Simulations computed with the $p10, p50$ and $p90$ values of $a^{*}_i$ in Table\,\ref{tab:C1}, respectively}
\end{center}
\end{figure*}

\begin{figure*}
\begin{center}
    \includegraphics[width=\textwidth,height=0.30\textheight]{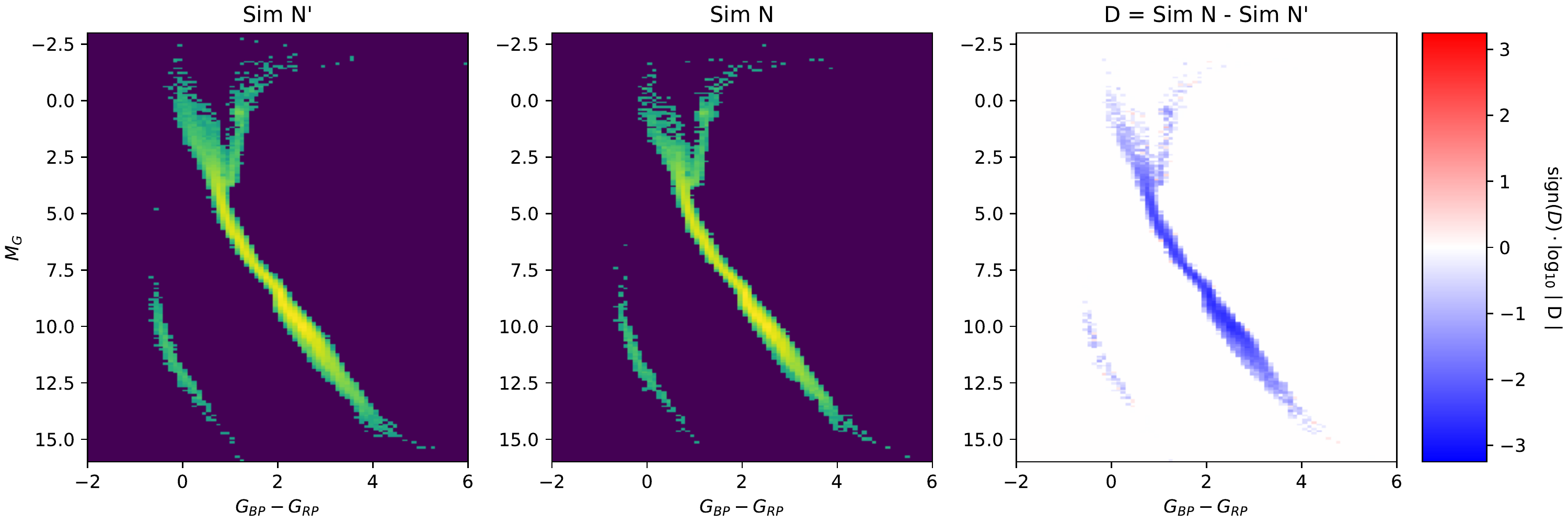}
    \includegraphics[width=\textwidth,height=0.30\textheight]{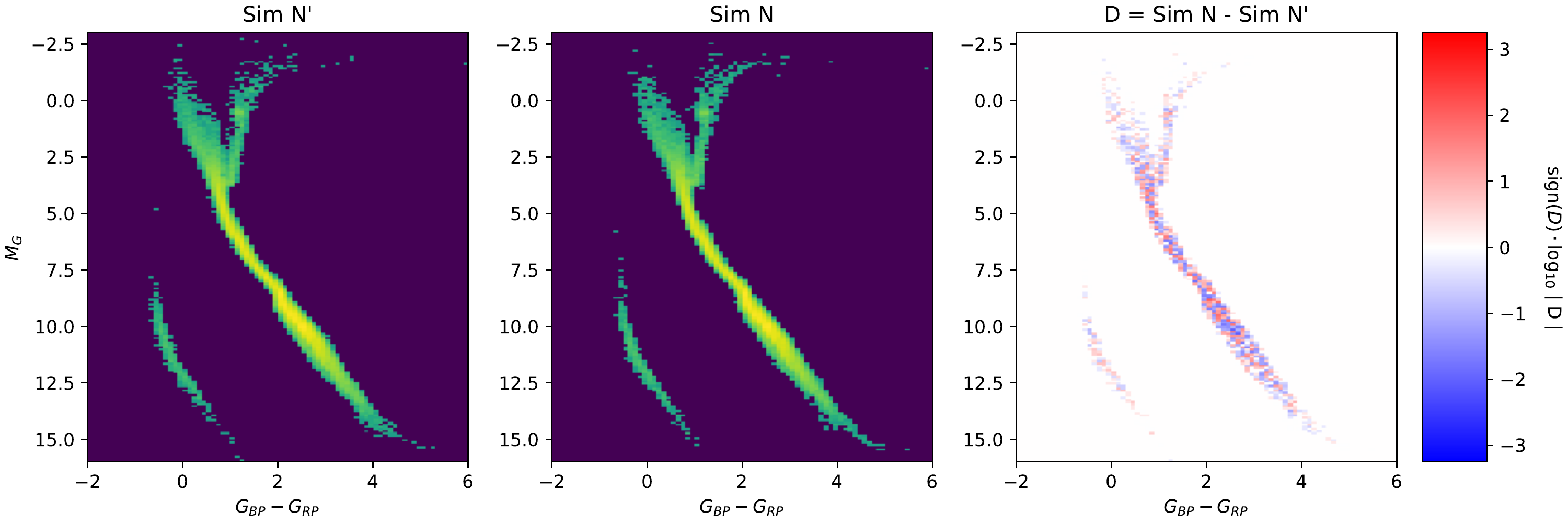}
    \includegraphics[width=\textwidth,height=0.30\textheight]{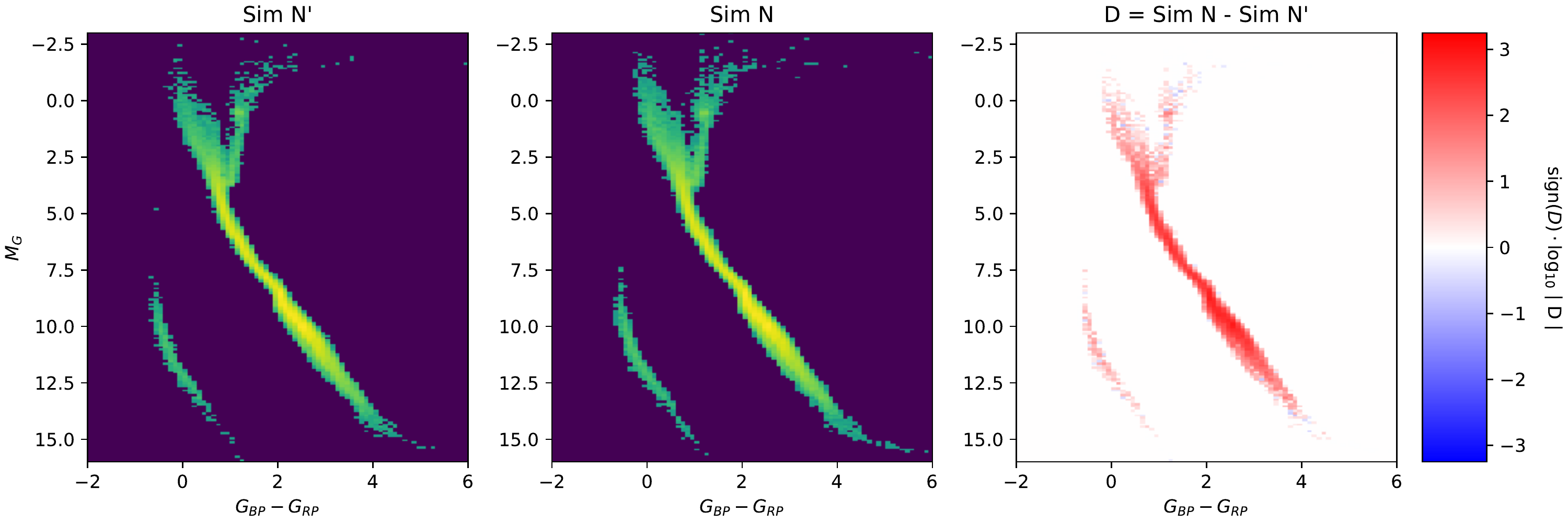}
    \caption{\label{fig:D2} CMDs from simulations \simn\ and \simn'. The CMDs are binned in 0.1 mag bins.
    The rightmost panel shows the residual \simn\,-\,\simn', colour coded as indicated in the auxiliary axis.
    The residuals show the flat behaviour characteristic of unbiased samples.
    ({\it Top, middle} and {\it bottom row}) Simulations computed with the $p10, p50$ and $p90$ values of $a^{*}_i$ in Table\,\ref{tab:C1}, respectively}
\end{center}
\end{figure*}

For completeness, in Fig.\,\ref{fig:C2} ({\it top row}) we compare the CMDs resulting from the \simn\ and \simm\ simulations. 
The residuals \simn\,-\,\simm\ show an excess of stars on the red side of the MS and a deficit on the blue side. 
In the {\it bottom row} we compare two \simn\ simulations to illustrate the flat residuals expected when comparing unbiased samples. 
Even though in \simn\ there are not enough stars on the red side of the single star MS to explain the observations, there are more of these stars in \simn\ than in \simm.
Figs.\,\ref{fig:9} and \ref{fig:C2} show CMDs for simulations computed with the $p50$ values of $a^{*}_i$ in Table\,\ref{tab:C1}.
CMDs for $p10$, $p50$ and $p90$ are shown in Figs.\,\ref{fig:D3},\,\ref{fig:D1} and\,\ref{fig:D2}.



\begin{table*}
  \begin{center}
  \caption{\label{tab:d1}Inference results for grid A (42 isochrones).}
  \renewcommand{\arraystretch}{1.70} 
  \begin{tabular}{ccccccccr}
  \hline
  Age     & \mc{7}{c}{$a_i \times 100$} \\
  (Gyr)   & \mc{1}{c}{Z=0.001}           & \mc{1}{c}{Z=0.002}           & \mc{1}{c}{Z=0.004}           & \mc{1}{c}{Z=0.008}           & \mc{1}{c}{Z=0.014}           & \mc{1}{c}{Z=0.017}            & \mc{1}{c}{Z=0.030}            & \mc{1}{c}{Total}   \\
  \hline
  0.2     & $0.0014_{-0.0012}^{+0.0031}$ & $0.0023_{-0.002 }^{+0.0052}$ & $0.0029_{-0.0024}^{+0.0064}$ & $0.0036_{-0.0031}^{+0.0083}$ & $0.0053 _{-0.0045}^{+0.0124}$ & $0.0069 _{-0.0058}^{+0.0163}$ & $2.9480_{-0.2054}^{+0.2116}$ &  2.9704   \\
  1.0     & $0.0018_{-0.0015}^{+0.0039}$ & $0.0036_{-0.0031}^{+0.0083}$ & $0.0049_{-0.0041}^{+0.0104}$ & $0.0075_{-0.0063}^{+0.0170}$ & $0.0139 _{-0.0118}^{+0.0318}$ & $0.0216 _{-0.0184}^{+0.0501}$ & $6.5276_{-0.4000}^{+0.4068}$ &  6.5809   \\
  2.0     & $0.0021_{-0.0018}^{+0.0044}$ & $0.0052_{-0.0044}^{+0.0115}$ & $0.0071_{-0.0060}^{+0.0140}$ & $0.0131_{-0.0111}^{+0.0298}$ & $0.0364 _{-0.0307}^{+0.0827}$ & $0.0351 _{-0.0296}^{+0.0774}$ & $9.1680_{-0.5109}^{+0.5188}$ &  9.2670   \\
  4.0     & $0.0026_{-0.0022}^{+0.0054}$ & $0.0076_{-0.0065}^{+0.0176}$ & $0.0073_{-0.0061}^{+0.0146}$ & $0.0152_{-0.0129}^{+0.0356}$ & $1.9902 _{-1.0210}^{+1.0688}$ & $7.5476 _{-1.5069}^{+1.4728}$ & $1.4354_{-0.5296}^{+0.5306}$ & 11.0059   \\
  6.0     & $0.0035_{-0.0029}^{+0.0068}$ & $0.0129_{-0.0108}^{+0.0269}$ & $0.0084_{-0.0070}^{+0.0164}$ & $0.0207_{-0.0175}^{+0.0466}$ & $0.0705 _{-0.0597}^{+0.1650}$ & $30.4803_{-1.3681}^{+1.3670}$ & $0.0292_{-0.0248}^{+0.0677}$ & 30.6255   \\
  13.0    & $0.0057_{-0.0047}^{+0.0094}$ & $0.0574_{-0.0333}^{+0.0287}$ & $0.0078_{-0.0066}^{+0.0155}$ & $2.2336_{-0.1338}^{+0.1287}$ & $36.9861_{-0.9941}^{+0.9665}$ & $0.0656 _{-0.0554}^{+0.1536}$ & $0.0076_{-0.0064}^{+0.0173}$ & 39.3638   \\
  \hline
   Total: & 0.0171                       & 0.0890                       & 0.00380                       & 2.2937                      &  39.1024                      &  38.1571                      &  20.1158                     & 99.8135   \\
  \hline
  \end{tabular}
  \end{center}
\end{table*}

\begin{table*}
  \begin{center}
  \caption{\label{tab:d2}Inference results for grid B (50 isochrones).}
  \renewcommand{\arraystretch}{1.70} 
  \begin{tabular}{ccccccr}
  \hline
  Age      & \mc{5}{c}{$a_i \times 100$} \\
  (Gyr)  &   \mc{1}{c}{Z=0.008}             & \mc{1}{c}{Z=0.010}               & \mc{1}{c}{Z=0.014}                & \mc{1}{c}{Z=0.017}                & \mc{1}{c}{Z=0.030}               & \mc{1}{c}{Total}  \\
  \hline
  0.2    &   $0.0072_{-0.0061}^{+0.0166}$   &   $0.0046_{-0.0039}^{+0.0108}$   &   $0.0059 _{-0.0050}^{+0.0139}$   &   $0.0081 _{-0.0069}^{+0.0189}$   &   $1.3049_{-0.2314}^{+0.2456}$   &    1.3307         \\
  0.5    &   $0.0094_{-0.0080}^{+0.0216}$   &   $0.0062_{-0.0053}^{+0.0146}$   &   $0.0091 _{-0.0077}^{+0.0211}$   &   $0.0142 _{-0.0121}^{+0.0330}$   &   $4.2322_{-0.4073}^{+0.4125}$   &    4.2711         \\
  1.0    &   $0.0384_{-0.0307}^{+0.0572}$   &   $0.0115_{-0.0097}^{+0.0267}$   &   $0.0211 _{-0.0179}^{+0.0484}$   &   $0.0290 _{-0.0247}^{+0.0653}$   &   $4.4716_{-0.4410}^{+0.4459}$   &    4.5716         \\
  1.6    &   $0.0674_{-0.0557}^{+0.1186}$   &   $0.0194_{-0.0165}^{+0.0449}$   &   $0.0515 _{-0.0438}^{+0.1151}$   &   $0.0461 _{-0.0389}^{+0.1019}$   &   $4.3435_{-0.6082}^{+0.6334}$   &    4.5279         \\
  2.0    &   $0.0296_{-0.0250}^{+0.0664}$   &   $0.0175_{-0.0148}^{+0.0400}$   &   $0.0573 _{-0.0485}^{+0.1285}$   &   $0.0980 _{-0.0825}^{+0.2146}$   &   $4.7836_{-0.6573}^{+0.6522}$   &    4.9860         \\
  3.0    &   $0.0453_{-0.0380}^{+0.0995}$   &   $0.0262_{-0.0222}^{+0.0607}$   &   $0.7648 _{-0.6264}^{+1.1614}$   &   $9.1772 _{-1.4153}^{+1.1579}$   &   $0.1039_{-0.0872}^{+0.2163}$   &   10.1174         \\
  5.0    &   $0.4052_{-0.3263}^{+0.4943}$   &   $0.0480_{-0.0407}^{+0.1099}$   &   $0.0763 _{-0.0647}^{+0.1776}$   &   $15.2982_{-1.5818}^{+1.5821}$   &   $0.0180_{-0.0152}^{+0.0419}$   &   15.8457         \\
  7.0    &   $0.5429_{-0.4286}^{+0.5377}$   &   $0.0995_{-0.0845}^{+0.2264}$   &   $8.9117 _{-2.7139}^{+2.7240}$   &   $10.3234_{-2.7615}^{+2.7759}$   &   $0.0140_{-0.0118}^{+0.0321}$   &   19.8915         \\
  10.0   &   $0.3642_{-0.3007}^{+0.5538}$   &   $1.7058_{-0.9989}^{+0.7116}$   &   $19.6952_{-3.0896}^{+2.9857}$   &   $9.4716 _{-2.5909}^{+2.5881}$   &   $0.0083_{-0.0070}^{+0.0192}$   &   31.2451         \\
  13.0   &   $0.1015_{-0.0856}^{+0.2280}$   &   $0.5889_{-0.4949}^{+1.0099}$   &   $1.4178 _{-1.1560}^{+2.0899}$   &   $0.0523 _{-0.0442}^{+0.1208}$   &   $0.0074_{-0.0062}^{+0.0172}$   &    2.1679         \\
  \hline
  Total: &  1.6111                          &    2.5276                        &   31.0107                         &    44.5181                        &   19.2874                        &   98.9549         \\
  \hline
  \end{tabular}
  \end{center}
\end{table*}

\begin{table*}
  \begin{center}
  \caption{\label{tab:d3}Inference results for grid C (44 isochrones).}
  \renewcommand{\arraystretch}{1.70} 
  \setlength{\tabcolsep}{12pt}       
  \begin{tabular}{cccccr}
           & \mc{4}{c}{No extinction correction} \\
  \hline
  Age      & \mc{4}{c}{$a_i \times 100$} \\
  (Gyr)    & \mc{1}{c}{Z=0.010}            & \mc{1}{c}{Z=0.014}              & \mc{1}{c}{Z=0.017}               & \mc{1}{c}{Z=0.030}            & \mc{1}{c}{Total} \\
  \hline
   0.1     & $0.0047_{-0.0040}^{+0.0109}$  &  $0.0050 _{-0.0042}^{+0.0118}$  &  $0.0067 _{-0.0057}^{+0.0154}$  &  $0.1529_{-0.1201}^{+0.1681}$  &   0.1693         \\
   0.2     & $0.0056_{-0.0047}^{+0.0129}$  &  $0.0059 _{-0.0050}^{+0.0139}$  &  $0.0083 _{-0.0070}^{+0.0192}$  &  $0.2899_{-0.2248}^{+0.2971}$  &   0.3097         \\
   0.5     & $0.0070_{-0.0059}^{+0.0166}$  &  $0.0076 _{-0.0064}^{+0.0175}$  &  $0.0125 _{-0.0107}^{+0.0288}$  &  $4.4201_{-0.3628}^{+0.3457}$  &   4.4472         \\
   1.0     & $0.0959_{-0.0574}^{+0.0815}$  &  $0.0337 _{-0.0285}^{+0.0734}$  &  $0.0401 _{-0.0340}^{+0.0909}$  &  $5.0174_{-0.4342}^{+0.4453}$  &   5.1871         \\
   1.7     & $0.0422_{-0.0358}^{+0.0947}$  &  $0.0450 _{-0.0381}^{+0.0984}$  &  $0.3496 _{-0.2790}^{+0.4443}$  &  $5.9293_{-0.5933}^{+0.5905}$  &   6.3661         \\
   2.5     & $0.0726_{-0.0610}^{+0.1609}$  &  $0.1501 _{-0.1273}^{+0.3463}$  &  $3.8322 _{-0.7885}^{+0.7647}$  &  $2.9486_{-0.5671}^{+0.5956}$  &   7.0035         \\
   4.0     & $0.0791_{-0.0673}^{+0.1796}$  &  $0.6118 _{-0.5094}^{+1.0409}$  &  $3.4156 _{-1.3235}^{+1.2782}$  &  $0.0214_{-0.0182}^{+0.0489}$  &   4.1279         \\
   4.8     & $0.1266_{-0.1068}^{+0.2841}$  &  $0.0800 _{-0.0680}^{+0.1853}$  &  $18.679 _{-1.6871}^{+1.7001}$  &  $0.0153_{-0.0130}^{+0.0351}$  &  18.9009         \\
   6.5     & $0.2182_{-0.1839}^{+0.4694}$  &  $0.1369 _{-0.1162}^{+0.3092}$  &  $13.3578_{-1.9682}^{+1.9719}$  &  $0.0128_{-0.0109}^{+0.0297}$  &  13.7257         \\
   8.0     & $0.3547_{-0.3000}^{+0.7379}$  &  $1.5769 _{-1.2617}^{+2.0533}$  &  $5.2775 _{-2.2878}^{+2.2301}$  &  $0.0104_{-0.0088}^{+0.0248}$  &   7.2195         \\
   10.0    & $6.0980_{-0.9025}^{+0.6392}$  &  $19.7228_{-2.1539}^{+1.8542}$  &  $5.7568 _{-2.2257}^{+2.3502}$  &  $0.0079_{-0.0067}^{+0.0186}$  &  31.5855         \\
  \hline
   Total:  & 7.1046                        & 22.3757                         & 50.7361                         & 18.826                         &  99.0424         \\
  \hline
  \\ \\ \\
             \mc{6}{c}{Extinction correction using the Stilism tool$^a$} \\
  \hline
  Age      & \mc{4}{c}{$a_i \times 100$} \\
  (Gyr)    & \mc{1}{c}{Z=0.010}            & \mc{1}{c}{Z=0.014}              & \mc{1}{c}{Z=0.017}               & \mc{1}{c}{Z=0.030}            & \mc{1}{c}{Total} \\
  \hline
   0.1     & $0.0050_{-0.0043}^{+0.0113}$  &  $0.0052 _{-0.0044}^{+0.0125}$  &  $0.0072 _{-0.0061}^{+0.0166}$  &  $0.2053_{-0.1527}^{+0.1871}$  &   0.2227         \\
   0.2     & $0.0058_{-0.0049}^{+0.0138}$  &  $0.0062 _{-0.0052}^{+0.0145}$  &  $0.0087 _{-0.0074}^{+0.0203}$  &  $0.3171_{-0.2495}^{+0.3411}$  &   0.3378         \\
   0.5     & $0.0075_{-0.0063}^{+0.0168}$  &  $0.0080 _{-0.0068}^{+0.0184}$  &  $0.0135 _{-0.0114}^{+0.0303}$  &  $4.5449_{-0.3870}^{+0.3681}$  &   4.5739         \\
   1.0     & $0.1056_{-0.0622}^{+0.0883}$  &  $0.0357 _{-0.0302}^{+0.0793}$  &  $0.0441 _{-0.0372}^{+0.0969}$  &  $5.2737_{-0.4669}^{+0.4634}$  &   5.4591         \\
   1.7     & $0.0479_{-0.0406}^{+0.1026}$  &  $0.0480 _{-0.0405}^{+0.1052}$  &  $0.5494 _{-0.4011}^{+0.5255}$  &  $5.6663_{-0.5996}^{+0.6092}$  &   6.3116         \\
   2.5     & $0.0805_{-0.0679}^{+0.1734}$  &  $0.1503 _{-0.1272}^{+0.3402}$  &  $5.0367 _{-0.8244}^{+0.8053}$  &  $2.0529_{-0.5869}^{+0.6133}$  &   7.3204         \\
   4.0     & $0.0870_{-0.0734}^{+0.1923}$  &  $0.7358 _{-0.6042}^{+1.1397}$  &  $3.8083 _{-1.4189}^{+1.3415}$  &  $0.0212_{-0.0181}^{+0.0490}$  &   4.6523         \\
   4.8     & $0.1309_{-0.1105}^{+0.2886}$  &  $0.0832 _{-0.0703}^{+0.1931}$  &  $19.7555_{-1.7915}^{+1.7711}$  &  $0.0154_{-0.0130}^{+0.0355}$  &   19.9850        \\
   6.5     & $0.2582_{-0.2178}^{+0.5401}$  &  $0.1497 _{-0.1272}^{+0.3461}$  &  $13.8344_{-1.9959}^{+2.0057}$  &  $0.0131_{-0.0111}^{+0.0308}$  &   14.2554        \\
   8.0     & $0.3737_{-0.3151}^{+0.7690}$  &  $2.5154 _{-1.7888}^{+2.2383}$  &  $5.4309 _{-2.2342}^{+2.1144}$  &  $0.0108_{-0.0092}^{+0.0254}$  &   8.3308         \\
   10.0    & $6.1622_{-0.9599}^{+0.6903}$  &  $19.6321_{-2.1886}^{+1.8906}$  &  $1.7232 _{-1.3614}^{+2.1215}$  &  $0.0082_{-0.0070}^{+0.0187}$  &  27.5257         \\
  \hline
  Total:  & 7.2643                         & 23.3696                         & 50.2119                         & 18.1289                        &  98.9747         \\
  \hline
  \mc{6}{l}{$^a${\citet[][\url{https://stilism.obspm.fr}]{lallement19}}}\\
  \end{tabular}
  \end{center}
\end{table*}

\bsp	
\label{lastpage}
\end{document}